\documentclass[seceqn]{elsart}
\journal{Annals of Physics}

\usepackage{graphicx}
\usepackage{amsmath}
\usepackage{amssymb}
\usepackage{citesort}
\newcommand{\<}[1]{\hspace{-0.11111em}#1\hspace{-0.11111em}}
\newcommand{\rtrim}[1]{#1\hspace{-0.11111em}}

\DeclareRobustCommand{\abs}[1]{\lvert#1\rvert}
\DeclareRobustCommand{\scrE}{\mathcal{E}}
\DeclareRobustCommand{\scrH}{\mathcal{H}}
\DeclareRobustCommand{\rootstinline}{\sqrt{7}/2}

\DeclareRobustCommand{\Lhat}{\hat{L}}
\DeclareRobustCommand{\Mhat}{\hat{M}}
\DeclareRobustCommand{\Nhat}{\hat{N}}
\DeclareRobustCommand{\nhat}{\hat{n}}
\DeclareRobustCommand{\Qhat}{\hat{Q}}

\DeclareRobustCommand{\grp}[1]{\mathrm{#1}}

\DeclareRobustCommand{\grpusix}{\grp{U}(6)}
\DeclareRobustCommand{\grpusixp}{\grp{U}_{\pi}(6)}
\DeclareRobustCommand{\grpusixn}{\grp{U}_{\nu}(6)}

\DeclareRobustCommand{\grpufivep}{\grp{U}_{\pi}(5)}
\DeclareRobustCommand{\grpufiven}{\grp{U}_{\nu}(5)}
\DeclareRobustCommand{\grpufivepn}{\grp{U}_{\pi\nu}(5)}

\DeclareRobustCommand{\grpsosix}{\grp{SO}(6)}
\DeclareRobustCommand{\grpsosixp}{\grp{SO}_{\pi}(6)}
\DeclareRobustCommand{\grpsosixn}{\grp{SO}_{\nu}(6)}
\DeclareRobustCommand{\grpsosixpn}{\grp{SO}_{\pi\nu}(6)}

\DeclareRobustCommand{\grpsuthree}{\grp{SU}(3)}
\DeclareRobustCommand{\grpsuthreep}{\grp{SU}_{\pi}(3)}
\DeclareRobustCommand{\grpsuthreen}{\grp{SU}_{\nu}(3)}
\DeclareRobustCommand{\grpsuthreepn}{\grp{SU}_{\pi\nu}(3)}

\DeclareRobustCommand{\grpsuthreenbar}{\overline{\grp{SU}_{\nu}(3)}}
\DeclareRobustCommand{\grpsuthreepnbar}{\overline{\grp{SU}_{\pi\nu}(3)}}

\DeclareRobustCommand{\grpsuthreepnstar}{\grp{SU}_{\pi\nu}^*(3)}
\DeclareRobustCommand{\grpsuthreepnstarbar}{\overline{\grp{SU}_{\pi\nu}^*(3)}}

\DeclareRobustCommand{\grpsothree}{\grp{SO}(3)}

\DeclareRobustCommand{\grpsothreepn}{\grp{SO}_{\pi\nu}(3)}

\DeclareRobustCommand{\betap}{\beta_{\pi}}
\DeclareRobustCommand{\betan}{\beta_{\nu}}
\DeclareRobustCommand{\betar}{\beta_{\rho}}
\DeclareRobustCommand{\gammap}{\gamma_{\pi}}
\DeclareRobustCommand{\gamman}{\gamma_{\nu}}
\DeclareRobustCommand{\gammar}{\gamma_{\rho}}

\DeclareRobustCommand{\Np}{N_{\pi}}
\DeclareRobustCommand{\Nn}{N_{\nu}}
\DeclareRobustCommand{\Nr}{N_{\rho}}
\DeclareRobustCommand{\epsilonp}{\varepsilon_{\pi}}
\DeclareRobustCommand{\epsilonn}{\varepsilon_{\nu}}
\DeclareRobustCommand{\epsilonr}{\varepsilon_{\rho}}
\DeclareRobustCommand{\ep}{e'_{\pi}}  
\DeclareRobustCommand{\en}{e'_{\nu}}
\DeclareRobustCommand{\er}{e'_{\rho}}
\DeclareRobustCommand{\kappapp}{\kappa_{\pi\pi}}
\DeclareRobustCommand{\kappapn}{\kappa_{\pi\nu}}
\DeclareRobustCommand{\kappann}{\kappa_{\nu\nu}}
\DeclareRobustCommand{\kapparr}{\kappa_{\rho\rho}}
\DeclareRobustCommand{\kapparrp}{\kappa_{\rho\rho'}}
\DeclareRobustCommand{\kpp}{k'_{\pi\pi}}
\DeclareRobustCommand{\kpn}{k'_{\pi\nu}}
\DeclareRobustCommand{\knn}{k'_{\nu\nu}}

\DeclareRobustCommand{\krrp}{k'_{\rho\rho'}}
\DeclareRobustCommand{\chip}{{\chi_{\pi}}}
\DeclareRobustCommand{\chin}{{\chi_{\nu}}}
\DeclareRobustCommand{\chir}{{\chi_{\rho}}}
\DeclareRobustCommand{\chis}{{\chi_{S}}}
\DeclareRobustCommand{\chiv}{{\chi_{V}}}
\DeclareRobustCommand{\ap}{a_{\pi}}
\DeclareRobustCommand{\an}{a_{\nu}}
\DeclareRobustCommand{\ar}{a_{\rho}}

\DeclareRobustCommand{\esqrbsqr}{e^2\mathrm{b}^2}

\begin{document}

\begin{frontmatter}

\title{Phase structure of a two-fluid bosonic system}

\author{M. A. Caprio}
and
\author{F. Iachello}
\address{Center for Theoretical Physics, Sloane Physics Laboratory, 
Yale University, New Haven, Connecticut 06520-8120, USA}

\begin{abstract}
The phase diagram of a two-fluid bosonic system is investigated.
The proton-neutron interacting boson model (IBM-2) possesses a rich
phase structure involving three control parameters and multiple order
parameters.  The surfaces of quantum phase transition between
spherical, axially-symmetric deformed, and $\grpsuthreepnstar$ triaxial
phases are determined, and the evolution of classical equilibrium
properties across these transitions is investigated.  Spectroscopic
observables are considered in relation to the phase diagram.
\end{abstract}


\end{frontmatter}

\section{Introduction}
\label{secintro}

The phase structure of quantum many-body systems has in recent years
been a subject of great experimental and theoretical interest.  Models
based upon algebraic Hamiltonians are well-suited to the study of
phase transitions.  They possess a well-defined classical
limit~\cite{gilmore1978:coherent}, allowing classical order parameters
to be determined.  And for certain specific forms of their Hamiltonians, 
algebraic models exhibit dynamical symmetries, which
correspond to qualitatively distinct ground-state equilibrium
configurations.  These constitute the phases of the
system~\cite{feng1981:ibm-phase}.  Algebraic models have found
extensive application to the spectroscopy of many-body systems,
including nuclei~\cite{iachello1987:ibm} and
molecules~\cite{iachello1995:vibron}.

In the present work, the phase structure of a system comprised of two
interacting fluids is investigated.  The phase structure of one-fluid
algebraic models, especially the interacting boson
model~(IBM)~\cite{iachello1987:ibm} for nuclei, has been studied in
detail~\cite{dieperink1980:ibm-classical,feng1981:ibm-phase}.
However, algebraic models may also be used to describe multi-fluid
systems, with multiple interacting constituent species.  While
one-fluid systems are described by a single elementary Lie algebra,
usually $\grp{U}(n)$, multi-fluid systems are described by a coupling
of such Lie algebras,
$\grp{U}_1(n)\otimes\grp{U}_2(n)\otimes\cdots$~\cite{iachello1987:ibm,iachello1995:vibron}.
A richer phase structure arises for multi-fluid models, not just from
the greater number of control and order parameters afforded by the
more complicated model, but more fundamentally from the coupling of
multiple subsystems, each of which can exist in a different phase or
can drive phase transitions of the composite system.

Here we consider the proton-neutron interacting boson
model~(IBM-2)~\cite{arima1977:ibm2-shell,otsuka1978:ibm2-shell,otsuka1978:ibm2-shell-details,iachello1987:ibm},
in which proton pairs and neutron pairs are treated as distinct
constituents.  The one-fluid IBM, with $\grpusix$ algebraic structure,
has three dynamical symmetries, separated by first and second order
phase
transitions~\cite{dieperink1980:ibm-classical,feng1981:ibm-phase}.
The IBM-2, with $\grpusix\otimes\grpusix$ algebraic structure,
supports four dynamical
symmetries~\cite{vanisacker1986:ibm2-limits,dieperink1982:ibm2-triax}.
Its phase diagram is therefore more involved and is found to possess
qualitatively new features.  Due to the complexity of the problem, a
combination of analytic and numerical methods have been applied in the
present work.  Preliminary results were reported in
Refs.~\cite{caprio2004:ibmpn-icnpls04,caprio2004:ibmpn}.  Numerical
studies of the IBM-2 phase structure have also been carried out by
Arias, Dukelsky, and
Garc\'ia-Ramos~\cite{arias2004:ibmpn-icnpls04,arias2004:ibmpn}.

The IBM-2 and its classical limit are summarized in
Sec.~\ref{secdefs}.  The phase diagram of the IBM-2 is first
investigated for an essential Hamiltonian with few parameters, for
which the most complete analytic results are obtained
(Sec.~\ref{secschematic}).  The treatment is then extended to a more
general Hamiltonian, incorporating realistic quadrupole and Majorana
interactions (Sec.~\ref{secgeneral}).  Connection of the IBM-2 phase
diagram with experimental data requires knowledge of the spectroscopic
predictions across the phase transitions.  These are addressed in
Sec.~\ref{secspectro}.  

The results of Sec.~\ref{secspectro} are of relevance in the search
for triaxial shapes in nuclei.  Specific signatures of two-fluid
triaxial deformation, and of the phase transition between axially
symmetric and triaxial structure, are presented.  Those signatures
involving low-lying states are applicable both to current experiments
and to experiments planned for next generation radioactive beam
facilities. Those involving the high-lying magnetic dipole mode might
be most directly investigated through resonance fluorescence
experiments at high intensity gamma-ray facilities.

\section{IBM-2 definition and classical limit}
\label{secdefs}

\subsection{Hamiltonian}
\label{subsecmodel}

Let us first summarize the IBM-2 Hamiltonian and the dynamical
symmetries it supports.  Operators in the IBM-2 are constructed from
the generators of the group $\grpusixp\otimes\grpusixn$, realized in
terms of the boson creation operators $s_{\rho,0}^{\dag}$ and
$d_{\rho,\mu}^{\dag}$ (where $\rho$ represents $\pi$ or $\nu$, and
$\mu$$=$$-2,\ldots,2$) and their associated annihilation operators,
acting on a basis of good boson numbers $N_\pi$ and $N_\nu$.  The
physically dominant interactions are contained in a Hamiltonian
\begin{equation}
\label{eqnHgeneralraw}
H =
\varepsilon_\pi \nhat_{d\pi}+\varepsilon_\nu\nhat_{d\nu}
+\kappa_{\pi\pi}\Qhat_\pi^{\chi_\pi}\cdot\Qhat_\pi^{\chi_\pi}
+\kappa_{\pi\nu}\Qhat_\pi^{\chi_\pi}\cdot\Qhat_\nu^{\chi_\nu}
+\kappa_{\nu\nu}\Qhat_\nu^{\chi_\nu}\cdot\Qhat_\nu^{\chi_\nu} +\lambda
\Mhat,
\end{equation}
where $\nhat_{d\rho}\<\equiv d_\rho^{\dag}\<\cdot\tilde{d}_\rho$,
$\Qhat_\rho^{\chi_\rho}\<\equiv(s_\rho^{\dag}\<\times\tilde{d}_\rho +
d_\rho^{\dag}\<\times\tilde{s}_\rho)^{(2)} +
\chi_\rho(d_\rho^{\dag}\<\times\tilde{d}_\rho)^{(2)}$, 
$\Mhat\<\equiv-2\sum_{k=1,3}(d_\pi^\dag\times
d_\nu^\dag)^{(k)}\cdot(\tilde{d}_\pi\times \tilde{d}_\nu)^{(k)}
+(s_\pi^\dag\times d_\nu^\dag - s_\nu^\dag\times
d_\pi^\dag)^{(2)}\cdot(\tilde{s}_\pi\times \tilde{d}_\nu -
\tilde{s}_\nu\times
\tilde{d}_\pi)^{(2)}$, 
and conventional spherical tensor coupling notation (\textit{e.g.},
Ref.~\cite{deshalit1963:shell}) has been used.  This Hamiltonian
contains one-fluid contributions, arising from like-nucleon pairing
($\nhat_{d\rho}$) and quadrupole ($\Qhat_\rho\cdot
\Qhat_\rho$) interactions, as well as two-fluid coupling terms, arising from 
proton-neutron quadrupole ($\Qhat_\pi\cdot\Qhat_\nu$) and Majorana
($\Mhat$) interactions.  The physically relevant ranges of the
Hamiltonian parameters are $\varepsilon_\rho\<\geq0$,
$\kappa_{\rho\rho'}\<\leq0$,
$-\sqrt{7}/2\<\leq\chi_\rho\<\leq\sqrt{7}/2$, and
$\lambda\geq0$~\cite{iachello1987:ibm}.

A dynamical symmetry occurs when, for certain values of the
parameters, the Hamiltonian is constructed from the Casimir operators
of a chain of subalgebras of $\grpusixp\otimes\grpusixn$.  Three of the IBM-2 dynamical symmetries
occur for $\chi_\pi\<=\chi_\nu$ and have direct analogues in the
one-fluid IBM~\cite{vanisacker1986:ibm2-limits}.  When
$\kappapp\<=\kappapn\<=\kappann\<=0$, the $\grpufivepn$ dynamical
symmetry
is realized, with subalgebra chain
\begin{equation}
\grpusixp\otimes\grpusixn
\supset\grpufivep\otimes\grpufiven
\supset\grpufivepn
\supset\grpsothreepn.
\end{equation}
The geometric interpretation is that the proton and neutron
fluids undergo oscillations about a spherical equilibrium configuration.  Parameter values
$\epsilonp\<=\epsilonn\<=0$ with $\chi_\pi\<=\chi_\nu\<=0$ produce the
$\grpsosixpn$ dynamical symmetry
\begin{equation}
\grpusixp\otimes\grpusixn
\supset\grpsosixp\otimes\grpsosixn
\supset\grpsosixpn
\supset\grpsothreepn,
\end{equation}
yielding deformed, $\gamma$-unstable structure.  And 
$\epsilonp\<=\epsilonn\<=0$ with $\chi_\pi\<=\chi_\nu\<=-\rootstinline$
gives the $\grpsuthreepn$ dynamical symmetry
\begin{equation}
\grpusixp\otimes\grpusixn
\supset\grpsuthreep\otimes\grpsuthreen
\supset\grpsuthreepn
\supset\grpsothreepn,
\end{equation}
for which prolate axially symmetric structure is obtained.  The
complementary case $\chi_\pi\<=\chi_\nu\<=+\rootstinline$, giving
oblate axially symmetric structure, is distinguished by the notation
$\grpsuthreepnbar$.

\begin{figure}
\begin{center}
\includegraphics[width=0.5\hsize]{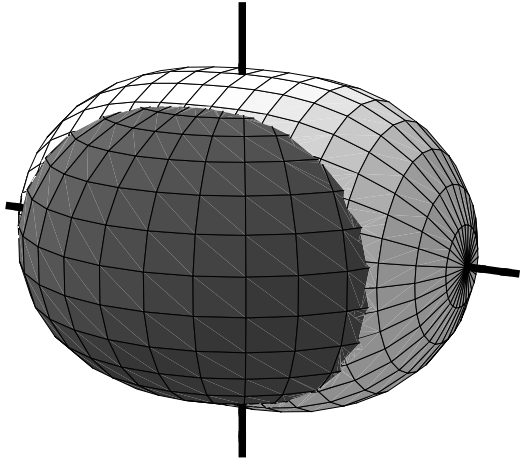}
\end{center}
\caption
{Geometrical interpretation of the equilibrium structure for the
$\grpsuthreepnstar$ dynamical symmetry.  A prolate deformed proton fluid
(light gray) and oblate deformed neutron fluid (dark gray) are coupled
with orthogonal symmetry axes.  Figure adapted from
Ref.~\cite{caprio2005:ibmpn-nmp04}.
\label{figortho}
}
\end{figure}

However, a symmetry special to the IBM-2, denoted $\grpsuthreepnstar$, is
obtained for $\epsilonp\<=\epsilonn\<=0$ with $\chi_\pi\<=-\rootstinline$ and
$\chi_\nu\<=+\rootstinline$~\cite{dieperink1982:ibm2-triax,dieperink1983:ibm2-su3star-operators,walet1987:ibm2-su3star-details,sevrin1987:ibm2-su3star-1,sevrin1987:ibm2-su3star-2}.
In this case the Hamiltonian is constructed from Casimir operators of
the subalgebra chain
\begin{equation}
\label{eqnchainsu3star}
\grpusixp\otimes\grpusixn
\supset\grpsuthreep\otimes\grpsuthreenbar
\supset\grpsuthreepnstar
\supset\grpsothreepn.
\end{equation}
The equilibrium configuration consists of a proton fluid with axially
symmetric prolate deformation coupled to a neutron fluid with axially
symmetric oblate deformation, with their symmetry axes orthogonal to
each
other~\cite{dieperink1982:ibm2-triax,leviatan1990:ibm2-modes,ginocchio1992:ibm2-shapes}.
This yields an overall composite nuclear shape with triaxial
deformation, as shown in Fig.~\ref{figortho}.  To avoid ambiguity, we
shall adopt here the notation $\grpsuthreepnstarbar$ for the
complementary case $\chi_\pi\<=+\rootstinline$ and
$\chi_\nu\<=-\rootstinline$, in which the proton and neutron
deformations are interchanged.

\subsection{Coherent state energy surface}
\label{subseccoherent}

The phase structure of an algebraic model is determined from the
classical limit of the model, which is obtained through the coherent
state
formalism~\cite{gilmore1975:multilevel-coherent,gilmore1978:coherent,zhang1990:coherent}.
A coherent state of good boson number is constructed, and the
parameters of this state are directly related to the classical
coordinates of the system.  This coherent state is used as a
variational trial state: the global minimum of the energy surface
$\scrE\equiv\langle H\rangle$ is used to determine the ground state
energy and equilibrium classical coordinates of the system.  In the
limit of infinite boson number, the coherent state variational energy
converges to the true ground state
energy~\cite{gilmore1975:multilevel-coherent,gilmore1978:coherent}.
Detailed coherent state analyses of the one-fluid IBM may be found in
Refs.~\cite{ginocchio1980:ibm-classical,ginocchio1980:ibm-coherent-bohr,dieperink1980:ibm-classical,dieperink1980:ibm-phase,feng1981:ibm-phase,vanisacker1981:ibm-triax,leviatan1987:ibm-intrinsic,lopezmoreno1996:ibm-catastrophe}.

Once the equilibrium properties of the model have been established,
ground state phase transitions can be identified and categorized
according to the Ehrenfest
classification~\cite{ehrenfest1933:phase-trans}: a phase transition is
first order if the first derivative of the system's energy is
discontinuous with respect to the control parameter being varied,
second order if the second derivative is discontinuous,
\textit{etc.}  Where the system's energy is obtained, as in the present classical
analysis, as the global minimum of an energy function $\scrE$, a first
order transition is usually associated with a discontinuous jump in
the equilibrium coordinates (``order parameters'') between distinct
competing minima.  Second or higher order transitions are associated
instead with a continuous evolution of the equilibrium coordinates, as
when an initially solitary global minimum becomes unstable (possessing
a vanishing second derivative with respect to some coordinate) and
evolves into two or more minima.  It should be noted that, whenever
the order of a phase transition is obtained by numerical analysis,
application of the Ehrenfest criterion is limited by the ability to
numerically resolve sufficiently small discontinuities.  This is
especially a consideration for points of first-order transition very
close to a point of second-order transition.  Moreover, problems with
the classification scheme, not addressed here, may arise at the
boundaries of the parameter space or when the Hamiltonian possesses
additional symmetries.

For a two-fluid model, the boson number for each constituent fluid is
conserved separately.  The coherent state is constructed not only with
good total boson number but with good boson number for each fluid.
For the IBM-2, the number operators are $\hat{N_\rho}\equiv
s_{\rho,0}^\dag s_{\rho,0}+\sum_{\mu=-2}^{2}d_{\rho,\mu}^\dag
d_{\rho,\mu}$.  The IBM-2 coherent
state~\cite{ginocchio1980:ibm-coherent-bohr,dieperink1984:ibm2,bijker1985:ibm2-coherent,vanegmond1985:ibm2-generator-coordinate}
is defined in terms of proton and neutron condensate bosons
\begin{equation}
\label{eqnBrc}
B_{\rho c}^\dag
\equiv
\frac{1}{(1+|\alpha_\rho^{(2)}|^2)^{1/2}}
\Bigl(
s_{\rho, 0}^\dag +\sum_{\mu=-2}^{2} \alpha_{\rho,\mu}^{(2)}
d_{\rho,\mu}^\dag
\Bigr),
\end{equation}
where
$|\alpha_\rho^{(2)}|\equiv(\sum_{\mu=-2}^{2}\alpha_{\rho,\mu}^{(2)\,*}\alpha_{\rho,\mu}^{(2)})^{1/2}$,
as
\begin{equation}
\label{eqncoherentpn}
|N_\pi,\alpha^{(2)}_\pi;N_\nu,\alpha^{(2)}_\nu\rangle
\equiv
\frac{1}{(N_\pi!)^{1/2}(N_\nu!)^{1/2}}\left(B_{\pi c}^\dag\right)^{N_\pi}\left(B_{\nu c}^\dag\right)^{N_\nu}
\, |0\rangle.
\end{equation}
The $\alpha^{(2)}_{\rho,\mu}$ are interpreted geometrically as
quadrupole shape variables~\cite{bohr1998:v2} for the proton and
neutron fluids.  They are related to the four deformation parameters
($\beta_\pi$, $\gamma_\pi$, $\beta_\nu$, and $\gamma_\nu$) and to the
six Euler angles ($\theta_{1\pi}$, $\theta_{2\pi}$, $\theta_{3\pi}$,
$\theta_{1\nu}$, $\theta_{2\nu}$, and $\theta_{3\nu}$) specifying the
orientations of the proton and neutron intrinsic frames by
\begin{multline}
\label{eqnalpharho}
\alpha_{\rho,\mu}^{(2)} =
\beta_\rho\cos\gamma_\rho D^{2\,*}_{\mu0}(\theta_{1\rho},\theta_{2\rho},\theta_{3\rho})\\
+\frac{1}{\sqrt{2}}\beta_\rho\sin\gamma_\rho
\bigl[D^{2\,*}_{\mu2}(\theta_{1\rho},\theta_{2\rho},\theta_{3\rho})+D^{2\,*}_{\mu-2}(\theta_{1\rho},\theta_{2\rho},\theta_{3\rho})\bigr],
\end{multline}
where $D_{M'M}^J$ is the Wigner $D$ function~\cite{rose1957:am}.

Calculation of the expectation value of $H$ with respect to the
coherent state~(\ref{eqncoherentpn}) yields an energy surface
$\scrE\equiv
\langle N_\pi,\alpha^{(2)}_\pi;N_\nu,\alpha^{(2)}_\nu| H
|N_\pi,\alpha^{(2)}_\pi;N_\nu,\alpha^{(2)}_\nu\rangle$.  By rotational
invariance, $\scrE$ can only depend upon the
\textit{relative} Euler angles $\vartheta_i$ between the proton and
neutron fluid intrinsic frames, not the $\theta_{i\pi}$ and
$\theta_{i\nu}$ separately.  For evaluation of $\scrE$, the Euler
angles may thus simply be chosen to be $\theta_{i\pi}\<=0$ and
$\theta_{i\nu}\<=\vartheta_{i}$.  Investigations of the IBM-2 coherent
state energy surface have been carried out in
Refs.~\cite{balantekin1983-ibm2-coherent,leviatan1990:ibm2-modes,ginocchio1992:ibm2-shapes}.

The expectation value with respect to the IBM-2 coherent
state of an operator constructed from the boson operators
of one fluid only (\textit{e.g.}, $ \nhat_{d\pi}$ or $ \Qhat_\pi
\cdot \Qhat_\pi$) can be calculated as in the one-fluid IBM,
by the methods of
Refs.~\cite{ginocchio1980:ibm-coherent-bohr,vanisacker1981:ibm-triax}.
The expectation values of the one-fluid operators appearing
in~(\ref{eqnHgeneralraw}) are~\cite{vanisacker1981:ibm-triax}
\begin{align}
\label{eqnndrho}
\langle \hat{n}_{d\rho} \rangle &= \frac{N_\rho\beta_\rho^2}{1+\beta_\rho^2}
\\
\label{eqnQrhoQrho}
\langle \Qhat^{\chi_\rho}_\rho \cdot \Qhat^{\chi_\rho}_\rho \rangle 
&=\frac{N_\rho}{1+\beta_\rho^2}\Bigl[5+(1+\chi_\rho^2)\beta_\rho^2\Bigr]\\
&\qquad \qquad \qquad +\frac{N_\rho(N_\rho-1)}{(1+\beta_\rho^2)^2}
\Bigl[ 4\beta_\rho^2-4\sqrt\frac{2}{7}\chi_\rho\beta_\rho^3\cos
3\gamma_\rho +\frac{2}{7}\chi_\rho^2\beta_\rho^4\Bigr]. \notag
\end{align}
The expectation value of a two-fluid operator constructed as the
scalar product of one-fluid operators (\textit{e.g.}, $ \Qhat_\pi
\cdot \Qhat_\nu$) can be obtained using a factorization result~\cite[(C1)]{ginocchio1992:ibm2-shapes}.  
However, for more complicated operators, the general method presented
in Appendix~\ref{appcoherentme} provides a convenient means of
calculation.  The expectation value of $\Qhat_\pi\cdot \Qhat_\nu$ is a
function of all seven possible coordinates ($\beta_\pi$, $\gamma_\pi$,
$\beta_\nu$, $\gamma_\nu$, $\vartheta_1$, $\vartheta_2$, and
$\vartheta_3$),
\begin{multline}
\label{eqnQpQn}
\langle \Qhat_\pi^{\chi_\pi} \cdot \Qhat_\nu^{\chi_\nu} \rangle 
=
\frac{N_\pi N_\nu}{(1+\beta_\pi^2)(1+\beta_\nu^2)}
[\alpha_\pi^{(2)\,*}+\tilde\alpha_\pi^{(2)}+\chi_\pi
(\alpha_\pi^{(2)\,*}\times\tilde\alpha_\pi^{(2)})^{(2)}]
\\
\cdot
[\alpha_\nu^{(2)\,*}+\tilde\alpha_\nu^{(2)}+\chi_\nu(\alpha_\nu^{(2)\,*}\times\tilde\alpha_\nu^{(2)})^{(2)}]
,
\end{multline}
with the $\alpha_\rho^{(2)}$ as defined in Eqn.~(\ref{eqnalpharho}),
which for vanishing Euler angles simplifies to
\begin{multline}
\label{eqnQpQnaligned}
\langle \Qhat_\pi^{\chi_\pi} \cdot \Qhat_\nu^{\chi_\nu} \rangle 
=
\frac{2 N_\pi N_\nu \beta_\pi\beta_\nu }{7(1+\beta_\pi^2)(1+\beta_\nu^2)}
\Bigl[
14 \cos(\gamma_\pi-\gamma_\nu)
+\chi_\pi\chi_\nu\beta_\pi\beta_\nu\cos(2\gamma_\pi-2\gamma_\nu)\\
\qquad \qquad + \sqrt{14}[\chi_\pi\beta_\pi\cos(2\gamma_\pi+\gamma_\nu)+\chi_\nu\beta_\nu\cos(\gamma_\pi+2\gamma_\nu)]
\Bigr].
\end{multline}
The expectation value of the Majorana operator is a complicated
function of the coordinates.  For vanishing Euler angles, it reduces
to
\begin{equation}
\label{eqnMaligned}
\langle \Mhat \rangle = \frac{N_\pi N_\nu}{(1+\beta_\pi^2)(1+\beta_\nu^2)}
\bigl[  
\beta_\pi^2 +\beta_\nu^2 -2
\beta_\pi\beta_\nu\cos(\gamma_\pi-\gamma_\nu)
+\beta_\pi^2\beta_\nu^2 \sin^2(\gamma_\pi-\gamma_\nu)
\bigr]
.
\end{equation}

A detailed study of the possible equilibrium Euler angle and
$\gamma_\rho$ values for the most general two-body IBM-2 Hamiltonian
has been presented in Ref.~\cite{ginocchio1992:ibm2-shapes}.  So long
as the multipole decomposition of the proton-neutron interaction
contains no hexadecapole component
[$(d_\pi^\dag\times\tilde{d}_\pi)^{(4)}\cdot(d_\nu^\dag\times\tilde{d}_\nu)^{(4)}$],
it is found that the equilibrium configuration must have Euler angles
which are vanishing or multiples of $\pi/2$.  Thus, the proton and
neutron intrinsic frames are ``aligned'', to within a possible
relabeling of axes.  [For clarity, we note that in the $\grpsuthreepnstar$
equilibrium configuration (Fig.~\ref{figortho}), even though the
proton and neutron \textit{symmetry} axes are orthogonal, the
intrinsic frame
\textit{coordinate} axes are actually parallel, \textit{i.e.}, the Euler
angles vanish.  The proton ($\gamma_\pi\<=0$) symmetry axis is the $z$
axis, while the neutron ($\gamma_\nu\<=\pi/3$) symmetry axis is the
$y$ axis.]  If, instead, a hexadecapole contribution is present,
``oblique'' equilibrium configurations are possible.  The operator
$\Qhat_\pi\cdot\Qhat_\nu$ contains no hexadecapole component, but the
Majorana operator does (Sec.~\ref{subsecmajorana}).  The
IBM-2 is thus characterized by either four order parameters
($\beta_\pi$, $\gamma_\pi$, $\beta_\nu$, and $\gamma_\nu$) or seven
order parameters ($\beta_\pi$, $\gamma_\pi$, $\beta_\nu$,
$\gamma_\nu$, $\vartheta_1$, $\vartheta_2$, and $\vartheta_3$),
depending upon the form of the proton-neutron interaction.

\section{Phase structure for an essential Hamiltonian}
\label{secschematic}

\subsection{The $F$-spin invariant Hamiltonian}
\label{subsecdieperink}

A simple, schematic Hamiltonian which retains the essential dynamical
symmetry features of the IBM-2, obtained as a special case of
Eqn.~(\ref{eqnHgeneralraw}),  is
\begin{equation}
\label{eqnHdieperinkraw}
H=\varepsilon ( \nhat_{d\pi}+\nhat_{d\nu})+
\kappa(\Qhat_\pi^{\chi_\pi}+\Qhat_\nu^{\chi_\nu})\cdot(\Qhat_\pi^{\chi_\pi}+\Qhat_\nu^{\chi_\nu}).
\end{equation}
This Hamiltonian is often preferred for theoretical studies due to its
invariance under $F$-spin rotations (\textit{e.g.},
Ref.~\cite{lipas1990:ibm2-fspin}) for $\chip\<=\chin$.  To make the
analysis of the IBM-2 phase structure as transparent as possible, let
us first consider in detail this $F$-spin invariant Hamiltonian before
proceeding in Section~\ref{secgeneral} to the more general form.  It
is also convenient for the moment to restrict attention to the case
$N_\pi=N_\nu$.

To obtain the classical limit, it is necessary to take
$N\<\rightarrow\infty$, where $N\<=N_\pi+N_\nu$.  In this limit, the
lower-order terms with respect to $N$ in Eqn.~(\ref{eqnQrhoQrho})
become negligible, and it is seen that the quantities $\langle
\nhat_{d\rho}\rangle$ are linear in $N$, while the $\langle \Qhat_\rho
\cdot \Qhat_{\rho'} \rangle$ are all quadratic in $N$.  To prevent the $\langle
\nhat_{d\rho}\rangle$ contributions from vanishing identically when the limit is taken, 
it is necessary to rescale the Hamiltonian coefficients by appropriate
powers of $N$.  It is thus convenient to reparametrize the
Hamiltonian~(\ref{eqnHdieperinkraw}) as
\begin{equation}
\label{eqnHdieperink}
H=\frac{1-\xi'}{N} ( \nhat_{d\pi}+\nhat_{d\nu})
-\frac{\xi'}{N^2}(\Qhat_\pi^{\chi_\pi}+\Qhat_\nu^{\chi_\nu})\cdot(\Qhat_\pi^{\chi_\pi}+\Qhat_\nu^{\chi_\nu}),
\end{equation}
so that the energy function $\scrE$ is independent of $N$.  This
definition also condenses the full range of possible ratios
$\varepsilon/\kappa$ onto the finite interval $0\<\leq\xi'\<\leq1$.
An overall normalization parameter for $H$ has been discarded as
irrelevant to the extremum structure of the energy surface.

\begin{figure}
\begin{center}
\includegraphics[width=0.7\hsize]{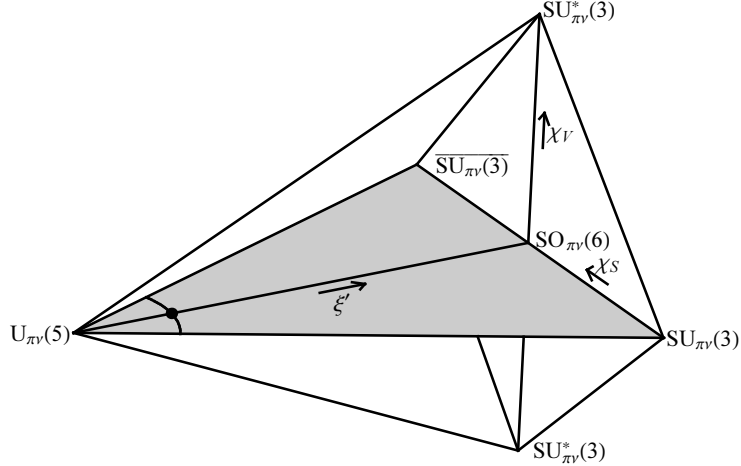}
\end{center}
\caption
{Parameter space for the $F$-spin invariant Hamiltonian of
Eqn.~(\ref{eqnHdieperink}).  For $\chi_V\<=0$ (shaded plane) the
analysis reduces to that for the one-fluid IBM (see text), yielding a
point of second-order phase transition embedded in a
curve~(\ref{eqnxipccurve}) of first-order phase transition points.
The axes cover ranges $0\<\leq\xi'\<\leq 1$,
$-\sqrt{7}/2\<\leq\chi_S\<\leq\sqrt{7}/2$, and
$-\sqrt{7}/2\<\leq\chi_V\<\leq\sqrt{7}/2$.  The $\chi_S$ and $\chi_V$
axes are scaled by $\xi'$, so as to converge to a point at the
$\grpufivepn$ limit, where the Hamiltonian is independent of $\chi_S$ and
$\chi_V$.  Figure from Ref.~\cite{caprio2005:ibmpn-nmp04}.  
}
\label{figtetraspace}
\end{figure}

There are three control parameters~--- $\xi'$, $\chi_\pi$, and
$\chi_\nu$~--- for the Hamiltonian~(\ref{eqnHdieperink}). It is
convenient to alternatively introduce ``scalar'' and ``vector''
parameters $\chi_S$$\equiv$$\frac{1}{2}(\chi_\pi+\chi_\nu)$ and
$\chi_V$$\equiv$$\frac{1}{2}(\chi_\pi-\chi_\nu)$.  The parameter space
is outlined in Fig.~\ref{figtetraspace}.

For $\chi_V\<=0$, indicated by the shaded triangle in
Fig.~\ref{figtetraspace}, the analysis essentially reduces to that for
the one-fluid IBM.  The equilibrium configuration occurs for
$\beta_\pi\<=\beta_\nu(\rtrim\equiv\beta)$ and
$\gamma_\pi\<=\gamma_\nu\<=0$ and is identical to that obtained for
the IBM Hamiltonian $H_\text{IBM}=[(1-\xi')/N] \nhat_d
-(\xi'/N^2)\Qhat^{\chi}\cdot \Qhat^{\chi}$, the phase structure of
which is well
known~\cite{dieperink1980:ibm-classical,feng1981:ibm-phase}, with
$\chi\<=\chi_S$.  For $\xi'\<\leq1/5$, the energy surface has a
minimum at $\beta\<=0$.  At $\xi'\<=1/5$, this minimum becomes
unstable with respect to $\beta$,
\textit{i.e.}, $\partial^2\scrE/\partial\beta^2\<=0$.  If $\chi\<=0$, this
instability leads to a second-order transition between undeformed
($\beta\<=0$) and deformed ($\beta\<\neq0$) structure at $\xi'\<=1/5$.
However, for any other value of $\chi$, the minimum at
$\beta\<=0$ is preempted as global minimum by a distinct minimum with
nonzero $\beta$ before $\xi'\<=1/5$.
This leads to a first-order phase transition, at a parameter value
$\xi'=\xi'_c(\chi)$ given by
\begin{equation}
\label{eqnxipccurve}
\xi'_c(\chi)=\frac{1}{5+\frac{2}{7}\chi^2},
\end{equation}
easily derived from
Ref.~\cite[(2.9b)]{leviatan1987:ibm-intrinsic}.  Thus, the point of
second-order phase transition at $\xi'\<=1/5$ and $\chi\<=0$ lies on a
trajectory~(\ref{eqnxipccurve}) of first-order transition points
(Fig.~\ref{figtetraspace}).

It simplifies the analysis of the phase diagram for the remainder of
the parameter space to note the presence of reflection symmetries with
respect to the $\chi_\rho$ parameters.  Each term contributing to
$\scrE$ is seen by inspection of Eqns.~(\ref{eqnndrho}),
(\ref{eqnQrhoQrho}), and (\ref{eqnQpQnaligned}) to be invariant under
a simultaneous transformation of coordinates and Hamiltonian
parameters
\begin{equation} 
\label{eqnchiinversion}
\chi_\pi\rightarrow-\chi_\pi,
\qquad
\chi_\nu\rightarrow-\chi_\nu,
\qquad
\gamma_\pi\rightarrow \frac{\pi}{3}-\gamma_\pi,
\qquad
\gamma_\nu\rightarrow \frac{\pi}{3}-\gamma_\nu.
\end{equation}
Thus, the equilibrium deformation at a point $(\chi_\pi,\chi_\nu)$ in
parameter space is simply related to that at the point
$(-\chi_\pi,-\chi_\nu)$ by Eqn.~(\ref{eqnchiinversion}).  The phase
diagram is symmetric under simultaneous negation of
$\chi_\pi$ and $\chi_\nu$ or, equivalently, simultaneous negation of
$\chi_S$ and $\chi_V$.  Moreover, for the $F$-spin invariant
Hamiltonian~(\ref{eqnHdieperink}) with
$N_\pi\<=N_\nu$, the numerical coefficients on corresponding proton
and neutron terms in $\scrE$ are equal.  Thus, $\scrE$ is symmetric
under interchange of all proton and neutron variables, and the phase
diagram is symmetric under interchange of $\chi_\pi$ and $\chi_\nu$
or, equivalently, negation of $\chi_V$.

\subsection{The $\grpsuthreepn-\grpsosixpn-\grpsuthreepnstar$ plane}
\label{subsecbackplane}

Let us begin the investigation of the phase structure of the
Hamiltonian~(\ref{eqnHdieperink}) with an analytic study for
$\xi'\<=1$, corresponding to the rightmost plane of the parameter
space diagram in Fig.~\ref{figtetraspace}.  This plane encompasses the
$\grpsosixpn$, $\grpsuthreepn$, and $\grpsuthreepnstar$ dynamical symmetries.

Surrounding the $\grpsuthreepn$ dynamical symmetry is a region of parameter
space in which the equilibrium deformations are axially symmetric
($\gamma_\pi\<=\gamma_\nu\<=0$), and a similar region surrounds the
$\grpsuthreepnbar$ dynamical symmetry ($\gamma_\pi\<=\gamma_\nu\<=\pi/3$).
In either of these cases ($\gamma_\pi\<=\gamma_\nu\<=0$ or $\pi/3$),
the energy surface can be written in an especially simple form in
terms of the function $F$ defined in Eqn.~(\ref{eqnf}),
\begin{equation}
\label{eqnscrEaxial}
\scrE(\beta_\pi,\beta_\nu)=-4\left[ 
F\left(-\frac{1}{\sqrt{14}}\chi_\pi;\sigma\beta_\pi\right)
+F\left(-\frac{1}{\sqrt{14}}\chi_\nu;\sigma\beta_\nu\right)
\right]^2,
\end{equation}
where $\sigma\equiv\cos 3\gamma_\rho=\pm1$.  The energy function
in~(\ref{eqnscrEaxial}) is minimized when the magnitude of the
quantity within brackets is maximized.  In the vicinity of $\grpsuthreepn$,
including in the entire quadrant for which $\chi_\pi$ and $\chi_\nu$
are both negative, it follows from the results of Appendix~\ref{appf}
that the minimum in $\scrE(\beta_\pi,\beta_\nu)$ occurs at coordinate
values
\begin{equation}
\label{eqnbetaprolate}
\beta_\rho= -\frac{\chi_\rho}{\sqrt{14}}+ \Biggl[\Biggl( \frac{\chi_\rho}{\sqrt{14}}\Biggr)^2+1\Biggr]
^{1/2}
\end{equation}
with $\sigma=+1$ (\textit{i.e.}, $\gamma_\pi=\gamma_\nu=0$).
Similary, in the vicinity of $\grpsuthreepnbar$, including in the entire
quadrant for which $\chi_\pi$ and $\chi_\nu$ are both positive, the
minimum in $\scrE(\beta_\pi,\beta_\nu)$ occurs at coordinate values
\begin{equation}
\label{eqnbetaoblate}
\beta_\rho= -\frac{\chi_\rho}{\sqrt{14}}- \Biggl[\Biggl( \frac{\chi_\rho}{\sqrt{14}}\Biggr)^2+1\Biggr]
^{1/2}
\end{equation}
with $\sigma=-1$ (\textit{i.e.}, $\gamma_\pi=\gamma_\nu=\pi/3$).  In
either case, the value of the energy function at the global minimum is
\begin{equation}
\label{eqnscrEequil}
\scrE=-\frac{1}{4}(\beta_\pi+\beta_\nu)^2,
\end{equation}
with $\beta_\pi$ and $\beta_\nu$ given by Eqn.~(\ref{eqnbetaprolate})
or~(\ref{eqnbetaoblate}) as appropriate.

\begin{figure}
\begin{center}
\includegraphics[width=0.5\hsize]{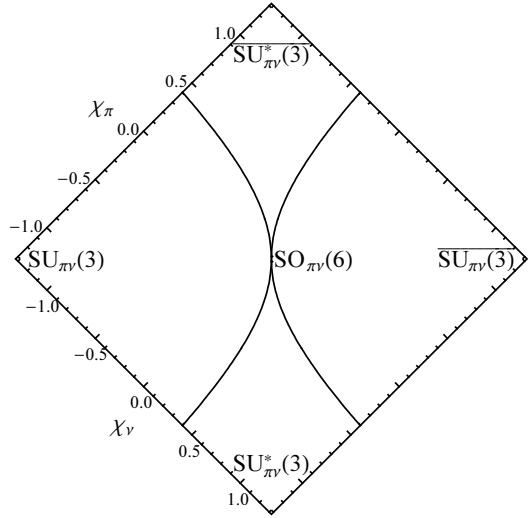}
\end{center}
\caption
{Phase diagram for the
$\grp{SU}_{\pi\nu}(3)$-$\grp{SO}_{\pi\nu}(6)$-$\grp{SU}_{\pi\nu}^*(3)$
plane ($\xi'\<=1$) in the parameter space of the Hamiltonian
of~(\ref{eqnHdieperink}), for $N_\pi\<=N_\nu$, showing the
curve~(\ref{eqnboundary}) of second-order phase transition.  The
diagram is rotated to allow more direct comparison with
Figs.~\ref{figtetraspace} and~\ref{figquadrant}.  Figure adapted from
Ref.~\cite{caprio2004:ibmpn}.}
\label{figboundary}
\end{figure}

As $\chi_\pi$ and $\chi_\nu$ are varied away from their $\grpsuthreepn$
values, eventually axial equilibrium deformation gives way to triaxial
deformation, with $\gamma_\pi$ and/or $\gamma_\nu$ nonzero.  This
transition occurs on the locus of points at which the minimum given
by~(\ref{eqnbetaprolate}) first becomes unstable with respect to
$\gamma$ deformation.  Since $\scrE$ depends upon both $\gamma_\pi$
and $\gamma_\nu$, instablility occurs when the
\textit{directional} second derivative of $\scrE$ first vanishes along
some ``direction'' in $(\gamma_\pi,\gamma_\nu)$ coordinate space,
which may generally happen before either
$\partial^2\scrE/\partial\gamma_\pi^2$ or
$\partial^2\scrE/\partial\gamma_\nu^2$ vanishes individually.  The
equation describing the boundary curve in $\chi_\pi$ and $\chi_\nu$ is
most compactly expressed in terms of the corresponding equilibrium
values $\beta_\pi$ and $\beta_\nu$ from Eqn.~(\ref{eqnbetaprolate})
or~~(\ref{eqnbetaoblate}), as
\begin{equation}
\label{eqnboundary}
1-\beta_\pi^2-\beta_\nu^2
-2\beta_\pi\beta_\nu^3+3\beta_\pi^2\beta_\nu^2-2\beta_\pi^3\beta_\nu+2\beta_\pi^3\beta_\nu^3
=0.
\end{equation}
This curve is shown in Fig.~\ref{figboundary}.  The direction in
$(\gamma_\pi,\gamma_\nu)$ space in which the instability occurs, at
any given point on this boundary, is given by
\begin{equation}
\label{eqngammadir}
\frac{\gamma_\pi}{\gamma_\nu} 
= - \frac{1+\beta_\pi^2}{1+\beta_\nu^2} \frac{B}{2A} = -
\frac{1+\beta_\pi^2}{1+\beta_\nu^2} \frac{2C}{B},
\end{equation}
where
\begin{equation}
\begin{aligned}
A&\equiv 9 \beta_\pi^2
(\beta_\pi^2-1)+\beta_\pi\beta_\nu(2\beta_\pi^2-1)(\beta_\pi^2+1)\\
B&\equiv -4 \beta_\pi \beta_\nu (\beta_\pi^2-2)(\beta_\nu^2-2)\\
C&\equiv 9 \beta_\nu^2
(\beta_\nu^2-1)+\beta_\pi\beta_\nu(2\beta_\nu^2-1)(\beta_\nu^2+1).
\end{aligned}
\end{equation}
Along the $\grpsuthreepn$-$\grpsuthreepnstar$ line in particular, for which
$\chi_\pi\<=-\rootstinline$, the transition occurs at
$\chi_\nu\<\approx0.4035$, obtained from Eqn.~(\ref{eqnboundary}) as
the root of a quartic equation.  In this special case, the global
minimum becomes soft purely with respect to $\gamma_\nu$ at fixed
$\gamma_\pi\<=0$.

The analytic results of this section indicate a continuous transition
between axially symmetric and triaxial structure on the
curve~(\ref{eqnboundary}).  In Sec.~\ref{subsecnumerical} it is verified
numerically that this transition is second, rather than higher, order,
and that no prior phase transition to a distinct minimum occurs before
the curve~(\ref{eqnboundary}) is reached.

\subsection{The transition between spherical and deformed equilibrium}
\label{subsecspherical}

The transition between spherical and deformed equilibrium for the
IBM-2 for arbitrary $\chi_V$ is closely related to the transition
which occurs in the special plane $\chi_V\<=0$ (shaded in
Fig.~\ref{figtetraspace}).  The energy function
$\scrE(\xi',\chi_S,\chi_V;\beta_\pi,\gamma_\pi,\beta_\nu,\gamma_\nu)$
obeys the identity
\begin{equation}
\label{eqnscrEslice}
\scrE(\xi',\chi_S,\chi_V;\beta,0,\beta,0)=\scrE(\xi',\chi_S,0;\beta,0,\beta,0),
\end{equation}
\textit{i.e.}, for a deformation which is axial and has
$\beta_\pi\<=\beta_\nu$, the value of $\scrE$ is independent of
$\chi_V$.  Recall that the equilibrium deformation for $\chi_V\<=0$ is
always axial with $\beta_\pi\<=\beta_\nu$
(Sec.~\ref{subsecdieperink}).  It is straight-forward to demonstrate
from these observations (see below) that a nonzero equilibrium
deformation for the parameter point $(\xi',\chi_S)$ in the
$\chi_V\<=0$ plane implies a nonzero equilibrium deformation for all
the out-of-plane points $(\xi',\chi_S,\chi_V)$.  Thus, the $\xi'$
value $\xi'_c(\chi_S,\chi_V)$ for which the spherical-deformed
transition occurs at a given $\chi_S$ and $\chi_V$ obeys, by
Eqn.~(\ref{eqnxipccurve}),
\begin{equation}
\label{eqnxipcbound}
\xi'_c(\chi_S,\chi_V) \leq \frac{1}{5+\frac{2}{7}\chi_S^2}.
\end{equation}

The detailed argument proceeds as follows.  Let the equilibium energy
at a general point in parameter space be denoted by
$\scrE_0(\xi',\chi_S,\chi_V)$, and let $\beta_0(\xi',\chi_S)$ be the
equilibrium value of $\beta_\pi$ and $\beta_\nu$ for $\chi_V\<=0$ as
determined, \textit{e.g.}, from
Ref.~\cite[(2.12)]{leviatan1987:ibm-intrinsic}.  Then
$\scrE_0(\xi',\chi_S,0)\<=\scrE[\xi',\chi_S,0;\beta_0(\xi',\chi_S),0,\beta_0(\xi',\chi_S),0]$.
The global minimum energy for arbritrary $\chi_V$ then is subject, by
Eqn.~(\ref{eqnscrEslice}), to the upper bound
\begin{equation}
\label{eqnscrEbound}
\begin{aligned}
\scrE_0(\xi',\chi_S,\chi_V)
&\leq\scrE[\xi',\chi_S,\chi_V;\beta_0(\xi',\chi_S),0,\beta_0(\xi',\chi_S),0]\\
&=\scrE[\xi',\chi_S,0;\beta_0(\xi',\chi_S),0,\beta_0(\xi',\chi_S),0]=\scrE_0(\xi',\chi_S,0).
\end{aligned}
\end{equation}
Note that $\scrE$ evaluated at zero deformation is zero for the
Hamiltonian considered here~(\ref{eqnHgeneralraw}), so $\scrE_0\<<0$
if and only if the configuration is deformed.  If
$\beta_0(\xi',\chi_S)\neq0$, then $\scrE_0(\xi',\chi_S,\chi_V)\<<0$
for all $\chi_V$ by Eqn.~(\ref{eqnscrEbound}), and the equilibrium
deformation for parameter point $(\xi',\chi_S,\chi_V)$ is also
nonzero.

To further investigate the nature of the spherical-deformed
transition, let us consider the stability of the minimum of $\scrE$ at
$\beta_\pi\<=\beta_\nu\<=0$.  Instability occurs when the second
derivative of $\scrE$ vanishes along some ``direction'', which we
denote $(u_\pi,u_\nu)$, in $(\beta_\pi,\beta_\nu)$ coordinate space,
for some fixed values of $\gamma_\pi$ and $\gamma_\nu$.  The second
derivative of $\scrE$ along the ray parametrized as $\beta_\pi\<=u_\pi
\beta$ and $\beta_\nu\<= u_\nu \beta$ is
\begin{equation}
\label{eqnbetasoft}
\left.\frac{d^2\scrE}{d\beta^2}\right|_{\beta=0}
=(1-3\xi')(u_\pi^2+u_\nu^2)-4\xi'u_\pi u_\nu
\cos(\gamma_\pi-\gamma_\nu).
\end{equation}
This is independent of $\chi_S$ and $\chi_V$ and depends upon the
$\gamma_\rho$ coordinates only through their difference
$\gamma_\pi-\gamma_\nu$.  The smallest $\xi'$ value at which the
second derivative~(\ref{eqnbetasoft}) vanishes is $\xi'\<=1/5$, with
$u_\pi\<=u_\nu$, indicating instability of the minimum against
deformations with $\beta_\pi\<=\beta_\nu$ and
$\gamma_\pi\<=\gamma_\nu(\rtrim\equiv\gamma)$.  As the second
derivative is independent of $\gamma$, the system is unstable against
deformations of all possible $\gamma$ values, representing prolate,
oblate, and intermediate triaxial deformations, simultaneously.

For $\chi_S\<\neq0$, it is apparent from Eqn.~(\ref{eqnxipcbound})
that the spherical-deformed phase transition occurs \textit{before}
the spherical minimum becomes unstable ($\xi'\<=1/5$), and so the
transition is a first-order transition to a distinct minimum.
Thus, the curve of first-order phase transition occuring in the plane
$\chi_V\<=0$ ``propagates'' out of this plane to form a surface of
first-order transition.  In the approximation that
$\beta_\pi\<\approx\beta_\nu$ for the deformed minimum, the surface
would be a vertical extension of the one-fluid IBM transition
trajectory in Fig.~\ref{figtetraspace}.  The deviation of the
surface from this limiting location is studied numerically in
Sec.~\ref{subsecnumerical}.

For $\chi_S\<=0$, however, it is possible for the minimum at zero
deformation to remain the global minimum until $\xi'\<=1/5$, leading
to a second-order phase transition.  Let us consider the likely
properties of any alternative lower, deformed global minimum (these
are confirmed numerically in Sec.~\ref{subsecnumerical}).  By symmetry
in the proton and neutron parameter magnitudes ($\chi_\pi=-\chi_\nu$),
it is reasonable to expect this minimum to have
$\beta_\pi\<=\beta_\nu$.  A minimum with axial deformation has been
excluded by Eqn.~(\ref{eqnxipccurve}), but a minimum with triaxial
deformation must be considered.  Since $\chi_S\<=0$ represents the
``reflection'' plane in parameter space between prolate and oblate
structure~(\ref{eqnchiinversion}), it is natural to expect the minimum
to have $(\gamma_\pi+\gamma_\nu)/2\<=\pi/6$.  Inspection of the
restricted energy function
$\scrE(\xi',0,\chi_V;\beta,\pi/6+\gamma_V,\beta,\pi/6-\gamma_V)$
reveals that a triaxial deformed minimum does preempt the minimum at
zero deformation as global minimum [at $\xi'_c(0,\chi_V) \<=
1/(3+7\chi_V^{-2}/2+2\chi_V^2/7)$] but that this only occurs for
$\abs{\chi_V}\<\geq\sqrt{7/2}$, which lies outside the parameter range
$\abs{\chi_V}\<\leq\sqrt{7}/2$ of interest in the present
study.

In conclusion, a line of second-order phase transition between
spherical and deformed equilibrium occurs at $\xi'\<=1/5$ and
$\chi_S\<=0$, subsuming the one-fluid IBM second-order transition
point.  This line is embedded in a surface of first-order
transition points.

\subsection{Numerical investigation of the full phase diagram}
\label{subsecnumerical}

The remainder of the phase diagram is obtained by numerical
minimization of the energy surface with respect to $\beta_\pi$,
$\gamma_\pi$, $\beta_\nu$, and $\gamma_\nu$.  This minimization
provides the equilibrium coordinate values at any point in the
parameter space of Fig.~\ref{figtetraspace}.

For robust identification of the global minimum, $\scrE$ is first
evaluated at each point on a fine mesh in the deformation coordinates
($\Delta\beta_\rho\<=0.02$ and $\Delta\gamma_\rho\<=1^\circ$ provides
sufficient resolution for most of the calculations shown here).  All
points which are discrete local minima of $\scrE$ relative to the
neighboring mesh points are identified.  The coordinate values for
these minima are then refined iteratively by a conjugate gradient
method.  The global minimum is identified from among the refined
values.

\begin{figure}
\begin{center}
\includegraphics[width=0.75\hsize]{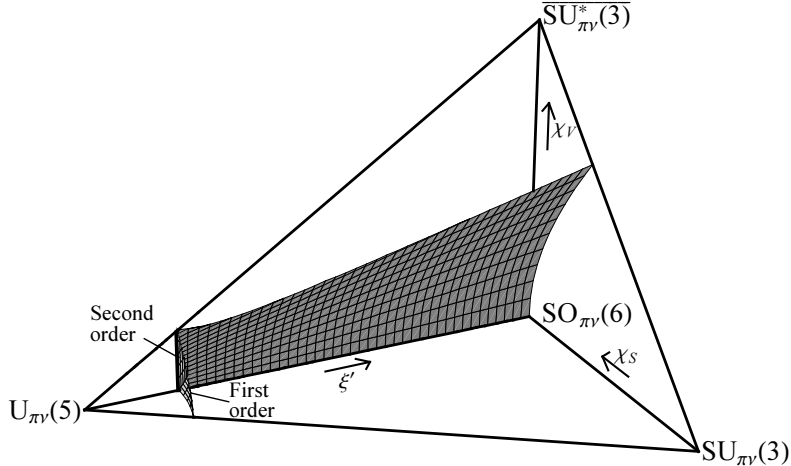}
\end{center}
\caption
{Phase diagram of the proton-neutron interacting boson model (IBM-2)
for the $F$-spin invariant Hamiltonian~(\ref{eqnHdieperink}) with
$N_\pi\<=N_\nu$, as obtained by numerical minimization of $\scrE$.
The surfaces of first-order and second-order transition
between regions of undeformed, axially symmetric deformed, and
triaxially deformed equilibria are shown.  Only one ``quadrant'' of
parameter space ($0\<\leq\xi'\<\leq 1$,
$-\sqrt{7}/2\<\leq\chi_S\<\leq0$, and $0\<\leq\chi_V\<\leq\sqrt{7}/2$)
is included in this plot, since the others may be obtained by
reflection.  The $\chi_S$ and $\chi_V$ axes are scaled by $\xi'$, so
as to converge to a point at the $\grpufivepn$ limit.  Figure from
Ref.~\cite{caprio2004:ibmpn}.}
\label{figquadrant}
\end{figure}

The phase diagram for the $F$-spin invariant Hamiltonian with
$\Np\<=\Nn$, obtained numerically in this fashion, is shown in
Fig.~\ref{figquadrant}.  [Only one quadrant of parameter space is
included in this plot, since the others may be obtained by reflection
(Sec.~\ref{subsecdieperink}).]  The line of second-order transition
between spherical and deformed equilibrium at $\xi'\<=1/5$ and
$\chis\<=0$ is the locus of simultaneous contact of four regions of
the phase diagram, those of spherical, prolate axially symmetric,
oblate axially symmetric (not shown, in adjacent quadrant), and
triaxial equilibria.  Numerical investigation of the behavior of
$\scrE$ at points on the axial-triaxial transition surface allows
the Ehrenfest criterion to be applied, and it appears that the
transition is everywhere second order.

\begin{figure}
\begin{center}
\includegraphics[width=0.5\hsize]{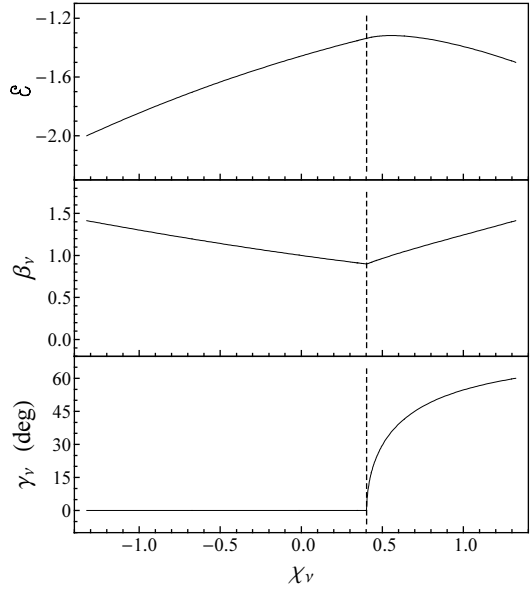}
\end{center}
\caption
{Evolution of the equilibrium values of (a)~$\scrE$, (b)~$\beta_\nu$,
and (c)~$\gamma_\nu$ along the line in parameter space between the
$\grpsuthreepn$ and $\grpsuthreepnstar$ dynamical symmetries
($\chi_\pi\<=-\sqrt{7}/2$,
$-\sqrt{7}/2\<\leq\chi_\nu\<\leq\sqrt{7}/2$), obtained numerically.
The dashed line indicates the second-order transition point,
$\chi_\nu\<\approx0.4035$.}
\label{figrayssu3su3star}
\end{figure}

We now consider in detail the evolution of the equilibrium energy and
coordinates at selected locations throughout the phase diagram.
Within the plane $\xi'\<=1$, the boundary curve separating axially
symmetric and triaxial equilibria was established in
Sec.~\ref{subsecbackplane}.  The equilibrium properties along the
$\grpsuthreepn$-$\grpsuthreepnstar$ transition line are shown in
Fig.~\ref{figrayssu3su3star}.  Several characteristics may be noted.
The second derivative of $\scrE$ with respect to the control parameter
$\chin$ is discontinuous, as in an Ehrenfest second-order phase
transition.  The first derivative of one order parameter, $\gamman$,
is infinite at the critical point, with an approximately square-root
dependence upon the control parameter after this point, as in a Landau
second-order phase transition~\cite{landau1980:statistical1}.  The
first derivative of $\betan$ at this point is, in contrast,
discontinous but finite.  In the special case of the
$\grpsuthreepn$-$\grpsuthreepnstar$ line, as already discussed, the two other
order parameters $\betap$ and $\gammap$ completely decouple from the
transition process, remaining constant throughout.

\begin{figure}
\begin{center}
\includegraphics[width=\hsize]{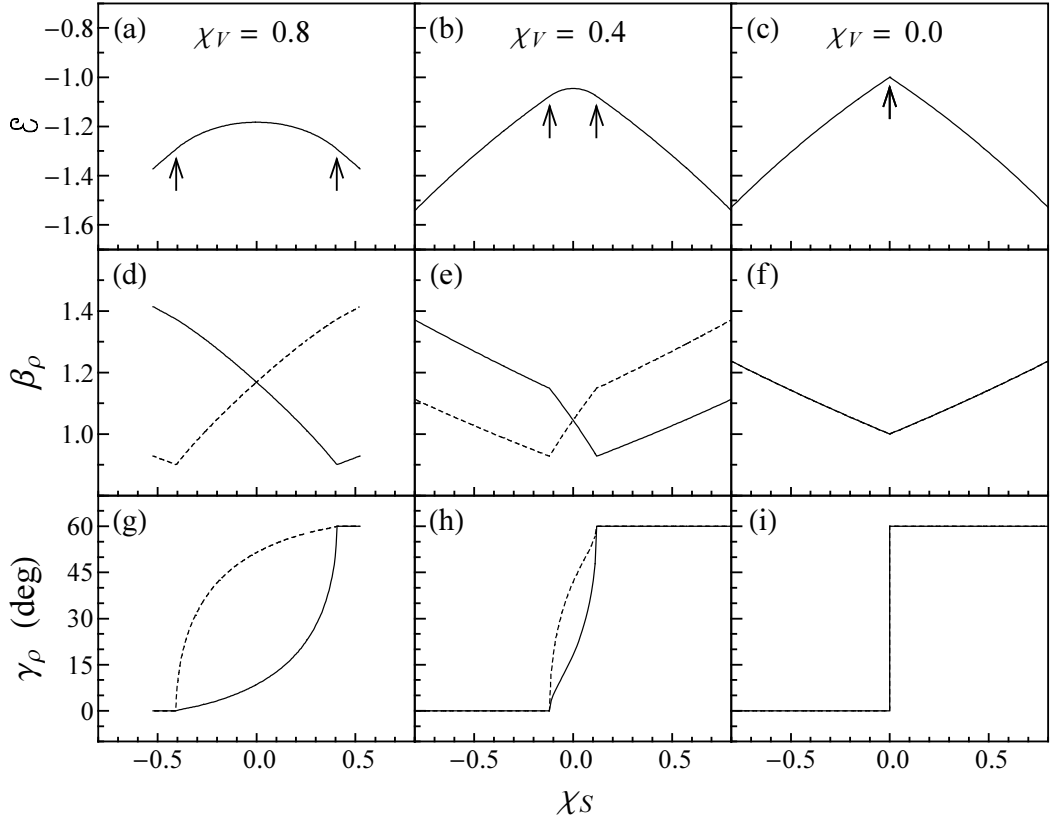}
\end{center}
\caption
{Evolution of the equilibrium properties along rays of constant
$\chiv$ in the $\xi'\<=1$ plane of parameter space, passing from the
region of axially symmetric prolate equilibrium to that of axially
symmetric oblate equilibrium.  Results are shown for the $F$-spin
invariant Hamiltonian~(\ref{eqnHdieperink}) with $\Np\<=\Nn$, along
the rays of (left)~$\chiv\<=0.8$, (center)~$\chiv\<=0.4$, and
(right)~$\chiv\<=0$.  Proton fluid variables ($\beta_\pi$ and
$\gamma_\pi$) are represented by solid curves, while neutron fluid
variables ($\beta_\nu$ and $\gamma_\nu$) are represented by dashed
curves.  The two points of second order phase transition (left,
middle), converging towards the single point $\chis\<=0$ in the limit
$\chiv\<=0$~(right), are marked with arrows.  }
\label{figrayshorizontal}
\end{figure}

In Fig.~\ref{figrayshorizontal}, the equilibrium properties on lines
of constant $\chiv$ in the plane $\xi'\<=1$, which would appear as
horizontal lines across the diagram of Fig.~\ref{figboundary}, are
shown. These lines cross both the prolate-triaxial and triaxial-oblate
second-order transition curves.  The equilibrium coordinate values for
positive $\chis$ are related to those for negative $\chis$ by the
transformations of Sec.~\ref{subsecdieperink}.  Observe that the order
parameters ($\beta_\pi$, $\gamma_\pi$, $\beta_\nu$, and $\gamma_\nu$)
are coupled in their behavior at the transition points, all
simultaneously exhibiting discontinuities in their derivatives at each
transition point.  In the limiting case $\chiv\<=0$
[Fig.~\ref{figrayshorizontal}(right)], the two points of second order
transition at positive and negative $\chis$ converge to a single point
at $\chis\<=0$.  The discontinuities in the second derivative of
$\scrE$ with respect to $\chis$ at the two transition points
[Fig.~\ref{figrayshorizontal}(left,middle)] combine to form a ``cusp''
in $\scrE$.  The discontinuity of $\partial\scrE/\partial\chi$ at
$\chi\<=0$ was noted in the context of the one-fluid IBM by
Jolie~\textit{et al.}~\cite{jolie2001:ibm-o6-phase}.  Such a
discontinuity in the slope of $\scrE$ would, according to the
Ehrenfest criterion, indicate a first-order phase transition.
However, the present case provides an example of the great variety of
phenomena possible in multi-parameter problems, as is does not
naturally fit the one-dimensional Laundau model for a first-order
phase transition.  The conventional Landau first-order phase
transition~\cite{landau1980:statistical1} involves ``competition''
between two distinct local minima simultantously present in the energy
surface, one overtaking the other as global minimum; whereas, in the
present example, a single minimum with $\gammap\<=\gamman\<=0$ is
present for $\chis\<<0$, a single minimum with
$\gammap\<=\gamman\<=\pi/3$ is present for $\chis\<>0$, and a
continuous trajectory of minima of arbitary
$\gamma(\rtrim=\gammap\<=\gamman)$, arising from the
$\grp{SO}_{\pi\nu}(5)$ symmetry of the Hamiltonian, are present for
$\chis\<=0$.  This example suggests the need for a comprehensive
extension of the classification schemes for single-parameter phase
transitions to cover phase transitions with multiple order parameters
(see also Ref.~\cite[Chapter 17]{gilmore1981:catastrophe}).

\begin{figure}
\begin{center}
\includegraphics[width=0.8\hsize]{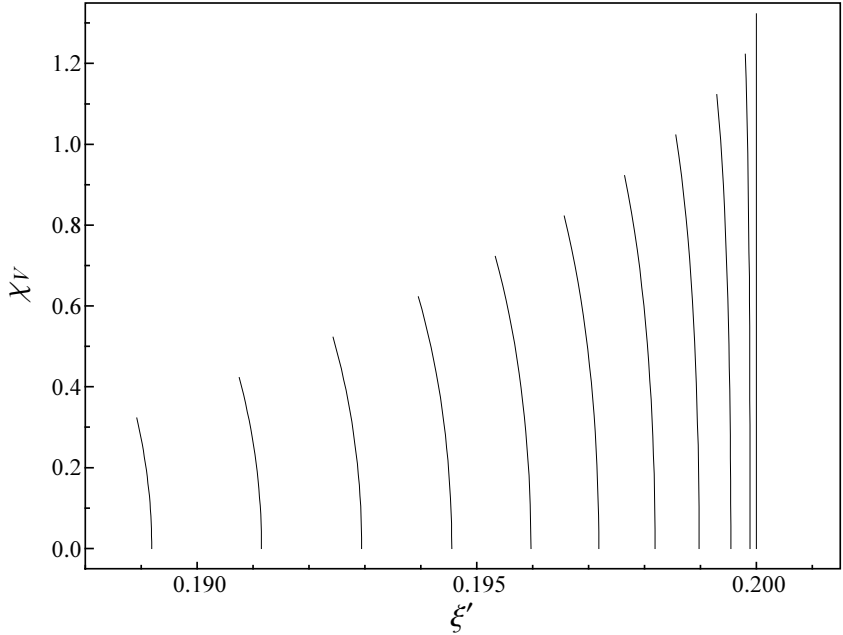}
\end{center}
\caption
{ Slices through the boundary surface between the regions of spherical
and deformed equilibrium at fixed $\chis$.  The $\xi'$ value at
$\chiv\<=0$ is the value $\xi'_c(\chis)$ from the one-fluid
IBM~(\ref{eqnxipccurve}), and the recession of each curve towards smaller
$\xi'$ above $\chiv\<=0$ indicates the extent to which
$\xi'_c(\chis,\chiv)$ recedes below the upper
limit~(\ref{eqnxipcbound}) obtained using $\betap\<=\betan$.  The
curves shown are for $\chis\<=0$ to $1$ in steps of $0.1$ (right to
left).}
\label{figu5cuts}
\end{figure}

Finally, we consider the transition between spherical and deformed
equilibrium configurations.  In Sec.~\ref{subsecspherical}, an upper
bound, $\xi'_c(\chi_S,\chi_V)\<\leq\xi'_c(\chi_S,0)$ was placed upon
the location of the spherical-deformed transition surface [see
Eqn.~(\ref{eqnxipcbound})], equality holding in the case
$\betap\<=\betan$.  The actual shape of this surface, visible on a
coarse scale in Fig.~\ref{figquadrant}, is shown in detail in
Fig.~\ref{figu5cuts}.  Several ``vertical'' slices through the
surface, at constant $\chis$, are plotted.  The line of second-order
phase transition at $\xi'\<=1/5$ appears rightmost in the figure.  The
curves vary little in $\xi'$ ($\lesssim0.2\%$) below their upper
bound.

\begin{figure}
\begin{center}
\includegraphics[width=\hsize]{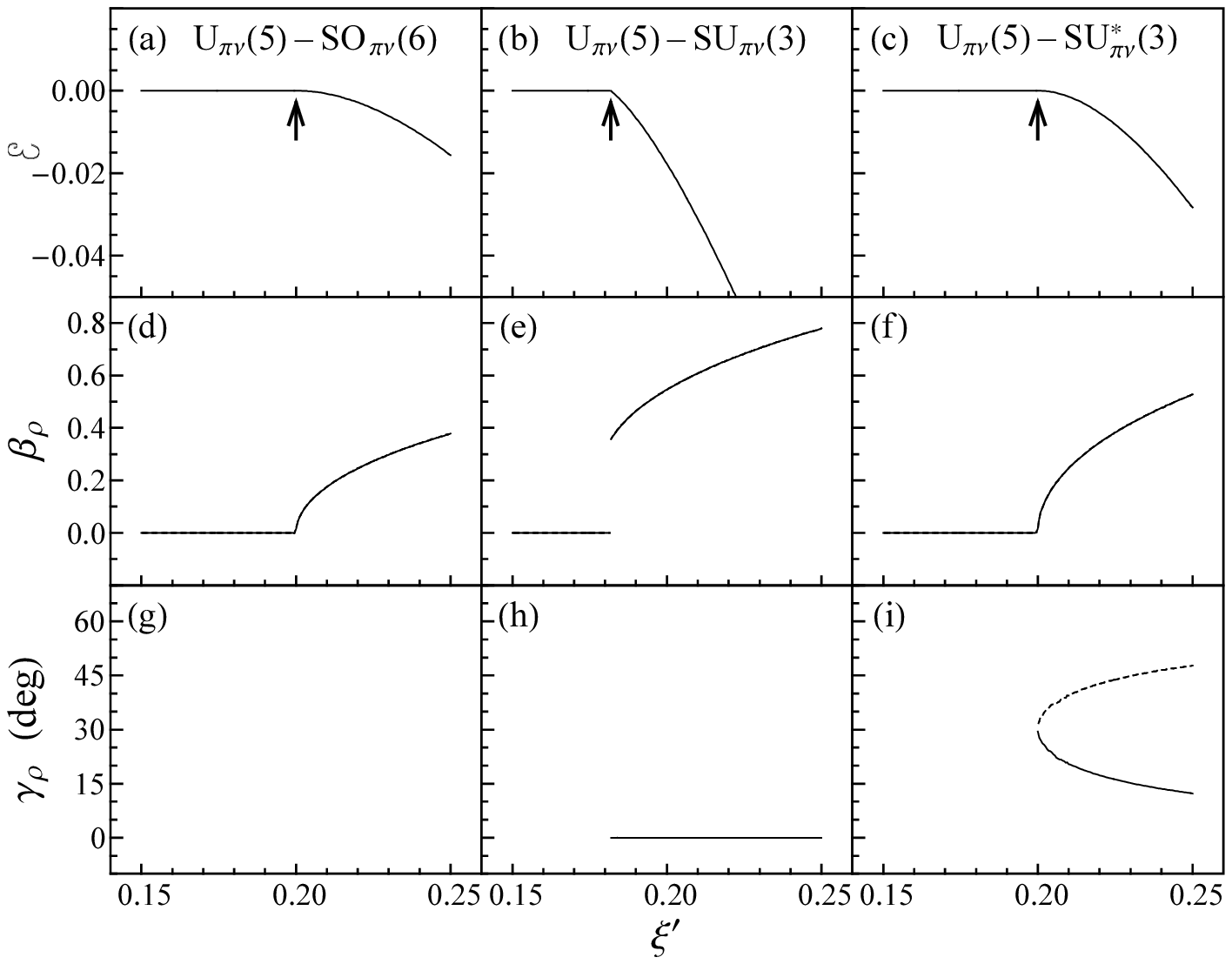}
\end{center}
\caption
{Evolution of the equilibrium properties in the vicinity of the
transition between spherical and deformed equilibria.  Results are
shown for the $F$-spin invariant Hamiltonian~(\ref{eqnHdieperink})
with $\Np\<=\Nn$, along (left)~the $\grpufivepn-\grpsosixpn$ transition line
($\chis\<=0$, $\chiv\<=0$), (middle)~the $\grpufivepn-\grpsuthreepn$ transition
line ($\chis\<=-\sqrt{7}/2$, $\chiv\<=0$), and (right)~the
$\grpufivepn-\grpsuthreepnstar$ transition line ($\chis\<=0$,
$\chiv\<=-\sqrt{7}/2$). Proton fluid variables are represented by
solid curves, while neutron fluid variables are represented by dashed
curves.  Equilibrium $\gammar$ values are undefined, and therefore not
shown, when $\betar\<=0$~(h,i) and along the entire $\grpufivepn-\grpsosixpn$
transition line~(g). Points of phase transition are marked with
arrows.  }
\label{figraysu5}
\end{figure}

The evolution of equilibrium properties across the spherical-deformed
transition is shown in Fig.~\ref{figraysu5}.  The second-order transition along
the $\grpufivepn$-$\grpsosixpn$ line and the first-order transition along the
$\grpufivepn$-$\grpsuthreepn$ line, quantitatively identical to the
corresponding transitions in the one-fluid IBM, are shown for
reference [Fig.~\ref{figraysu5}(left,middle)].  The transition along the
$\grpufivepn$-$\grpsuthreepnstar$ line [Fig.~\ref{figraysu5}(right)] is strongly
constrained by the proton-neutron interchange symmetry about
$\chis\<=0$, with $\betap\<=\betan$ and with $\gammap$ and $\gamman$
symmetric to each other about $\pi/6$.  

\section{Phase structure for a general Hamiltonian}
\label{secgeneral}

\subsection{Energy surface parameters}
\label{subsecparams}

In the preceding section, the general features of the IBM-2 phase
diagram were established using a schematic Hamiltonian, and only the
case $N_\pi\<=N_\nu$ was considered.  We now address the more general
Hamiltonian of Eqn.~(\ref{eqnHgeneralraw}) and also consider arbitrary
values of the ratio $N_\pi/N_\nu$.  These two generalizations are
closely related, as the coefficients in the energy surface $\scrE$
depend upon the Hamiltonian coefficients and the boson numbers in
combination.  To make explicit the dependence of each term contributing to $\scrE$
upon parameters, boson numbers, and coordinates, we express $\scrE$ as
\begin{multline}
\label{eqnEgeneralraw}
\scrE =
\varepsilon_\pi N_\pi \,f_\pi(\beta_\pi) 
+ \varepsilon_\nu N_\nu \,f_\nu(\beta_\nu) 
+\kappa_{\pi\pi} N_\pi^2 \,f_{\pi\pi}(\chi_\pi;\beta_\pi,\gamma_\pi)\\
+\kappa_{\pi\nu} N_\pi N_\nu \,f_{\pi\nu}(\chi_\pi,\chi_\nu;\beta_\pi,\gamma_\pi,\beta_\nu,\gamma_\nu,\vartheta_1,\vartheta_2,\vartheta_3)
+\kappa_{\nu\nu} N_\nu^2 \,f_{\nu\nu}(\chi_\nu;\beta_\nu,\gamma_\nu)\\
+\lambda N_\pi N_\nu \,f_{M}(\beta_\pi,\gamma_\pi,\beta_\nu,\gamma_\nu,\vartheta_1,\vartheta_2,\vartheta_3).
\end{multline}
The functions $f$ (\textit{e.g.},
$f_\pi\<\equiv\langle\nhat_{d\pi}\rangle/N_\pi$) may be read off
directly from Eqns.~(\ref{eqnndrho})--(\ref{eqnMaligned}).  The limit
of both large $N_\pi$ and large $N_\nu$ has been taken, so that in $\langle\Qhat_\rho\cdot
\Qhat_\rho\rangle$ the
term linear in $\Nr$ is suppressed.  The energy surface~(\ref{eqnEgeneralraw}) is
entirely determined by the boson-number-weighted coefficients
$\varepsilon_{\rho}'\<\equiv \varepsilon_\rho N_\rho$,
$\kappa_{\rho\rho'}'\<\equiv \kappa_{\rho\rho'} N_{\rho} N_{\rho'}$,
and $\lambda'\<\equiv \lambda N_{\pi} N_{\nu}$, together with the
$\chi_\rho$, in terms of which
\begin{multline}
\label{eqnEgeneralprime}
\scrE =
\varepsilon_\pi'  f_\pi(\beta_\pi) 
+ \varepsilon_\nu'  f_\nu(\beta_\nu) 
+\kappa_{\pi\pi}'  f_{\pi\pi}(\chi_\pi;\beta_\pi,\gamma_\pi)\\
+\kappa_{\pi\nu}'   f_{\pi\nu}(\chi_\pi,\chi_\nu;\beta_\pi,\gamma_\pi,\beta_\nu,\gamma_\nu,\vartheta_1,\vartheta_2,\vartheta_3)
+\kappa_{\nu\nu}'  f_{\nu\nu}(\chi_\nu;\beta_\nu,\gamma_\nu)\\
+\lambda'  f_{M}(\beta_\pi,\gamma_\pi,\beta_\nu,\gamma_\nu,\vartheta_1,\vartheta_2,\vartheta_3).
\end{multline}

\subsection{Quadrupole interaction coupling coefficients}
\label{subsecquadrupole}

The dominant role of the proton-neutron quadrupole interaction in
producing collective nuclear deformation has been well
established~\cite{talmi1993:shell-ibm}.  Consequently, the like-nucleon quadrupole
interactions are often neglected in the IBM-2 Hamiltonian (see
Refs.~\cite{otsuka1978:ibm2-shell,novoselsky1986:ibm2-o6-xe-ba-pt,lipas1990:ibm2-fspin}).
Microsopic estimates suggest that the shell model proton-proton and
neutron-neutron quadrupole interactions each have $\sim$1/10 to 1/5
the strength of the proton-neutron quadrupole
interaction~\cite{dobaczewski1988:hf-pn-quadrupole}.  However, within
the IBM-2, significant further strength is added to the effective
like-nucleon interactions by renormalization effects arising from
elimination of $g$-wave nucleon pairs from the model space.  This may
yield like-nucleon coupling strengths comparable to the proton-neutron
coupling strength~\cite{scholten1982:ibm2-like-qq}.  The actual
coupling strengths are a subject for further phenomenological study.
Examples spanning a considerable range of values for
$\kapparr/\kappapn$ are included in the following analysis.

For investigation of the phase diagram for the
Hamiltonian~(\ref{eqnHgeneralraw}), it is convenient to again define a
transition parameter $\xi'$ controling the relative weights of the
$\nhat_d$ operator and quadrupole operator in the Hamiltonian.  This
provides a coordinate system for the parameter space like that in
Fig.~\ref{figtetraspace}.  The contributions of the different terms
within each of these operators can then be specified by parameters
$\er$ and $\krrp$ defined such that
\begin{multline}
\label{eqnHquad}
H =
(1-\xi')\,\left(
\frac{\ep}{\Np}\nhat_{d\pi}
+\frac{\en}{\Nn}\nhat_{d\nu}
\right) 
\\
\,-\,
\xi'\,\left(
\frac{\kpp}{\Np^2}\Qhat_\pi^{\chi_\pi}\cdot\Qhat_\pi^{\chi_\pi}
+\frac{\kpn}{\Np\Nn}\Qhat_\pi^{\chi_\pi}\cdot\Qhat_\nu^{\chi_\nu}
+\frac{\knn}{\Nn^2}\Qhat_\nu^{\chi_\nu}\cdot\Qhat_\nu^{\chi_\nu} 
\right),
\end{multline}
yielding a $\Nr$-independent expression
\begin{equation}
\label{eqnEquad}
\scrE =
(1-\xi')\left(
\ep f_\pi
+ \en f_\nu
\right)
-
\xi'\left(
\kpp f_{\pi\pi}
+\kpn f_{\pi\nu}
+\knn f_{\nu\nu}
\right) 
\end{equation}
for the energy surface.  To unambiguously define $\xi'$ and avoid
redundancy in the parameters, we adopt the normalization convention
$\ep+\en\<=1$ and $\kpp+\kpn+\knn\<=1$.  This choice gives
$\er\<=\epsilonr'/(\epsilonp'+\epsilonn')$ and
$\krrp\<=\kapparrp'/(\kappapp'+\kappapn'+\kappann')$ and is consistent
with the definition of $\xi'$ in Eqn.~(\ref{eqnHdieperink}).  For the
$F$-spin invariant Hamiltonian with arbitrary
$\Np/\Nn$, the parameters are $\ep\<=\Np/N$, $\en\<=\Nn/N$,
$\kpp\<=\Np^2/N^2$, $\kpn\<=2\Np\Nn/N^2$, and $\knn\<=\Nn^2/N^2$.

Analytic results for the phase structure of the
Hamiltonian~(\ref{eqnHgeneralraw}) with arbitrary quadrupole coupling
coefficients can be obtained in the case $\xi'\<=1$ by a
straight-forward extension of the analysis described in
Sec.~\ref{subsecbackplane}.  Surrounding the $\grpsuthreepn$ and
$\grpsuthreepnbar$ dynamical symmetries are regions of parameter space in
which the equilibrium deformation is axially symmetric
($\gamma_\pi\<=\gamma_\nu\<=0$ or $\pi/3$).  The energy surface in
this case can be expressed in terms of the function $F$ (of
Appendix~\ref{appf}) as
\begin{multline}
\label{eqnscrEaxialquad}
\scrE(\beta_\pi,\beta_\nu)=4\Bigl[ 
\kappa_{\pi\pi}'F(-\chi_\pi/\sqrt{14};\sigma\beta_\pi)^2
\\
+\kappa_{\pi\nu}'F(-\chi_\pi/\sqrt{14};\sigma\beta_\pi)
F(-\chi_\nu/\sqrt{14};\sigma\beta_\nu)
+\kappa_{\nu\nu}'F(-\chi_\nu/\sqrt{14};\sigma\beta_\nu)^2
\Bigr],
\end{multline}
with $\sigma=\cos 3\gamma_\rho=\pm1$.  This expression is simply a
quadratic form in the quantities $F$, and the extremization problem
reduces to a constrained minimization of the quadratic form with
respect to these quantities.  Since the $\beta_\rho$ are limited to
positive values, the $F$ are restricted to the rectangular region $ 0
\<\leq
F(-\chi_\pi/\sqrt{14};\beta_\pi)
\<\leq
x_+(-\chi_\pi/\sqrt{14}) $ and $ 0
\<\leq
F(-\chi_\nu/\sqrt{14};\beta_\nu)
\<\leq
x_+(-\chi_\nu/\sqrt{14}) $ for the case $\sigma\<=+1$
[$\grpsuthreepn$-like] or to $ x_-(-\chi_\pi/\sqrt{14})
\<\leq
F(-\chi_\pi/\sqrt{14};-\beta_\pi)
\<\leq
0 $ and $ x_-(-\chi_\nu/\sqrt{14})
\<\leq
F(-\chi_\nu/\sqrt{14};-\beta_\nu)
\<\leq
0 $ for $\sigma\<=-1$ [$\grpsuthreepnbar$-like].  Provided
$\kappa_{\pi\pi}$, $\kappa_{\pi\nu}$, and $\kappa_{\nu\nu}$ are all
negative, the global minimum is again given by
Eqn.~(\ref{eqnbetaprolate}) for $\sigma\<=+1$ or
Eqn.~(\ref{eqnbetaoblate}) for $\sigma\<=-1$.  The energy at the
minimum is, in terms of the equilibrium $\beta_\rho$ values,
\begin{equation}
\label{eqnscrEquad}
\scrE = \kappa_{\pi\pi}'\beta_\pi^2 +
\kappa_{\pi\nu}'\beta_\pi\beta_\nu
+ \kappa_{\nu\nu}'\beta_\nu^2.
\end{equation}

The boundary curve separating axial and triaxial deformations is found
as in Sec.~\ref{subsecbackplane} from analysis of the second
derivatives of $\scrE$.  The axial minimum becomes unstable with
respect to $\gamma$ deformation along a curve in parameter space,
again most conveniently expressed in terms of the equilibrium
$\beta_\rho$ values, given by
\begin{multline}
\label{eqnboundaryquad}
1 =
\frac{
18\frac{\kappa_{\pi\pi}'}{\kappa_{\pi\nu}'}\betap(\betap^2-1)+\betan(2\betap^2-1)(\betap^2+1)
} {2\betap (\betap^2-2)^2}
\\
\times
\frac{
18\frac{\kappa_{\nu\nu}'}{\kappa_{\pi\nu}'}\betan(\betan^2-1)+\betap(2\betan^2-1)(\betan^2+1)
} {2 \betan (\betan^2-2)^2} .
\end{multline}
The instability occurs against deformations with a ratio
$\gamma_\pi/\gamma_\nu$ given
by Eqn.~(\ref{eqngammadir}) with the values
\begin{equation}
\begin{aligned}
A&\equiv 9 \kappa_{\pi\pi}' \beta_\pi^2
(\beta_\pi^2-1)+\beta_\pi\beta_\nu(2\beta_\pi^2-1)(\beta_\pi^2+1)\\
B&\equiv -2 \kappa_{\pi\nu}' \beta_\pi \beta_\nu
(\beta\pi^2-2)(\beta_\nu^2-2)\\ C&\equiv 9 \kappa_{\nu\nu}'
\beta_\nu^2
(\beta_\nu^2-1)+\beta_\pi\beta_\nu(2\beta_\nu^2-1)(\beta_\nu^2+1).
\end{aligned}
\end{equation}
Along the $\grpsuthreepn$-$\grpsuthreepnstar$ line in parameter space, the
location of the phase transition is obtained by solution of the
quartic equation
\begin{equation}
\label{eqnquarticquad}
18 \frac{\kappa_{\nu\nu}'}{\kappa_{\pi\nu}'}\beta_\nu(\beta_\nu^2-1)
+ \sqrt{2}(2\beta_\nu^2-1)(\beta_\nu^2+1)=0,
\end{equation}
and a similar relation with interchanged proton and neutron labels
holds for the $\grpsuthreepn$-$\grpsuthreepnstarbar$ line.  Observe that the
$\chi_\nu$ value at which the transition occurs depends only upon the
ratio $\kappa_{\nu\nu}'/\kappa_{\pi\nu}'$.

\begin{figure}
\begin{center}
\includegraphics[width=0.9\hsize]{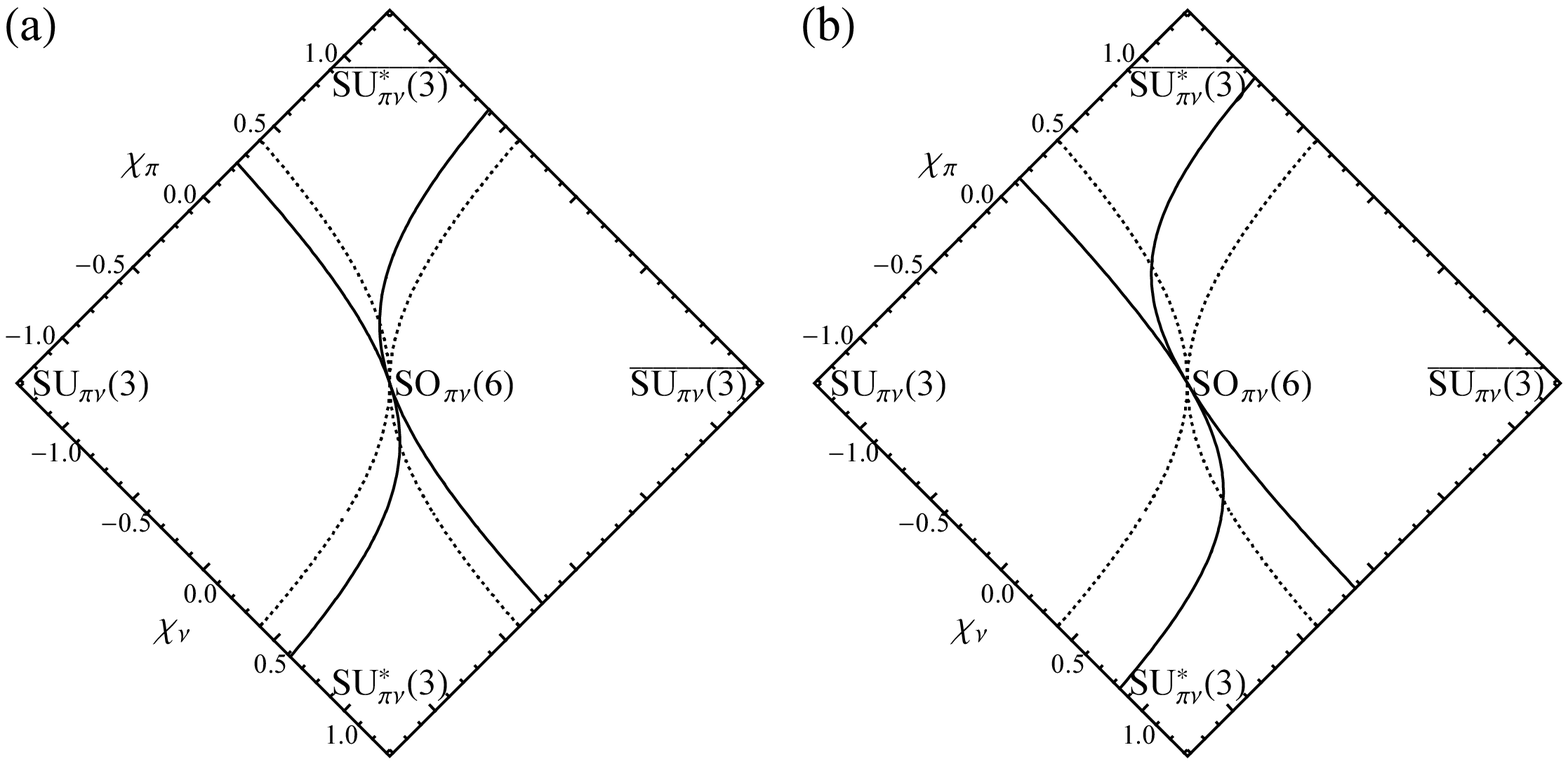}
\end{center}
\caption
{The curve of second-order phase transition between axial and triaxial
deformation in the plane $\xi'\<=1$, shown for the $F$-spin invariant
Hamiltonian with different boson number ratios $N_\pi/N_\nu$:
(a)~$N_\pi/N_\nu\<=2$ and (b)~$N_\pi/N_\nu\<=4$.  The boundary curve
for $N_\pi/N_\nu\<=1$ (Fig.~\ref{figboundary}) is shown in each panel
(dashed curve) for comparison.  }
\label{figboundaryquadnratio}
\end{figure}

The dependence of the second-order transition curve upon $N_\pi/N_\nu$
is illustrated in Fig.~\ref{figboundaryquadnratio}.  Since
$\kappapp'\<\neq\kappann'$, the phase diagram is not symmetric under
reflection with respect to $\chis$ or $\chiv$ but remains symmetric
under inversion~(\ref{eqnchiinversion}).  It is seen that increasing
$N_\pi/N_\nu$ from unity moves the $\grpsuthreepn$-$\grpsuthreepnstar$
transition farther away in parameter space from $\grpsuthreepn$ while
moving the $\grpsuthreepn$-$\grpsuthreepnstarbar$ transition closer.  The
$N_\pi/N_\nu$ depencence of the boundary curve raises the possibility
of considering phase transitions at a fixed point in the Hamiltonian
parameter space with $N_\pi/N_\nu$ as the control parameter,
\textit{i.e.}, phase transitions as a function of particle number.

For the $F$-spin invariant Hamiltonian considered in
Sec.~\ref{secschematic}, the transition between prolate axially
symmetric equilibrium [$\grpsuthreepn$-like] and oblate axially symmetric
equilibrium [$\grpsuthreepnbar$-like] always proceeds through an
intermediate stage of triaxial equilibrium, except at the point in
parameter space corresponding to $\grpsosixpn$.  However, if
$\kappa_{\pi\pi}'$ and $\kappa_{\nu\nu}'$ are sufficiently small
relative to $\kappa_{\pi\nu}'$, it is possible for a direct transition
to occur between the prolate and oblate equilibria of
Eqns.~(\ref{eqnbetaprolate}) and~(\ref{eqnbetaoblate}) without either
minimum becoming unstable with respect to triaxiality and thus for the
regions of prolate and oblate equilibrium in the phase diagram to
share a common boundary.  The boundary curve is given by
\begin{multline}
\label{eqnflipquad}
\kappapp'^2\ap^2(\ap^2+1)
-\kappann'^2\an^2(\an^2+1)
+\kappapn'\kappapp'\ap\an(\ap^2+1)
\\
-\kappapn'\kappann'\ap\an(\an^2+1)
-\frac{1}{4}\kappapn'^2(\ap^2-\an^2)
=0,
\end{multline}
where $\ar\<\equiv-\chir/\sqrt{14}$.  Over the range of $\chir$ and
parameter values considered, this curve differs little from its
small-$\chir$ approximation, the line
$\chis/\chiv\<=-(\kappapp'-\kappann')/(\kappapp'+\kappapn'+\kappann')$.
For any proton-neutron symmetric energy surface
($\kappapp'\<=\kappann'$), the boundary curve~(\ref{eqnflipquad})
reduces to the line $\chis\<=0$.  The general problem of determining
whether or not the prolate and oblate regions of the phase diagram
share a common boundary involves solving for an intersection of the
curves~(\ref{eqnboundaryquad}) and~(\ref{eqnflipquad}).  For the
proton-neutron symmetric case ($\kappapp'\<=\kappann'$), a
prolate-oblate boundary arises for
$\kapparr'/\kappapn'\<\leq(\sqrt{13}-2)/18\<\approx0.08919$.

\begin{figure}
\begin{center}
\includegraphics[width=0.9\hsize]{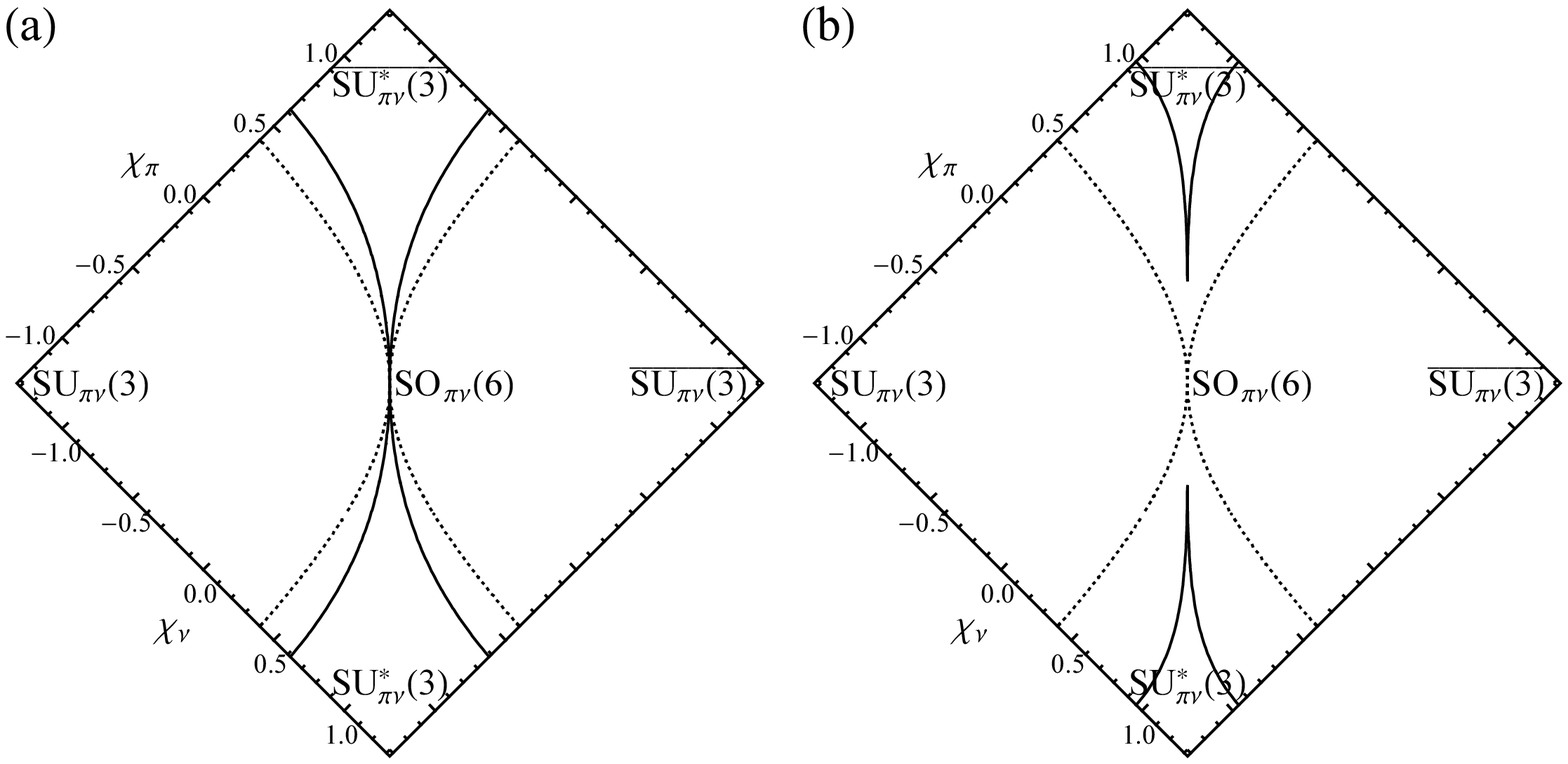}
\end{center}
\caption
{The curve of second-order phase transition between axial and triaxial
deformation in the plane $\xi'\<=1$, shown for different relative
strengths of like-nucleon and proton-neutron quadrupole interactions:
(a)~$\kappa_{\pi\pi}/\kappa_{\pi\nu}\<=\kappa_{\nu\nu}/\kappa_{\pi\nu}\<=1/4$
and
(b)~$\kappa_{\pi\pi}/\kappa_{\pi\nu}\<=\kappa_{\nu\nu}/\kappa_{\pi\nu}\<=1/12$.
Both curves are for $N_\pi\<=N_\nu$.  The boundary curve for
$\kappa_{\pi\pi}/\kappa_{\pi\nu}\<=\kappa_{\nu\nu}/\kappa_{\pi\nu}\<=1/2$
(Fig.~\ref{figboundary}) is shown in each panel (dashed curve) for
comparison.  }
\label{figboundaryquadkapparatio}
\end{figure}

\begin{figure}
\begin{center}
\includegraphics[width=\hsize]{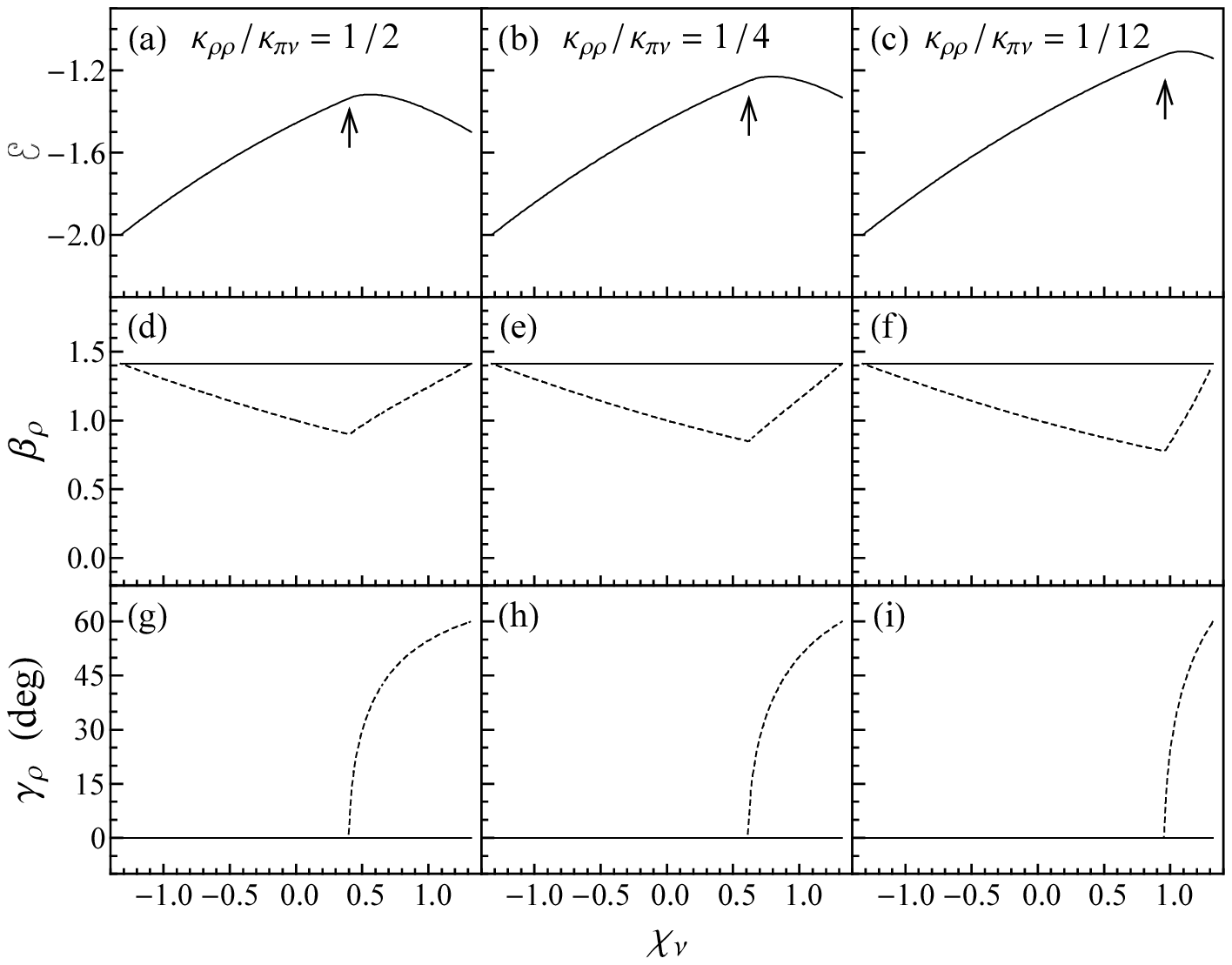}
\end{center}
\caption
{Evolution of the equilibrium properties between the $\grpsuthreepn$
and $\grpsuthreepnstar$ dynamical symmetries, for different relative
strengths of like-nucleon and proton-neutron quadrupole interactions:
(left)~$\kappa_{\pi\pi}/\kappa_{\pi\nu}\<=\kappa_{\nu\nu}/\kappa_{\pi\nu}\<=1/2$
(for reference),
(middle)~$\kappa_{\pi\pi}/\kappa_{\pi\nu}\<=\kappa_{\nu\nu}/\kappa_{\pi\nu}\<=1/4$
and
(right)~$\kappa_{\pi\pi}/\kappa_{\pi\nu}\<=\kappa_{\nu\nu}/\kappa_{\pi\nu}\<=1/12$.
The curves are for $N_\pi\<=N_\nu$.  Proton fluid variables are
represented by solid curves, while neutron fluid variables are
represented by dashed curves.  Points of phase transition are marked
with arrows.  }
\label{figraysquad}
\end{figure}

The dependence of the boundary curve upon the relative strengths of
like-nucleon and proton-neutron quadrupole coupling coefficients is
shown in Fig.~\ref{figboundaryquadkapparatio}.  As the $\kapparr'$ are
reduced in strength relative to $\kappapn'$, the region of triaxial
equilibrium contracts.  Its separation from the origin and the onset
of a direct prolate-oblate transition for
$\kapparr'/\kappapn'\<\leq(\sqrt{13}-2)/18$ is illustrated in
Fig.~\ref{figboundaryquadkapparatio}(b).  The parameter values chosen
for Fig.~\ref{figboundaryquadnratio}(a) and for
Fig.~\ref{figboundaryquadkapparatio}(a) are such that
$\kappa_{\nu\nu}'/\kappa_{\pi\nu}'$ is the same
($\kappa_{\nu\nu}'/\kappa_{\pi\nu}'\<=1/4$) in both cases.  Thus,
though the curves in these two figures are quite different overall, by
Eqn.~(\ref{eqnquarticquad}) they share the same endpoint
($\chi_\nu\<\approx0.6177$) on the $\grpsuthreepn$-$\grpsuthreepnstar$ line.
Numerical results for the equilibrium values of the energy and
coordinates along the $\grpsuthreepn$-$\grpsuthreepnstar$ line for different
relative strengths of like-nucleon and proton-neutron quadrupole
interactions are shown in Fig.~\ref{figraysquad}.  Although the
transition to triaxial equilibrium is progressively delayed as
$\kappa_{\nu\nu}'/\kappa_{\pi\nu}'$ decreases, the triaxial
configuration of Fig.~\ref{figortho}, with orthogonal symmetry axes
($\gammap\<=0$, $\gamman\<=\pi/3$), is still ultimately achieved in
the $\grpsuthreepnstar$ limit.

A qualitative understanding of the mechanism underlying the parameter
dependences observed in Figs.~\ref{figboundaryquadnratio}
and~\ref{figboundaryquadkapparatio} is easily obtained.  Among the
quadrupole terms in the energy surface, the
$\langle\Qhat_\pi\cdot\Qhat_\pi\rangle$ term has a dominant influence
on the proton fluid equilibrium deformation, the
$\langle\Qhat_\nu\cdot\Qhat_\nu\rangle$ terms drives the neutron fluid
deformation, and the $\langle\Qhat_\pi\cdot\Qhat_\nu\rangle$ term
couples the two deformations.  The $\grpsuthreepnstar$ configuration with
orthogonal symmetry axes (Fig.~\ref{figortho}) arises since the
$\langle\Qhat_\pi\cdot\Qhat_\pi\rangle$ term stabilizes the proton
fluid about a prolate deformation and the
$\langle\Qhat_\nu\cdot\Qhat_\nu\rangle$ term stabilizes the neutron
fluid about an oblate deformation, while the
$\langle\Qhat_\pi\cdot\Qhat_\nu\rangle$ term is responsible for the
relative orientation of the symmetry axes.  All three terms are
essential to producing the $\grpsuthreepnstar$ triaxial equilibrium
configuration.  If the strengths of the like-nucleon quadrupole
interactions are both reduced relative to that of the proton-neutron
interaction (Figs.~\ref{figboundaryquadkapparatio}
and~\ref{figraysquad}), the ability of the like-nucleon terms to
stabilize the proton and neutron fluids about distinct deformations is
reduced, against the tendency of the proton-neutron term to favor
equal deformations.  The onset of triaxiality is thus delayed.  For
$N_\pi/N_\nu$ greater than unity (Fig.~\ref{figboundaryquadnratio}),
the relative weight of the $\langle\Qhat_\pi\cdot\Qhat_\pi\rangle$
contribution to $\scrE$ increases, while that of the
$\langle\Qhat_\nu\cdot\Qhat_\nu\rangle$ contribution decreases.  Thus,
a larger larger positive $\chi_\nu$ (oblate tendency for the neutron
fluid) is needed for the neutrons to undergo a transition to an oblate
configuration against the restraining influence of the protons, while
only a small positive $\chi_\pi$ (oblate tendency for the proton
fluid) is necessary for the protons to undergo a transition to an
oblate configuration.

Limited analytic results can also be obtained for the transition
between spherical and deformed equilibrium.  Consider the stability of
the minimum at $\betap\<=\betan\<=0$.  The second derivative of $\scrE$ along
the ray $\beta_\pi\<=u_\pi
\beta$ and $\beta_\nu\<= u_\nu \beta$ is 
\begin{multline}
\label{eqnbetasoftquad}
\left.\frac{d^2\scrE}{d\beta^2}\right|_{\beta=0}
=2u_\pi^2[\ep(1-\xi')-4\kpp\xi']
\\
+2u_\nu^2[\en(1-\xi')-4\knn\xi']
-8u_\pi u_\nu\kpn\xi'\cos(\gamma_\pi-\gamma_\nu).
\end{multline}
As in the special case considered in Sec.~\ref{subsecspherical}, this
quantity is independent of $\chi_S$ and $\chi_V$ and depends upon the
$\gamma_\rho$ coordinates only through their difference
$\gamma_\pi-\gamma_\nu$.  As $\xi'$ is increased from 0, vanishing
of this second derivative for some value of $u_\pi/u_\nu$ and
$\gamma_\pi-\gamma_\nu$ indicates the onset of instability.  Since the
coefficient of $\cos(\gamma_\pi-\gamma_\nu)$ is negative, instability
must first occur for $\gamma_\pi-\gamma_\nu\<=0$.  With
$\gamma_\pi-\gamma_\nu$ set to zero,
$\partial^2\scrE/\partial\beta^2|_{\beta=0}$ vanishes at
\begin{equation}
\label{eqnxipinstabilityquad}
\xi'=\frac{\ep u_\pi^2+\en u_\nu^2}{(\ep u_\pi^2+\en u_\nu^2)+4(\kpp
u_\pi^2+\kpn u_\pi u_\nu+\knn u_\nu^2)}.
\end{equation}
Instability against deformations with \textit{equal} $\betap$ and
$\betan$  therefore occurs at $\xi'\<=1/5$ for
\textit{any} values of the energy surface parameters.
However, the minimum at $\beta\<=0$ can in general become unstable
against deformations with
\textit{unequal} $\betap$ and $\betan$ at a smaller value of $\xi'$.
The direction in $(\betap,\betan)$ space in which instability first sets in
is given by a quadratic equation for $u_\pi/u_\nu$,
\begin{equation}
\label{eqnuratioquad}
\en\kpn
+2(\en\kpp-\ep\knn)\left(\frac{u_\pi}{u_\nu}\right)-\ep\kpn\left(\frac{u_\pi}{u_\nu}\right)^2
=0.
\end{equation}
The $\xi'$ value at which instability occurs follows from this
$u_\pi/u_\nu$ value by Eqn.~(\ref{eqnxipinstabilityquad}).  There are
two important special cases in which instability first occurs at
$\xi'\<=1/5$: (1) for the $F$-spin invariant
Hamiltonian~(\ref{eqnHdieperink}) with arbitrary $\Np/\Nn$, and (2)
for energy surface parameters which are proton-neutron symmetric
(\textit{i.e.}, $\epsilonp'\<=\epsilonn'$ and
$\kappapp'\<=\kappann'$), as are obtained whenever a proton-neutron
symmetric Hamiltonian is considered with $\Np\<=\Nn$.

\begin{figure}
\begin{center}
\includegraphics[width=0.495\hsize]{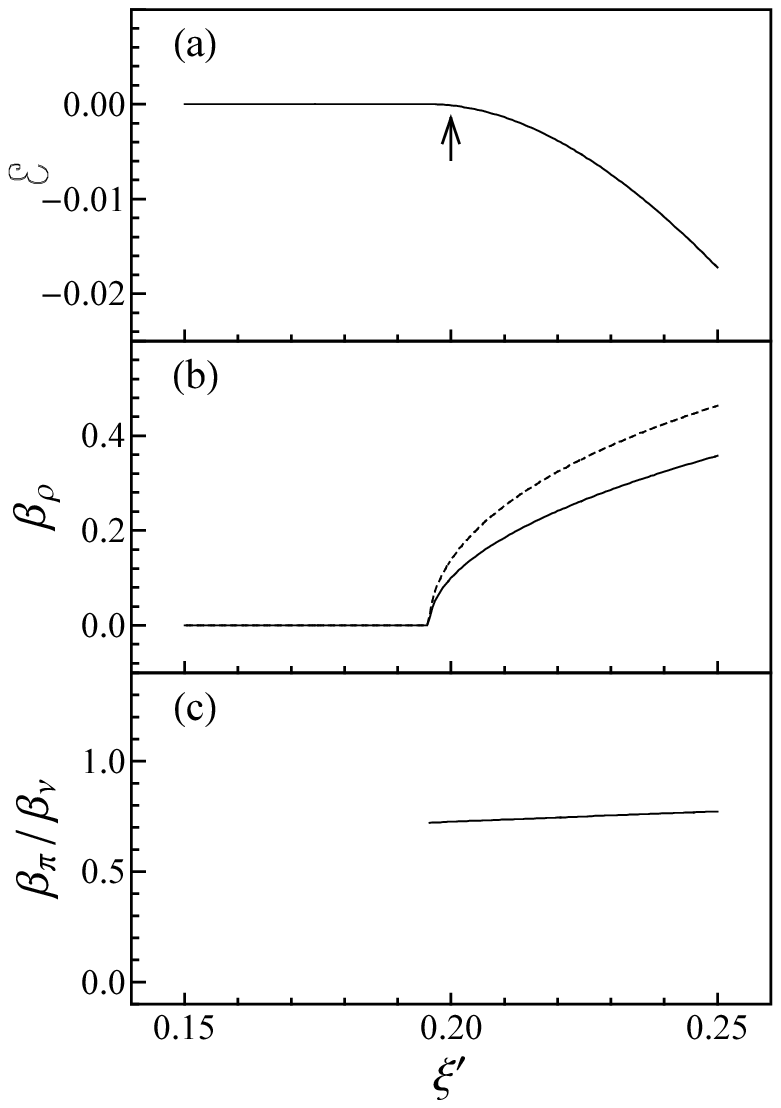}
\end{center}
\caption
{Evolution of the equilibrium properties along the
$\grpufivepn$-$\grpsosixpn$ line in parameter space, illustrating instability
against deformation with unequal $\betap$ and $\betan$.  Results are
shown for the Hamiltonian~(\ref{eqnHquad}) with
$\epsilonp\<=\epsilonn$ and $\kappapp\<=\kappann\<=\kappapn/4$
for~(left) the symmetric case $\Np/\Nn\<=1$ and~(right) the asymmetric
case $\Np/\Nn\<=4$.  The variable $\beta_\pi$ is represented
by the solid curve, while $\beta_\nu$ is represented by the dashed
curve.  The point of phase transition is marked with an arrow.  }
\label{figraysunequal}
\end{figure}

The evolution of equilibrium properties across the
$\grpufivepn$-$\grpsosixpn$ second-order transition point for a case
asymmetric in $\betap$ and $\betan$ is shown in
Fig.~\ref{figraysunequal}.  From Eqns.~(\ref{eqnxipinstabilityquad})
and~(\ref{eqnuratioquad}), the phase transition for the parameter
values used in the figure ($\epsilonp\<=\epsilonn$,
$\kappapp\<=\kappann=\kappapn/4$, and $\Np/\Nn\<=4$) occurs at
$\xi'\<\approx0.1959$, with $\betap/\betan\<\approx0.7215$.  The
evolution of the ratio $\betap/\betan$ past this point is shown in
Fig.~\ref{figraysunequal}(c).  Observe that the proton and neutron
equilibrium properties in this example are unequal even within the
plane $\chiv\<=0$, unlike the case of Sec.~\ref{subsecdieperink},
illustrating that the the IBM-2 equilibrium problem in the general
proton-neutron asymmetric case does not reduce to the one-fluid IBM
problem.

\begin{figure}
\begin{center}
\includegraphics[width=0.75\hsize]{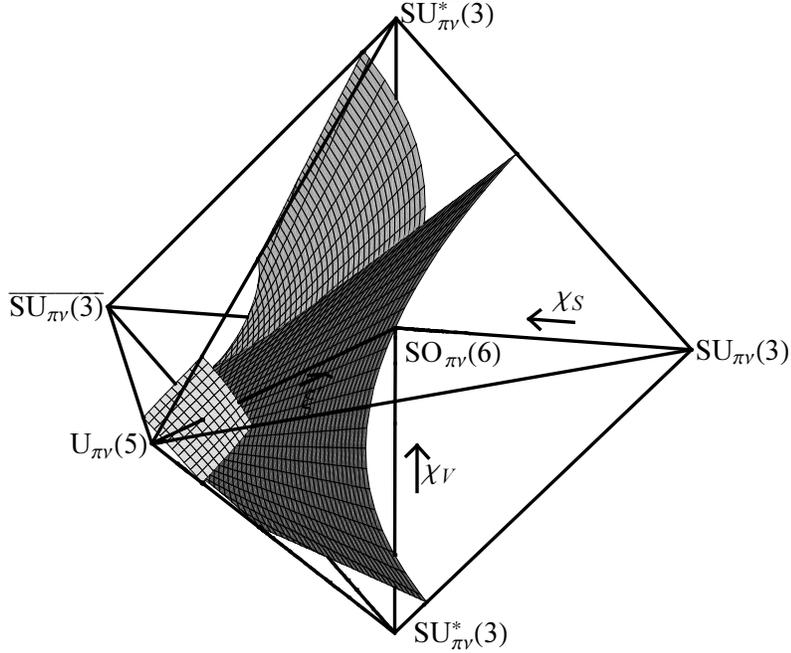}
\end{center}
\caption
{Phase diagram of the IBM-2 for the Hamiltonian~(\ref{eqnHquad}), with
$\kappa_{\pi\pi}/\kappa_{\pi\nu}\<=\kappa_{\nu\nu}/\kappa_{\pi\nu}\<=1/4$ and $N_\pi/N_\nu\<=4$, as obtained by
numerical minimization of $\scrE$.  The surfaces of first-order 
and second-order transition between regions of undeformed,
axially symmetric deformed, and triaxially deformed equilibria are
shown.  The axes are defined as in Fig.~\ref{figtetraspace}.  
}
\label{figtetraquad}
\end{figure}

The minimum at zero deformation always becomes unstable at the $\xi'$
value determined by Eqns.~(\ref{eqnxipinstabilityquad})
and~(\ref{eqnuratioquad}), independent of $\chis$ and $\chiv$.
However, this gives rise to a second order phase transition only if a
first-order transition to a distinct minimum has not already occured
at smaller $\xi'$ for those values of $\chis$ and $\chiv$, as
discussed in Sec.~\ref{subsecspherical}.  For the $F$-spin invariant
Hamiltonian~(\ref{eqnHdieperink}) with arbitrary $\Np/\Nn$, an
extension of relation~(\ref{eqnscrEslice}) is readily found, namely
that $\scrE(\xi',\chi_S,\chi_V;\beta,0,\beta,0)$ is invariant along
any line of constant $\Np\chip+\Nn\chin$.  By arguments analogous to
those of Sec.~\ref{subsecspherical}, a line of second order phase
transition occurs for $\xi'\<=1/5$ and $\chip/\chin\<=-\Nn/\Np$,
embedded in a surface of first order phase transition.  For a
Hamiltonian with arbitrary coupling constants, however, no such simple
result is obtained.  The full phase diagram of the IBM-2
Hamiltonian~(\ref{eqnHquad}) for a set of parameters involving
proton-neutron symmetric couplings
($\kappa_{\pi\pi}/\kappa_{\pi\nu}\<=\kappa_{\nu\nu}/\kappa_{\pi\nu}\<=1/4$)
but unequal boson numbers ($N_\pi/N_\nu\<=4$), obtained by numerical
mimimization, is shown in Fig.~\ref{figtetraquad}.

\subsection{Majorana operator}
\label{subsecmajorana}

The Majorana operator $\Mhat$ is the analogue in the IBM-2 Hamiltonian
to the proton-neutron symmetry energy of the liquid drop model.  This
operator arises through a combination of direct shell model effects
and renormalization
effects~\cite{scholten1983:ibm2-microscopic-majorana,vanegmond1984:ibm2-microscopic}.
As is evident from Eqn.~(\ref{eqnMaligned}), this operator
energetically discourages configurations with
$\beta_\pi\<\neq\beta_\nu$ or $\gamma_\pi\<\neq\gamma_\nu$.  The
approximate strength of the Majorana contribution can be deduced from
the energies of mixed symmetry
excitations~\cite{scholten1985:ibm2-mss}, including the scissors mode
excitation in axial rotor nuclei (\textit{e.g.},
Ref.~\cite{hartmann1987:ibm2-scissors-systematics}).  Comparison of
the $\beta$-vibrational or $\gamma$-vibrational energy scale
($\sim$1\,MeV) with the scissors mode energy scale ($\sim$2.5\,MeV)
indicates $\lambda/\abs{\kappapn}\<\approx 5$ to be a generally
reasonable estimate.

For axially symmetric configurations, the contribution of the Majorana
operator to the energy surface is, from Eqn.~(\ref{eqnMaligned}),
$\langle \Mhat \rangle = N_\pi
N_\nu(1+\beta_\pi^2)^{-1}(1+\beta_\nu^2)^{-1}(\betap-\betan)^2$.  If
the equilibrium configuration without the Majorana operator already
has $\betap\<=\betan$, as in the $\chis\<=0$ plane for the
proton-neutron symmetric case of Fig.~\ref{figtetraspace}, then
introduction of the Majorana term has no further effect.  Otherwise, the
effect of the Majorana contribution is, naturally, to bring the
equilibrium values $\betap$ and $\betan$ closer to each other.  In the
plane $\xi'\<=1$, this invalidates the simple
results~(\ref{eqnbetaprolate}) and~(\ref{eqnbetaoblate}) giving the
equilibrium $\betap$ purely as a function of $\chip$ and the
equilibrium $\betan$ as a function of $\chin$.  The derivation of the
simple equations~(\ref{eqnboundary}) and~(\ref{eqnboundaryquad}) for
the curve on which the axial minimum becomes unstable with respect to
the $\gammar$ depended upon the use of these results to eliminate
$\chir$ in favor of $\betar$ at the minimum, a simplification which is
no longer possible with a Majorana contribution.  

\begin{figure}[p]
\begin{center}
\includegraphics[width=\hsize]{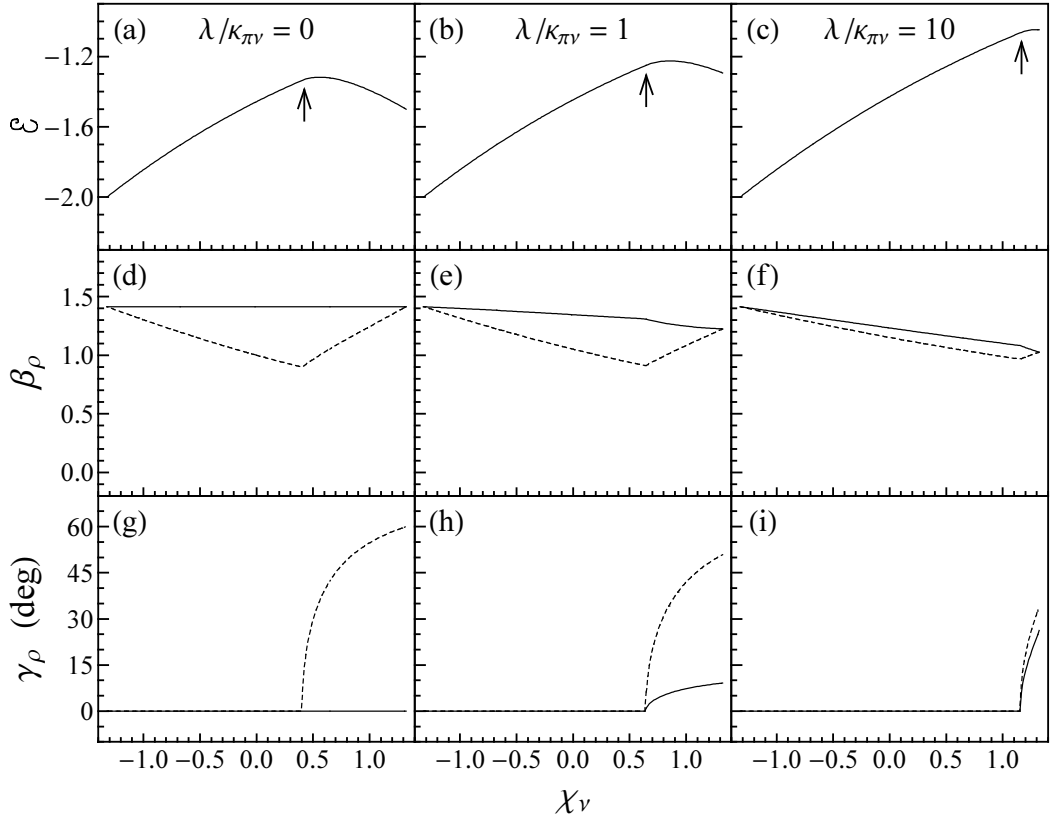}
\end{center}
\caption
{Evolution of the equilibrium properties between the $\grpsuthreepn$ and
$\grpsuthreepnstar$ dynamical symmetries, for different
Majorana operator strengths: (left)~$\lambda/\kappa_{\pi\nu}\<=0$ (for
reference), (middle)~$\lambda/\kappa_{\pi\nu}\<=1$, and
(right)~$\lambda/\kappa_{\pi\nu}\<=10$.  The curves are for the
quadrupole coefficient values
$\kappa_{\pi\pi}/\kappa_{\pi\nu}\<=\kappa_{\nu\nu}/\kappa_{\pi\nu}\<=1/2$
with $N_\pi\<=N_\nu$.  Proton fluid variables are represented by solid
curves, while neutron fluid variables are represented by dashed
curves.  Points of phase transition are marked with arrows.   }
\label{figraysmajorana}
\end{figure}

\begin{figure}[p]
\begin{center}
\includegraphics[width=0.5\hsize]{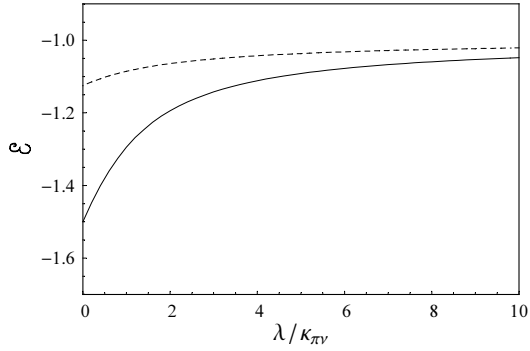}
\end{center}
\caption
{Global minimum energy (solid curve) and the lowest energy for axial
($\gammap\<=\gamman\<=0$) deformations (dashed curve), shown as a
function of $\lambda/\kappa_{\pi\nu}$, at the $\grpsuthreepnstar$
point in parameter space.  The difference is a measure of the
$\gamma$-stiffness of the triaxial minimum.  The curves are for
quadrupole coefficient values
$\kappa_{\pi\pi}/\kappa_{\pi\nu}\<=\kappa_{\nu\nu}/\kappa_{\pi\nu}\<=1/2$
with $N_\pi\<=N_\nu$.  }
\label{figmajoranasoftness}
\end{figure}

Investigation of the equilibrium properties in the presence of a
Majorana operator must therefore rely upon numerical minimization.
The evolution of the equilibrium properties between the
$\grpsuthreepn$ and $\grpsuthreepnstar$ points in parameter space, for
different strengths of the Majorana term in the Hamiltonian, is shown
in Fig.~\ref{figraysmajorana}.  The Majorana operator is seen to have
two main effects on the transition to triaxial equilibrium.  The
triaxial configuration of Fig.~\ref{figortho} is highly proton-neutron
asymmetric, and thus penalized energetically by the Majorana operator,
as seen from Eqn.~(\ref{eqnMaligned}).  Thus, first, the transition to
towards such triaxial structure is delayed by the Majorana operator.
Second, the Majorana operator has the effect of bringing the proton
and neutron coordinate values closer together throughout the
transition.  The $\grpsuthreepn$-$\grpsuthreepnstar$ line is an
extreme case.  Without the Majorana operator
[Fig.~\ref{figrayssu3su3star} or~\ref{figraysmajorana}(a)], the
evolution of the proton coordinates is completely decoupled from that
of the neutron coordinates, and their equilibrium values are constant
along the line.  A Majorana strength of
$\lambda/\kappa_{\pi\nu}\<\approx10$ is sufficent, however, to cause
the two fluids' $\betar$ and $\gammar$ coordinates to remain nearly
equal throughout the evolution.  The triaxial configurations produced
are thus essentially one-fluid, like those described by the Davydov
model~\cite{davydov1958:arm-intro} or by the one-fluid IBM with
three-body or four-body operators in the
Hamiltonian~\cite{ginocchio1980:ibm-coherent-bohr,vanisacker1981:ibm-triax,heyde1984:ibm-cubic-triax,casten1985:ibm-o6-triax,smirnov2000:ibm-triax}.
As the Majorana strength increases, the global minimum at triaxial deformation
becomes very shallow, and the difference in energy between the
$\gammap\<\approx\gamman\<\approx\pi/6$ triaxial minimum and axial
deformations ($\gammap\<=\gamman\<=0$) approaches zero, as shown in
Fig.~\ref{figmajoranasoftness}.
This indicates that the equilibrium structure for large Majorana strengths
is essentially one-fluid \textit{$\gamma$-unstable}, or
$\grpsosix$-like, rather than rigidly triaxial.

The decomposition of the Majorana operator into multipole components,
easily obtained from Ref.~\cite[(A2)]{ginocchio1992:ibm2-shapes},
includes a nonzero hexadecapole contribution.  The analysis of
Ref.~\cite{ginocchio1992:ibm2-shapes} therefore allows the existence
of ``oblique'' equilibrium configurations, in which the proton and
neutron intrinsic frames are not aligned.  
For axially symmetric aligned configurations ($\vartheta_i$ and
$\gammar$ all vanishing), instability with respect to the Euler angles
is simple to investigate.  The Hessian matrix of $\scrE$ with respect
to the coordinates $\beta_\pi$, $\beta_\nu$, $\gamma_\pi$,
$\gamma_\nu$, and $\vartheta_2$ decomposes as a direct sum
$\scrH=\scrH_\beta\oplus\scrH_\gamma\oplus\scrH_{\vartheta_2}$, with
$\scrH_\beta$ involving only derivatives with respect to the $\betar$,
\textit{etc.}.  The
dependence upon the azimuthal Euler angles $\vartheta_1$ and
$\vartheta_3$ vanishes by axial symmetry.  Thus, instability would be
indicated simply by a negative second derivative of $\scrE$ with
respect to the opening angle $\vartheta_2$ between the proton and
neutron frames.  For the parameter ranges considered ($\kappapn'\<<0$,
$\lambda'\<\geq0$, $\abs{\chir}\<\leq\rootstinline$) and coordinate
values encountered ($\betar\<\lesssim\sqrt{2}$) in the present study,
this quantity is always positive.  The possibility of a first order
transition to a distinct minimum with nonzero Euler angles must also
be considered.  Numerical searches provide no evidence for a such a
transition, but such searches are necessarily not exhaustive due to
the large number of parameters involved.

\section{Spectroscopic properties}
\label{secspectro}

\subsection{Basic properties}
\label{subsecspectrobasic}

To provide a connection between the IBM-2 phase structure considered
so far and observable quantities, let us now consider the evolution of
the predicted spectroscopic properties~--- eigenvalues and
electromagnetic transition strengths~--- across the phase transitions.
The present discussion emphasises the gross spectroscopic features
which emerge in the transitions between the different regions of the
phase diagram of Fig.~\ref{figquadrant}.  Numerical studies of several
observables in the vicinity of the $\grpsuthreepnstar$ dynamical
symmetry may also be found in
Refs.~\cite{sevrin1987:ibm2-su3star-1,sevrin1987:ibm2-su3star-2}.

Electric quadrupole and magnetic dipole transition matrix elements are
calculated using transition
operators~\cite{iachello1987:ibm}
\begin{align}
\label{eqnTE2}
T^{(E2)}&= e_\pi \Qhat_\pi^{\chi_\pi^{(E2)}} + e_\nu
\Qhat_\nu^{\chi_\nu^{(E2)}}
\\
\intertext{and}
\label{eqnTM1}
T^{(M1)}&= \sqrt\frac{3}{4\pi} \bigl[ 
g_\pi\Lhat_\pi + g_\nu\Lhat_\nu
\bigr],
\end{align}
where $\Lhat_\rho\equiv
\sqrt{10}(d_\rho^{\dag}\<\times\tilde{d}_\rho)^{(1)}$.  The transition
strengths are $B(\sigma\lambda;J_i\rightarrow
J_f)\<\equiv(2J_f+1)\,|\langle f || T^{(\sigma\lambda)} || i
\rangle|^2 /(2J_i+1)$.  Schematic values
$e_\pi\<=e_\nu\<=0.1\,e\mathrm{b}$, $g_\pi\<=\mu_N$, and $g_\nu\<=0$
are used for the effective charges (see
Ref.~\cite{vanisacker1986:ibm2-limits} for further discussion).  The
$\chir^{(E2)}$ parameters are taken equal to their counterparts
$\chir$ in the Hamiltonian, following the consistent quadrupole
formalism~\cite{warner1983:cqf}.  The Hamiltonian used for the present
discussion is the simple form~(\ref{eqnHdieperink}) with the addition
of the Majorana operator.  Diagonalization is carried out using the
computer code~\textsc{npbos}~\cite{otsuka1985:npbos-manual}, for boson
numbers $\Np\<=\Nn\<=5$.

\begin{figure}
\begin{center}
\includegraphics[width=0.9\hsize]{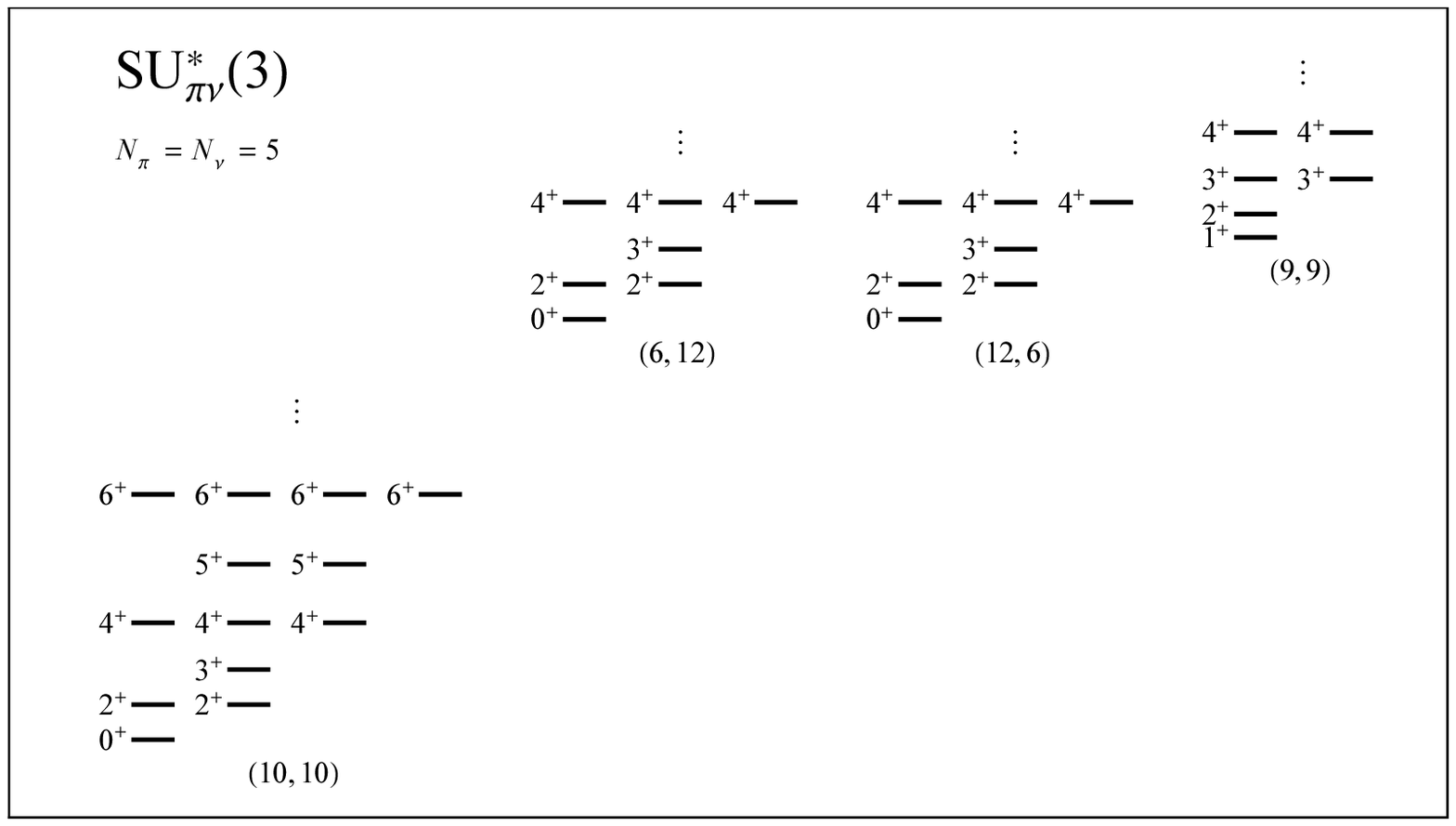}
\end{center}
\caption
{Level scheme for the $\grpsuthreepnstar$ dynamical symmetry,
following the energy relation~(\ref{eqnEsu3}).  The lowest-energy
$\grpsuthreepnstar$ representations, labeled by the $(\lambda,\mu)$
quantum numbers, are shown for $\Np\<=\Nn\<=5$.  }
\label{figrepnscheme}
\end{figure}

The energy spectrum for the $\grpsuthreepnstar$ symmetry is known
analytically~\cite{dieperink1982:ibm2-triax,walet1987:ibm2-su3star-details}.
The operator
$(\Qhat_\pi^\chip+\Qhat_\nu^\chin)\cdot(\Qhat_\pi^\chip+\Qhat_\nu^\chin)$
with $\chip\<=-\rootstinline$ and $\chin\<=+\rootstinline$ can be
reexpressed in terms of the quadratic Casimir
operators~\cite{iachello1987:ibm} of the subalgebra chain~(\ref{eqnchainsu3star}), as
$(3/4)C_2[\grpsuthreepnstar]-(3/16)C_2[\grpsothreepn]$.  For
$H=-\Qhat\cdot\Qhat$, the eigenvalues are thus given by
\begin{equation}
\label{eqnEsu3}
E(\lambda,\mu,L)
=-\frac{1}{2}(\lambda^2+\mu^2+\lambda\mu+ 3\lambda+3\mu)+\frac{3}{8}L(L+1),
\end{equation}
where $(\lambda,\mu)$ are the $\grpsuthreepnstar$ Elliott quantum
numbers~\cite{elliott1958:su3-shell-part1} and $L$ is the angular
momentum quantum number.  This is the $\grpsuthree$ energy formula,
and level energies within an $\grpsuthreepnstar$ representation follow
$L(L+1)$ rotational spacings.  However, the allowed $\grpsuthree$
representations (see
Refs.~\cite{dieperink1982:ibm2-triax,walet1987:ibm2-su3star-details})
are different from those for $\grpsuthreepn$.  The ground state
representation is the $(2\Np,2\Nn)$ representation, rather than the
usual $\grpsuthreepn$ ground state representation $(2\Np+2\Nn,0)$.
According to the $\grpsuthree\supset\grpsothree$ branching
rules~\cite{elliott1958:su3-shell-part1}, the ground state
representation thus contains multiple degenerate rotational bands,
with $K$ quantum numbers $0$, $2$, $4$,~$\ldots$, as shown in
Fig.~\ref{figrepnscheme}.  [It has become customary to label the bands
of an $\grpsuthree$ representation by the Elliott $K$ quantum
number~\cite{elliott1958:su3-shell-part1}, in part by analogy with the
rigid rotor spin-projection quantum number $K$~\cite{bohr1998:v2},
which yields a band of the same band head spin.  However, the Elliott
$\grpsuthree$ basis is not orthogonal, and the orthonormal states
obtained from diagonalization of a Hamiltonian are thus not in general
Elliott states.  Throughout this article, the label $K$ is used simply
to indicate the spin content of a band.]  Other $\grpsuthreepnstar$
representations appearing at low energy include the $(2\Np-4,2\Nn+2)$,
$(2\Np+2,2\Nn-4)$, and $(2\Np-1,2\Nn-1)$
representations~(Fig.~\ref{figrepnscheme}).  In the classical limit,
the first two of these correspond to coupled $\beta$ and $\gamma$
vibrations of the fluids~\cite{leviatan1990:ibm2-modes}.  The last
corresponds to an ``orthogonal'' scissors mode, in which the proton
and neutron symmetry axes oscillate about their equilibrium
perpendicular relative
orientation~\cite{leviatan1990:ibm2-modes,walet1987:ibm2-su3star-details}.
\begin{figure}[p]
\begin{center}
\includegraphics[width=\hsize]{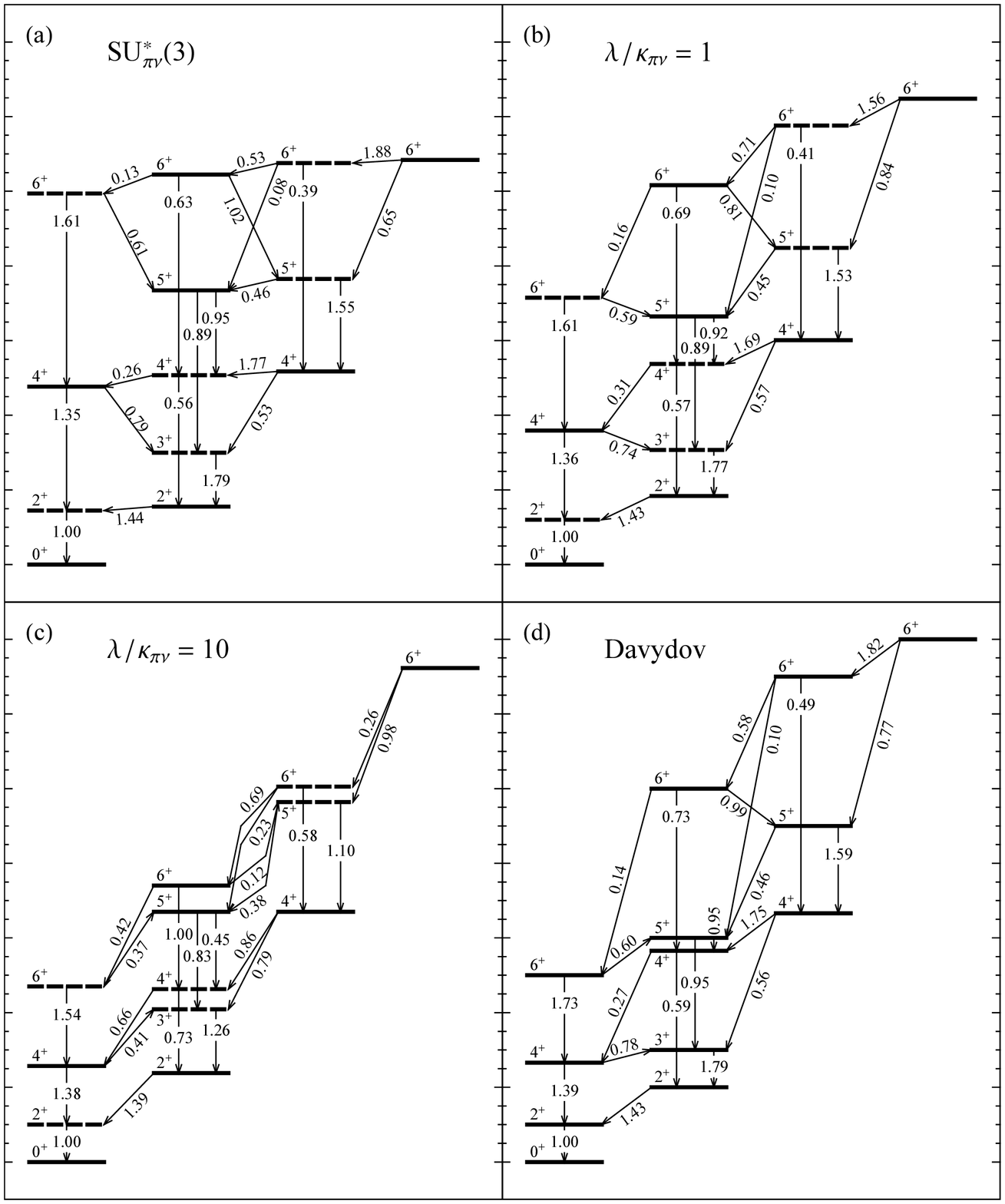}
\end{center}
\caption
{Level schemes and selected $B(E2)$ strengths for (a)~the
$\grpsuthreepnstar$ dynamical symmetry ($\chip\<=-\rootstinline$ and
$\chin\<=+\rootstinline$), (b)~$\lambda/\kappa_{\pi\nu}\<=1$, and
(c)~$\lambda/\kappa_{\pi\nu}\<=10$.  The level scheme for the Davydov
model with $\gamma\<=\pi/6$ is shown for comparison~(d).  Levels of
positive and negative $T$ parity are indicated by solid and dashed
lines, respectively.  All $B(E2)$ strengths are given for the
spin-descending transition direction, to allow direct comparison
between panels, regardless of the ordering of level energies, and are
normalized to $B(E2;2^+_1\rightarrow0^+_1)$.  Energies are normalized
separately in each panel.  A minimal Majorana term
($\lambda/\kappa_{\pi\nu}\<=0.125$) is used in the calculation of
part~(a) to lift the degeneracy of the $\grpsuthreepnstar$ multiplets.
Calculations are for $\Np\<=\Nn\<=5$.  }
\label{figtriaxschemes}
\end{figure}

The $B(E2)$ strengths for transitions between levels of the
$\grpsuthreepnstar$ ground state representation, calculated
numerically, are shown in Fig.~\ref{figtriaxschemes}(a).  These
strengths follow selection rules dictated by the presence of a
discrete, parity-like symmetry.  Consider the operation consisting of
negation of the $d_\rho$ bosons followed by interchange of all proton
and neutron bosons, together yielding $s_\pi\leftrightarrow s_\nu$ and
$d_\pi\leftrightarrow-d_\nu$.  This was denoted the $T$ parity
operation by Otsuka~\cite{otsuka1992:ibm2-mss-t-symmetry}.  For
$\chip\<=\-\chin$, the IBM-2 Hamiltonian~(\ref{eqnHgeneralraw}) with
$\epsilonp\<=\epsilonn$ and $\kappapp\<=\kappann$ is invariant under
the $T$ parity operation.  Thus $T$ parity is a good symmetry
throughout the central vertical plane of Fig.~\ref{figtetraspace},
including the $\grpsuthreepnstar$ dynamical symmetry.  The operator
$T^{(E2)}$ carries negative $T$ parity.  Therefore, $E2$ transitions
occur only between states of opposite $T$ parity, and all electric
quadrupole moments vanish.  The operator $T^{(M1)}$ decomposes into a
part of positive $T$ parity $\rtrim\propto(\Lhat_\pi+\Lhat_\nu)$,
which generates magnetic dipole moments, and a part of negative $T$
parity $\rtrim\propto(\Lhat_\pi-\Lhat_\nu)$, which induces transitions
between different states~\cite{otsuka1992:ibm2-mss-t-symmetry}.  Thus
$M1$ transitions follow the same $T$ parity selection rule as $E2$
transitions, but magnetic moments are allowed.  In applying the
selection rules, it should be noted that the $E2$ selection rule only
holds in its exact form for $e_\pi\<=e_\nu$.  Also, the $T$ parity
operation involves interchange of all proton and neutron bosons, so it
is only well defined for $\Np\<=\Nn$, but the selection rules persist
approximately even for $\Np\<\neq\Nn$.

The classical interpretation of the $T$ parity operation is obtained
by observing that, in the definition of the coherent
state~(\ref{eqncoherentpn}), negation of the $d_\rho$ boson is
equivalent to negation of the $\alpha_{\rho,\mu}^{(2)}$ deformation
coordinates.  In the geometric model, this is the $\gamma$ parity
operation of B\`es~\cite{bes1959:gamma}, which exchanges prolate and
oblate liquid drop deformations.  Thus, the $T$ parity operation
exchanges prolate and oblate deformations for each fluid and then
interchanges proton and neutron fluids. The $\grpsuthreepnstar$
triaxial configuration of Fig.~\ref{figortho} is seen to be invariant
under this combined transformation.

We now note the general characteristics of the $B(E2)$ strengths for
the $\grpsuthreepnstar$ symmetry shown in
Fig.~\ref{figtriaxschemes}(a).  While the energy levels for the
$\grpsuthreepnstar$ symmetry fall naturally into rotational
quasi-bands, the transition strengths are far from those for an
$\grpsuthreepn$ rotor.  Many of the interband transitions have
strenths of the same order as in-band transitions, while some of the
in-band transitions vanish due to the $T$ parity selection rule.
(Levels of positive and negative $T$ parity are indicated by solid and
dashed lines, respectively, in Fig.~\ref{figtriaxschemes}.)
Transitions between non-adjacent bands, not shown in the figure, are
weaker by an order of magnitude or more.

Magnetic dipole transitions arise in collective models from separation
between the proton and neutron fluid distributions (\textit{e.g},
Ref.~\cite{eisenberg1987:v1}).  The static asymmetry between these
distributions for the $\grpsuthreepnstar$ dynamical symmetry gives
rise to extremely large $M1$ admixtures whenever spin-allowed,
with $B(M1)/B(E2)\<\sim1\mu_N^2/(\esqrbsqr)$ for most of the
transitions in the ground state representation.  These $M1$ strengths
are comparable to the $M1$ decay strength of the scissors excitation.

The $\grpsuthreepnstar$ dynamical symmetry has previously been loosely
associated with the $\gamma\<=\pi/6$ triaxial rotor of the Davydov
model~\cite{davydov1958:arm-intro}, on the basis of two similarities:
the angular momentum content of the ground state representation and
the quadrupole moment components found in the classical
limit~\cite{dieperink1982:ibm2-triax}.  The level energies in the two
cases differ considerably, as the Davydov model bands exhibit a
characteristic $(2^+3^+)(4^+5^+)\ldots$ clustering or ``staggering''
of energies [Fig.~\ref{figtriaxschemes}(d)].  However, a detailed
comparison of the $E2$ transition strengths of the two models
[Fig.~\ref{figtriaxschemes}(a,d)] reveals an extraordinary similarity,
with many $E2$ strengths differing by only a few per cent.  The full
relationship between the models has not been established (an
approximate $\grpsuthree$ algebra underlying the dynamics of the
triaxial rotor has been discussed in Ref.~\cite{leschber:su3-rotor}).

If the Majorana operator is introduced into the $\grpsuthreepnstar$
Hamiltonian, the energy spectrum is radically altered, as depicted in
Fig.~\ref{figtriaxschemes}(b,c).  The
$F$-spin~\cite{otsuka1978:ibm2-shell-details}, formally analogous to
isospin, of a state provides a measure of its proton-neutron
asymmetry.  For sufficiently pure $F$-spin
$F\<=F_\text{max}\<\equiv\frac{1}{2}(\Np+\Nn)$, the IBM-2 effectively
reduces to the one-fluid IBM~\cite{harter1985:ibm2-projection}.  The
$\grpsuthreepnstar$ eigenstates are highly mixed in their $F$-spin
content ($\langle {\mathrm{\mathbf{\hat{F}}}}^2
\rangle^{1/2}\<\approx0.8F_\text{max}$), reflecting the 
proton-neutron asymmetry of the classical equilibrium configuration.
The Majorana operator, by energetically penalizing states of
$F\<<F_\text{max}$, purifies the $F$-spin contents of the low-energy
states.  \textit{E.g.}, the $\grpsuthreepnstar$ ground state is left
with an $F\<<F_\text{max}$ content of only about $\sim1\%$ for
$\lambda/\kappapn\<=10$ [Fig.~\ref{figtriaxschemes}(c)].  A small
Majorana contribution in the Hamiltonian,
$\lambda/\kappapn\<\lesssim1$, results in a $(2^+3^+)(4^+5^+)\ldots$
staggering of level energies [Fig.~\ref{figtriaxschemes}(b)] like that
of the Davydov model [Fig.~\ref{figtriaxschemes}(d)], as noted in
Ref.~\cite{sevrin1987:ibm2-su3star-1}.  A larger
Majorana contribution [Fig.~\ref{figtriaxschemes}(c)] leads to the
reverse $2^+(3^+4^+)(5^+6^+)\ldots$ staggering, characteristic of
$\gamma$-soft structure.  The $E2$ strengths evolve
toward $\grpsosix$-like values~\cite{iachello1987:ibm} with increasing
Majorana strength, and the $M1$ admixtures decrease rapidly from their
large $\grpsuthreepnstar$ values, with a typical scale $B(M1)/B(E2)\<\sim
10^{-1}\mu_N^2/(\esqrbsqr)$ for Fig.~\ref{figtriaxschemes}(b) and
$B(M1)/B(E2)\<\sim 10^{-2}\mu_N^2/(\esqrbsqr)$ for
Fig.~\ref{figtriaxschemes}(c).  The evolution observed of the apparent
structure with increasing
Majorana contribution~--- from $\grpsuthreepnstar$ triaxialiaty, through
one-fluid rigid triaxiality, to one-fluid $\gamma$-softness~--- is
as expected from the classical limit analysis of
Sec.~\ref{subsecmajorana}.

\subsection{The $\grpsuthreepn$--$\grpsuthreepnstar$ transition}
\label{subsecspectrosu3}

\begin{figure}
\begin{center}
\includegraphics[width=0.6\hsize]{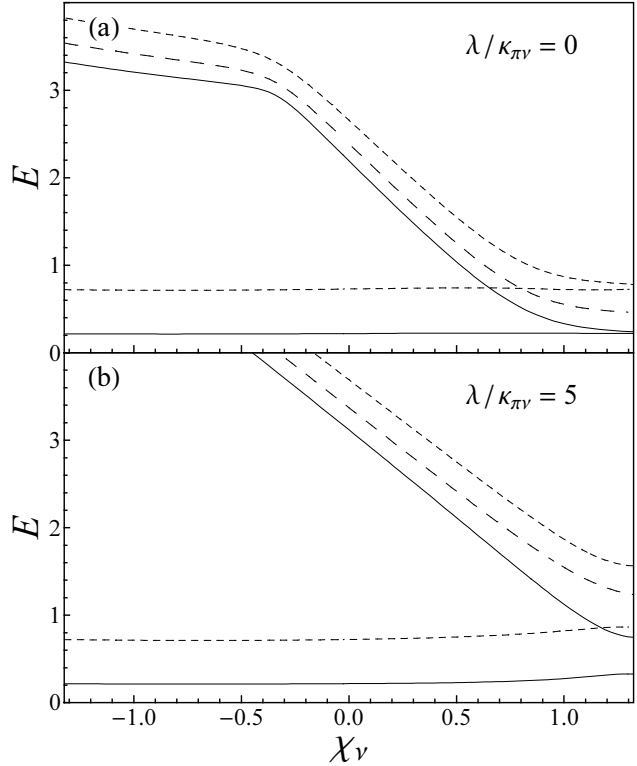}
\end{center}
\caption
{
Excitation energies of the lowest $2^+$ (solid), $3^+$ (dashed), and
$4^+$ (dotted) levels along the $\grpsuthreepn$--$\grpsuthreepnstar$
transition, (a)~with no Majorana operator and (b)~for
$\lambda/\kappapn\<=5$.  The change in slope for the excited
band level energies at $\chin\<\approx-0.4$ in part~(a) arises from
the crossing of the scissors and $\gamma$ bands. Calculations are for
$\chip\<=-\rootstinline$ and $\Np\<=\Nn\<=5$, with minimal
perturbations $\xi'\<=0.95$ and for part~(a)
$\lambda/\kappapn\<=0.125$ to remove numerical degeneracies.  
}
\label{figevolnsu3energies}
\end{figure}

\begin{figure}[p]
\begin{center}
\includegraphics[width=\hsize]{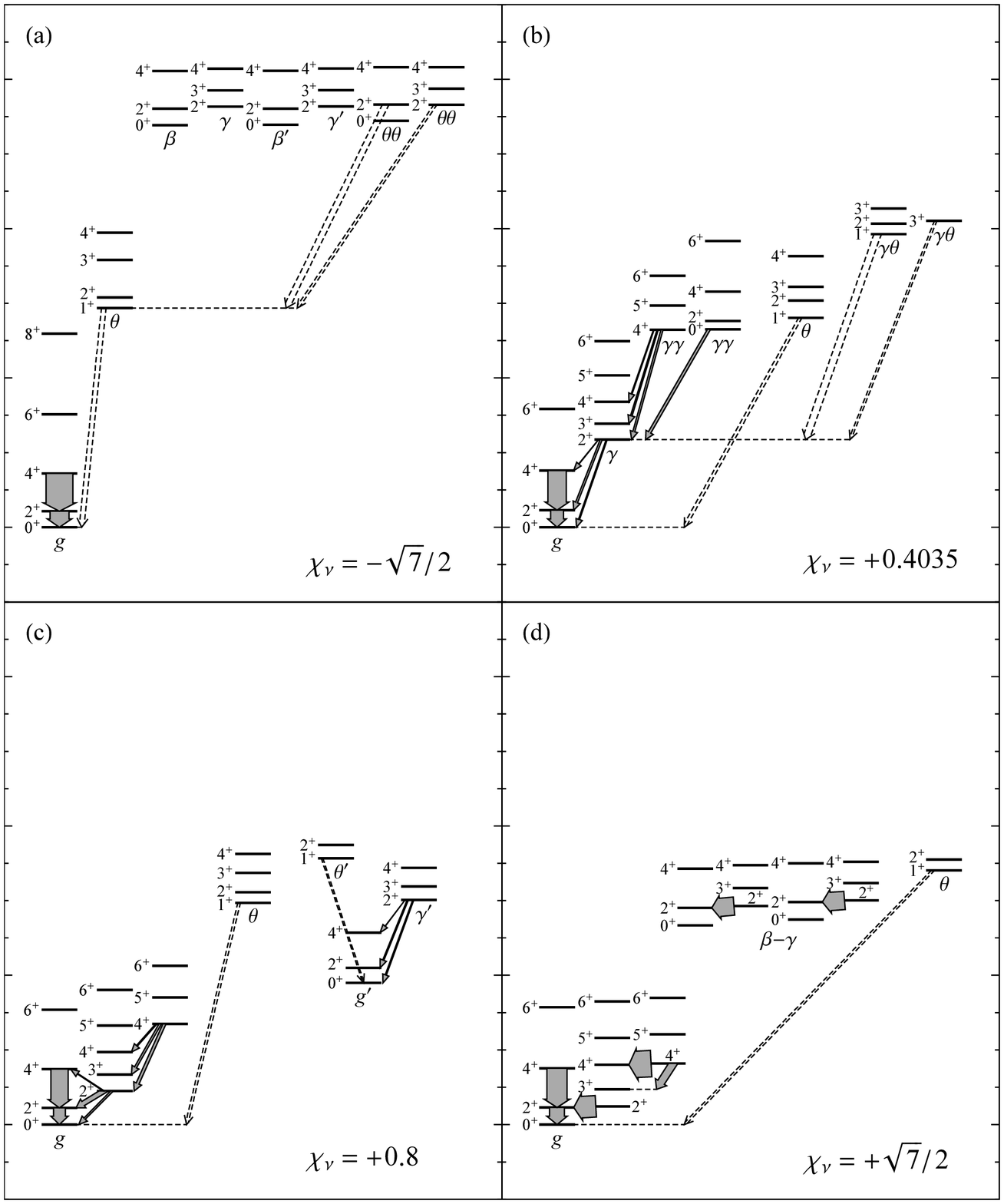}
\end{center}
\caption
{Level schemes for the $\grpsuthreepn$--$\grpsuthreepnstar$ transition
with no Majorana operator: (a)~the $\grpsuthreepn$ limit at
$\chin\<=-\rootstinline$, (b)~the second order transition point at
$\chin\<\approx0.4035$, (c)~the approximate ground state crossing
point at $\chin\<=0.8$, and (d)~the $\grpsuthreepnstar$ limit at
$\chin\<=+\rootstinline$.  The $E2$ branching patterns are shown for
selected band head states, and the ground state band
$4^+\rightarrow2^+$ and $2^+\rightarrow0^+$ transitions are included
for scale [shaded arrows, widths proportional to $B(E2)$ strength].
The $M1$ transitions deexciting the scissors band head states are
shown as well [dashed arrows, widths proportional to $B(M1)$
strength].  The $g$ (ground), $\beta$, $\gamma$, and $\theta$
(scissors) band designations are schematic (see text).  All panels
share a common energy scale.  Calculations are for
$\chip\<=-\rootstinline$ and $\Np\<=\Nn\<=5$, with minimal
perturbations $\xi'\<=0.95$ for part~(a) and
$\lambda/\kappa_{\pi\nu}\<=0.125$ for part~(d) to remove numerical
degeneracies.  }
\label{figevolnsu3}
\end{figure}

\begin{figure}
\begin{center}
\includegraphics[width=\hsize]{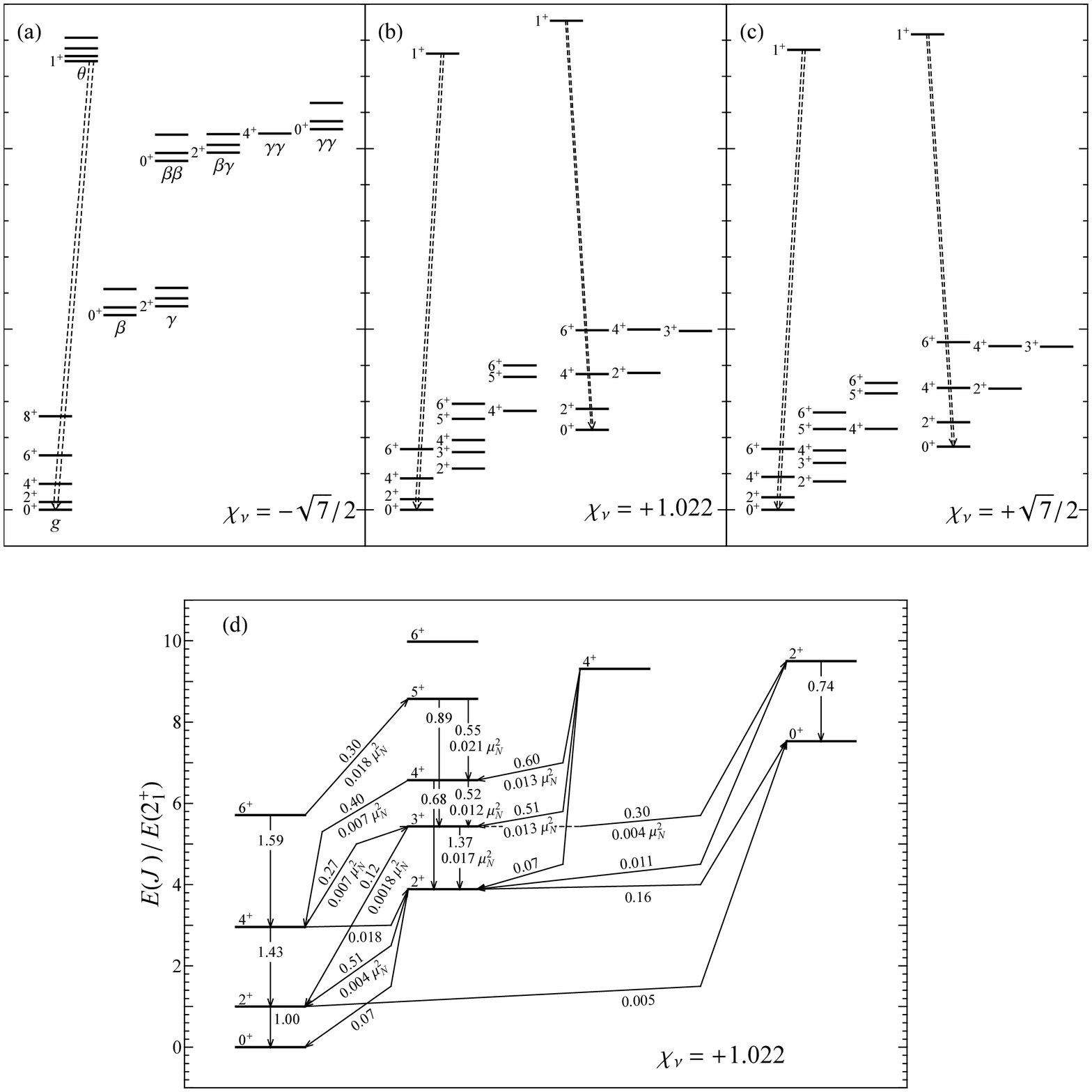}
\end{center}
\caption
{Level schemes for the $\grpsuthreepn$--$\grpsuthreepnstar$ transition
with Majorana contribution $\lambda/\kappa_{\pi\nu}\<=5$: (a)~the
$\grpsuthreepn$ limit at $\chin\<=-\rootstinline$, (b)~the second
order transition point at $\chin\<\approx1.022$, and (c)~the
$\grpsuthreepnstar$ limit at $\chin\<=+\rootstinline$.  For the
transition point, $B(E2)$ and $B(M1)$ strengths among selected
low-lying levels are given if part~(d).  Parts (a)--(c) share a common
energy scale, while energies in part~(d) are normalized to the $2^+_1$
level energy.  Dashed arrows in parts (a)--(c) indicate $M1$
transitions from the scissors excitations. $B(E2)$ and $B(M1)$
strengths in part~(d) are shown spin-descending, with $B(E2)$
strengths normalized to $B(E2;2^+_1\rightarrow0^+_1)$
[$0.283\,\esqrbsqr$].  The $g$ (ground), $\beta$, $\gamma$, and
$\theta$ (scissors) band designations are schematic.  
Calculations are for $\chip\<=-\rootstinline$ and $\Np\<=\Nn\<=5$,
with a minimal $\nhat_d$ perturbation $\xi'\<=0.95$ for part~(a) to remove
numerical degeneracies.  }
\label{figevolnsu3majorana}
\end{figure}

The evolution of the IBM-2 predictions along the
$\grpsuthreepn$--$\grpsuthreepnstar$ transition is shown in
Figs.~\ref{figevolnsu3energies}--\ref{figevolnsu3majorana}.  Since the extremely low lying $K=2$ band is
characteristic of the $\grpsuthreepnstar$ level scheme, the evolution
of the relevant energies is plotted in
Fig.~\ref{figevolnsu3energies}.  For simplicity, we first consider the
transition without a Majorana interaction.  In this case, rotational
$L(L+1)$ energy spacings within bands are almost exactly preserved
throughout the transition.

In the $\grpsuthreepn$ limit [Fig.~\ref{figevolnsu3}(a)], the lowest
excitation above the ground state band ($F\<=F_\text{max}$) is the
scissors mode ($F\<=F_\text{max}-1$), followed at higher energy by a
cluster of degenerate $K\<=0$ and $2$ excitations.
These arise from proton-neutron symmetric beta and gamma oscillations
($F\<=F_\text{max}$), asymmetric beta and gamma oscillations
($F\<=F_\text{max}-1$), and two-phonon scissors oscillations
($F\<=F_\text{max}-2$).  All $E2$ transitions from the excited bands to
the ground state band vanish exactly for the $\grpsuthreepn$ dynamical
symmetry, but for small breakings of the symmetry they approximately
follow the rotational Alaga rules.  $M1$ transitions vanish as well, except
those involving excitations of the scissors mode.

The level scheme for the $\grpsuthreepn$--$\grpsuthreepnstar$ second
order transition point ($\chin\<\approx0.4035$) is shown in
Fig.~\ref{figevolnsu3}(b).  As $\chin$ is increased from
$-\rootstinline$, one of the $\gamma$ bands (no longer of pure $F$
spin) rapidly descends to lower excitation energy.  At the transition
point, this band is connected to the ground state band by $E2$
transitions $\rtrim\sim1/10$ as strong as in-band transitions.  The
$M1$ admixtures are extremely large, with
$B(M1)/B(E2)\<\sim1\mu_N^2/(\esqrbsqr)$, both for the $\gamma$ to
ground and the in-band transitions.  The $\gamma$ band is followed in
its descent by a series of $K\<=4$, $6$,~$\ldots$ bands, originating
as two-phonon, three-phonon, \textit{etc.}, $\gamma$ excitations.
However, the $K\<=0$ two-phonon $\gamma$ excitation does not descend
as rapidly and thus has a positive anharmonicity, $\rtrim\sim2.8$ at
the transition point.  Positive anharmonicity of this band has also
recently been discussed as a signature of the axial-triaxial
transition both in the one-fluid
IBM~\cite{garciaramos1998:ibm-two-phonon} and in the $\grp{Y}(5)$
geometric model~\cite{iachello2003:y5}.  [The irregular level spacing
seen in this band in Fig.~\ref{figevolnsu3}(b) is not a fundamental
signature but rather arises from mixing with the nearly degenerate
scissors band for these parameter values.]  As the $\gamma$
vibrational energy scale decreases, the $K\<=1$ and $3$ $\gamma$
exitations of the scissors band similarly approach the scissors band
in energy.

Although the coherent state energy surface first becomes soft to
triaxial deformations at the second order transition point, it is only
for $\chin\<\gtrsim0.65$ that highly triaxial configurations like that
of Fig.~\ref{figortho} become lower in energy than axial
configurations.  A level scheme obtained in this regime, at
$\chin\<=0.8$, is shown in Fig.~\ref{figevolnsu3}(c).  The ground
state band and first $K\<=2$ and $4$ bands are close in energy,
resembling the $\grpsuthreepnstar$ ground state representation.  The
relative $E2$ transition strengths are intermediate between those for
the $\grpsuthreepn$ and $\grpsuthreepnstar$ symmetries.  The first
excited $K\<=0$ band and following $K\<=2$ band, in contrast, have
relative energies more appropriate to a ground state and its
associated $\gamma$ excitation, and $B(E2)$ values closely matching
the $\grpsuthreepn$ Alaga rules.  The energy spacing scales within
these bands is less (by $\rtrim\sim10\%$) than that of the ground
state family of bands, suggesting a different moment of inertia or
different triaxial nature to the rotation.  The excitation energy of
the $0^+_2$ state reaches its lowest value along the
$\grpsuthreepn$--$\grpsuthreepnstar$ transition at approximately this
$\chin$ value.  The observables are thus highly suggestive of an
avoided crossing between an $\grpsuthreepnstar$-like configuration and
an $\grpsuthreepn$-like configuration, although the actual eigenstate
structure is presumably more complicated.  An apparent avoided
crossing also occurs between $1^+$ scissors excitations based upon
these configurations.

We now briefly reconsider the $\grpsuthreepn$--$\grpsuthreepnstar$
transition for a realistic Majorana operator strength
($\lambda/\kappa_{\pi\nu}\<=5$) [Figs.~\ref{figevolnsu3energies}(b)
and~\ref{figevolnsu3majorana}].  At the $\grpsuthreepn$ limit, the
eigenstates are unchanged, but the levels with $F\<<F_\text{max}$ are
raised in energy [Fig.~\ref{figevolnsu3majorana}(a)].  The
lowest-lying excitations are thus proton-neutron symmetric $\beta$ and
$\gamma$ excitations, followed at higher energies by the symmetric
$\beta$ and $\gamma$ multiphonon excitations.  The second-order
transition between axial and triaxial equilibrium occurs at much
larger $\chin$ than without the Majorana operator,
$\chin\<\approx1.022$ (Fig.~\ref{figraysmajorana}).  The level scheme
for the transition point [Fig.~\ref{figevolnsu3majorana}(b)] shows
large differences between ground state and excited families of levels.
The lowest $K\<=0$, $2$, and $4$ quasi-bands have level spacings
intermediate between those for the $\grpsuthreepn$ and $\grpsosixpn$
symmetries.  The next $K\<=0$ quasi-band is far less rotational in
spacing and has a $2^+$-$0^+$ energy spacing about twice as large as
for the ground state.  The levels above it are assembled into an
approximate $\grpsosixpn$ multiplet structure.  All $M1$ strengths are
attenuated relative to the cases with no Majorana contribution, with a
typical admixture scale $B(M1)/B(E2)\<\sim0.05\mu_N^2/(\esqrbsqr)$.
Qualitatively, little further changes as $\chin$ increases to
$+\rootstinline$ [Fig.~\ref{figevolnsu3majorana}(c)], except that the
ground state family of levels takes on more distinctly
$\grpsosixpn$-like energy spacings and $B(E2)$ strengths.

\subsection{The $\grpufivepn$--$\grpsuthreepnstar$ transition}
\label{subsecspectrou5}
\begin{figure}[p]
\begin{center}
\includegraphics[width=0.9\hsize]{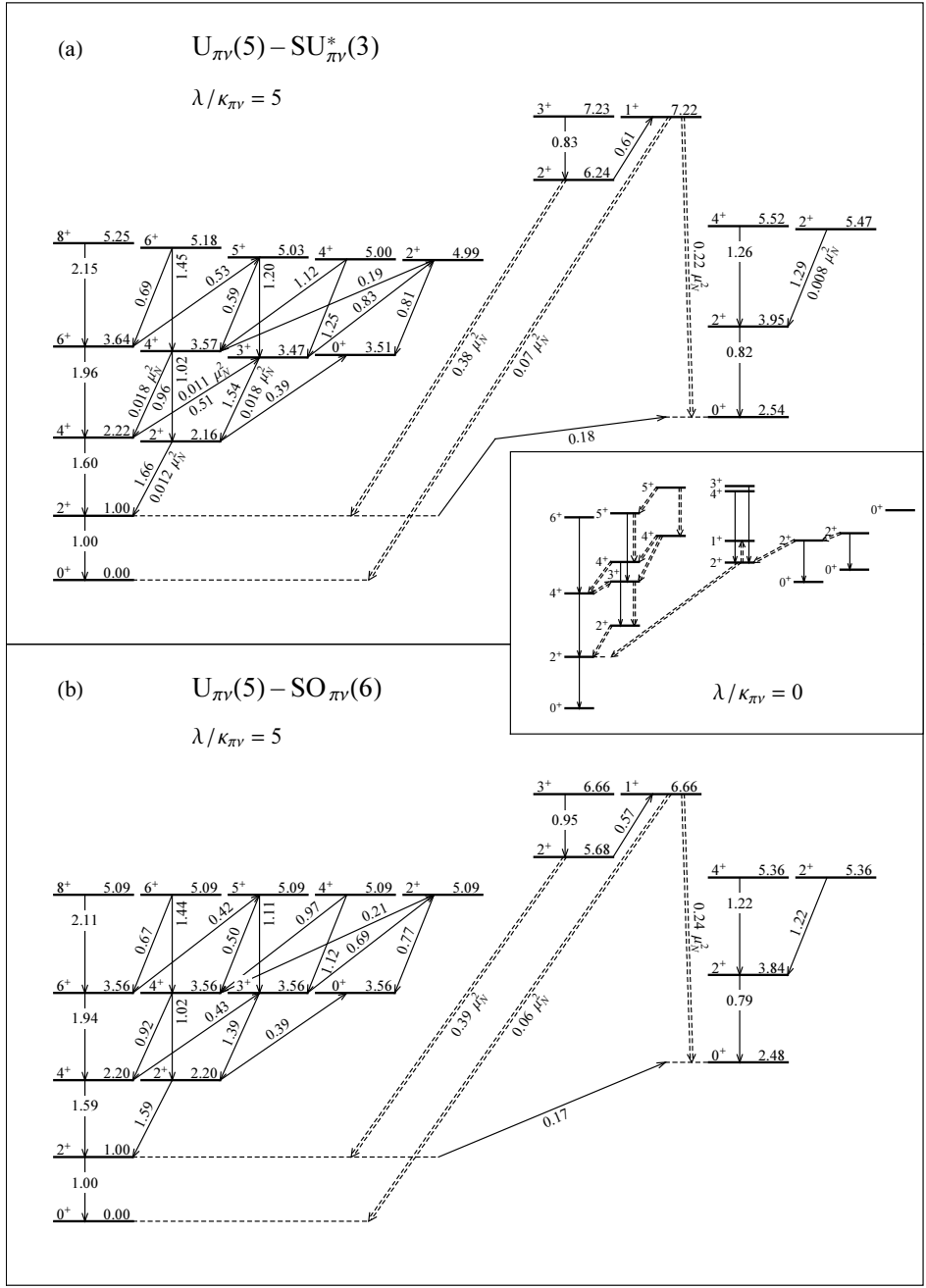}
\end{center}
\caption
{Level schemes for the second order transition points along (a)~the
$\grpufivepn$--$\grpsuthreepnstar$ transition and (b)~the
$\grpufivepn$--$\grpsosixpn$ transition, calculated for a Majorana
operator strength of $\lambda/\kappa_{\pi\nu}\<=5$.  The inset to
part~(a) shows the corresponding level scheme with no Majorana
operator.  Energies are indicated normalized to the $2^+_1$ level
energy.  $B(E2)$ strengths are shown spin-descending, normalized to
$B(E2;2^+_1\rightarrow0^+_1)$ separately in each panel
[$0.175\,\esqrbsqr$ for part~(a), $0.168\,\esqrbsqr$ for part~(b)],
and selected $B(M1)$ strengths are indicated as well.  Dashed double
arrows indicate selected strong $M1$ transitions.  Calculations are
for $\Np\<=\Nn\<=5$, with $\chip\<=-\rootstinline$,
$\chin\<=+\rootstinline$, and $\xi'\<=0.2$ for part~(a) and with
$\chip\<=0$, $\chin\<=0$, and $\xi'\<=0.2$ for part~(b).  }
\label{fige5schemes}
\end{figure}

The basic spectroscopic features along the
$\grpufivepn$--$\grpsuthreepnstar$ transition differ fundamentally
depending upon whether or not a realistic Majorana contribution is
included in the Hamiltonian.  The level scheme for the $\grpufivepn$
symmetry of the
IBM-2~\cite{iachello1987:ibm,vanisacker1986:ibm2-limits} exhibits
extreme degeneracies if no Majorana operator is present: there are two
degenerate one-phonon $2^+$ levels (one fully symmetric and one of
mixed symmetry) and eleven degenerate levels of various symmetry in
the two-phonon multiplet.  By the second order transition point
($\xi'\<=1/5$), these highly degenerate levels evolve into a cluster
of low-lying excitations of widely varying $F$-spin content, shown in
the inset to Fig.~\ref{fige5schemes}(a).  Among these levels,
incipient forms of the $\grpsuthreepnstar$ ground-state and scissors
quasi-band structures are apparent.  The electromagnetic transition
properties are dominated by the ubiquity of strong $M1$ transitions,
arising from the mixing of $\grpufivepn$ states of different $F$ spin.
For instance, both the $2^+_2\rightarrow2^+_1$ and
$2^+_3\rightarrow2^+_1$ transitions have $B(M1)$ strengths
$\rtrim>0.1\,\mu_N^2$, typical of mixed symmetry state decays.

If instead a realistic Majorana operator strength
$\lambda/\kappa_{\pi\nu}\<\approx5$ is considered, the low-energy
excitation spectrum simplifies considerably, as shown in
Fig.~\ref{fige5schemes}(a).  The low-lying levels have nearly pure
$F\<\approx F_\text{max}$, with a few per cent admixtures of other $F$
values, and are grouped into approximate $SO(5)$ multiplets.  Mixed
symmetry levels of $F\<\approx F_\text{max}-1$ are present at higher
energy ($1^+$, $2^+$, and $3^+$ levels at top of figure).  The $T$
parity is a good quantum number throughout the
$\grpufivepn$--$\grpsuthreepnstar$ transition, so $T$ parity selection
rules apply to the electromagnetic transitions.  The level scheme and
$E2$ transition strengths are nearly identical to those obtained for
the second order transition point between $\grpufivepn$ and
$\grpsosixpn$, shown for comparison in Fig.~\ref{fige5schemes}(b).
There is thus also considerable resemblance to the predictions of the
geometric $E(5)$ model~\cite{iachello2000:e5,casten2000:134ba-e5}.
Two features of the $\grpufivepn$--$\grpsuthreepnstar$ transition
point spectrum allow it to be distinguished from the
$\grpufivepn$--$\grpsosixpn$ case.  First, a slight breaking of the
degeneracy of the $SO(5)$ multiplets occurs, in the sense necessary to
create the $\grpsuthreepnstar$ quasi-bands.  Thus, the $2^+_2$ level
moves below the $4^+_1$ level, the $3^+_1$ and $4^+_2$ levels move
below the $6^+_1$ level, \textit{etc.}  This effect becomes much more
marked just past the transition point.  Second, sizeable $M1$
transition strengths, $B(M1)/B(E2)\<\approx0.05\,\mu_N^2/(\esqrbsqr)$
to $0.1\,\mu_N^2/(\esqrbsqr)$, are present among the low-lying
levels~[Fig.~\ref{fige5schemes}(a)], while these vanish when
$\chis\<=0$.  Such $M1$ admixtures in general play an essential role
in identifying deviations from
$\chiv\<=0$~\cite{vanisacker1988:ibm2-m1-rare-earth,kim1996:ibm2-pd,otsuka1992:ibm2-mss-t-symmetry}.

\section{Conclusion}
\label{secconcl}

The IBM-2 phase diagram investigated here provides a framework for
studying the transition between axial and triaxial structure in
nuclei.  Triaxial deformation might arise from several possible
sources: distinct deformations of the proton and neutron fluids as
considered in the present work, higher-order interactions in an
essentially one-fluid nucleus~\cite{vanisacker1981:ibm-triax}, or the
presence of configurations involving hexadecapole nucleon
pairs~\cite{heyde1983:ibm-g-boson-104ru}.  The main spectroscopic
features of proton-neutron triaxial structure are exemplified by the
$\grpsuthreepnstar$ dynamical symmetry, shown in
Fig.~\ref{figtriaxschemes}(a).  Proton-neutron triaxiality is
characterized by a low-lying $K\<=2$ band, as in the other forms of
nuclear triaxiality, but with level energies following a rotational
$L(L+1)$ sequence.  The $B(E2)$ strength pattern is remarkably similar
to that of the classic rigid triaxial rotor of the Davydov model,
including the unusual feature that $B(E2;2^+_2\rightarrow0^+_1)\<=0$
(this however only strictly holds for $\Np\<=\Nn$ and $e_\pi\<=e_\nu$).
The microscopic conditions leading to two-fluid triaxial structure,
which requires $\chip$ and $\chin$ of opposite sign, are found when
the proton bosons are particle like (below mid-shell) and the neutron
bosons are hole-like (above mid-shell) or \textit{vice versa}.  Nuclei
in this category include the heavy $\mathrm{Ru}$ and $\mathrm{Pd}$
isotopes~\cite{vanisacker1980:ibm2-ru-pd,kim1996:ibm2-pd,duarte1998:ibm2-ru-triax},
the light rare earth nuclei below the $N\<=82$ shell closure, and the
extremely light $\mathrm{W}$, $\mathrm{Os}$, and $\mathrm{Pt}$ nuclei.

The $\grpsuthreepnstar$ dynamical symmetry probably does not occur in
its pure form in any actual nuclei.  In this article, we have seen how
the features of $\grpsuthreepnstar$ proton-neutron triaxiality are
modified by the Majorana interaction
[Fig.~\ref{figtriaxschemes}(b,c)], which tends to alter the spectrum
towards that for the $\grpsosixpn$ dynamical symmetry.  Of most
importance from a phenomenological viewpoint, we have investigated the
transition from axially symmetric deformed [$\grpsuthreepn$-like] to
triaxially deformed [$\grpsuthreepnstar$-like] structure, either
without [Fig.~\ref{figevolnsu3}] or with
[Fig.~\ref{figevolnsu3majorana}] a Majorana interaction.  In
particular, Fig.~\ref{figevolnsu3majorana} represents a realistic
study of this transition, including the behavior of the lowest
$K\<=1$, or scissors, band.  It is found that, despite the attenuation
of proton-neutron triaxial features by the Majorana interaction,
two-fluid triaxiality can serve as the underlying mechanism driving
one-fluid triaxiality, and recognizable features of its two-fluid
nature can remain even in the presence of realistic Majorana
interactions.  The two-fluid phenomena are likely to take on more
importance as very neutron rich nuclei become accessible to
experimental study.  When the valence protons and neutrons occupy
well-separated orbitals, the proton-neutron interaction strengths are
reduced, yielding a situation seen in Sec.~\ref{secgeneral} to be much
more conducive to true two-fluid triaxiality.

The present analysis may also serve as a model for the study of other
multi-fluid bosonic systems.  Within nuclear physics, the
$\grp{U}_\text{core}(6)\otimes\grp{U}_\text{skin}(6)$ description of
core-skin collective modes in neutron rich
nuclei~\cite{warner1997:skin-soft-scissors} is most directly
analogous.  In molecular physics, the $U_1(4)\otimes U_2(4)$ vibron
model with two vibronic species~\cite{iachello1995:vibron} may be
studied similarly.  The appropriate coherent state formalism is
discussed in Ref.~\cite{leviatan1988:vibron-intrinsic}.  The phase
transition in this model is relevant to coupled vibronic bending modes
in ``floppy'' molecules such as acetylene~\cite{perez-bernal:INPREP}.
An extreme case of a multi-fluid algebraic model is found in
applications to polymers, in which an arbitarily large number of
vibronic fluids are coupled~\cite{iachello1999:polymer-algebraic}.
Yet another case is that of atomic condensates, for which scissors
modes of a single-constituent Bose-Einstein condensate relative to an
anisotropic potential have been
observed~\cite{marago2000:bec-scissors}.  Experiments have been
planned to produce condensates of two atomic species.  The exotic
features of the IBM-2 phase diagram are likely to be encountered for
other multi-fluid systems as well.  They highlight the need for a
classification scheme beyond the simple Ehrenfest or one-parameter
Landau models to address the phase structure of systems which
simultaneously possess multiple control parameters and multiple order
parameters.

\begin{ack}
Discussions with J.~M.~Arias, R.~Bijker, A.~Leviatan, and N.~Pietralla
are gratefully acknowledged.  This work was supported by the US DOE
under grant DE-FG02-91ER-40608 and was carried out in part at the
European Centre for Theoretical Studies in Nuclear Physics and Related
Areas.
\end{ack}

\appendix

\section{Matrix element of an arbitrary $m$-body operator between multi-fluid
coherent states}
\label{appcoherentme}

For calculations in the coherent state formalism involving multi-fluid
operators, it is useful to deduce a general formula for the matrix
element of an arbitrary $m$-body operator between arbitrary
multi-fluid coherent states.  Define $S$ different coherent boson
species $B_s^\dag$ ($s\<=1,\ldots,S$) in terms of the basic
boson operators $b_i^\dag$ ($i\<=1,\ldots,n$) as orthonormal linear
combinations
\begin{equation}
B_s^\dagger
\equiv 
\sum_{i=1}^{n} \alpha_{s,i} b_i^\dagger,
\end{equation}
where $\sum_{i=1}^n \alpha_{s',i}^*\alpha_{s,i}\<=\delta_{s',s}$.  Then the
multi-species coherent states
\begin{equation}
\label{eqnmultispecies}
\lvert N_1\cdots N_S\rangle 
\equiv \Bigl( \prod_{s=1}^S
\frac{1}{\sqrt{N_s!}}(B_s^\dagger)^{N_s}\Bigr) |0\rangle
\end{equation}
are normalized, are orthogonal to each other for different $N_s$
values, and have good total boson number $N\<=\sum_{s=1}^S N_s$, where
the total boson number operator is $\Nhat\<\equiv\sum_{i=1}^n
b_i^\dagger b_i$.  Coherent states of the type~(\ref{eqnmultispecies})
also arise in the study of intrinsic excitations in algebraic models.
In this context, the different coherent boson operators $B_s^\dag$
represent a ground state condensate and one or more orthogonal excitation
modes (see, \textit{e.g.},
Refs.~\cite{bohr1982:ibm-coherent-excited-168er,bijker1982:ibm-coherent-excited,leviatan1985:ibm-o5-coherent,leviatan1988:vibron-intrinsic,leviatan1990:ibm-ibm2-coherent-excited}).

The matrix element of an arbitrary $m$-body operator
($m\<\geq1$) between two arbitrary multi-species coherent states is~\cite{caprio2005:coherent}
\begin{multline}
\label{eqncoherentme}
\langle N_1'\cdots N_S'\rvert \Bigl(\prod_{i=1}^m b_{r_i'}^\dagger\Bigr)
\Bigl(\prod_{i=1}^m b_{r_i}\Bigr)|N_1\cdots N_S\rangle
\\
=
\sum_{\substack{t_1',\ldots,t_m'\\t_1,\ldots,t_m} =1}^S 
\Bigl[\prod_{s=1}^S
\delta_{N_s'-\nu_s',N_s-\nu_s}
\sqrt{N_s'^{\underline{\nu_s'}}N_s^{\underline{\nu_s}}} \Bigr]
\Bigl( \prod_{i=1}^m \alpha_{t_i',r_i'}^* \alpha_{t_i,r_i} \Bigr),
\end{multline}
where $\nu_s\<\equiv\sum_{i=1}^m\delta_{t_i,s}$,
$\nu_s'\<\equiv\sum_{i=1}^m\delta_{t_i',s}$, and an underlined
superscript indicates the falling factorial
[$m^{\underline{n}}\<\equiv m(m-1)\cdots(m-n+1)$]~\cite{graham1994:concrete}.  
The multiple sum in~(\ref{eqncoherentme}) nominally contains
$S^{2m}$ terms, but only those with $N_s'-\nu_s'\<=N_s-\nu_s$
and $N_s-\nu_s\<\geq0$ can yield nonzero contributions.  Three
stages are involved in evaluating the matrix element of a general
operator: reexpression of the operator in terms of normal-ordered
$m$-body terms, evaluation of the matrix elements of these by
(\ref{eqncoherentme}), and simplification of the result.  These steps
can all readily be carried out though computer-based symbolic
manipulation.

\section{Properties of $F(a;x)$}
\label{appf}

The IBM-2 energy surface considered in Secs.~\ref{secschematic}
and~\ref{secgeneral} involves terms of the form
\begin{equation}
\label{eqnf}
F(a;x)\equiv \frac{x(1+ax)}{1+x^2}.
\end{equation}
We summarize here the extremum properties of this expression.  For any
value of the parameter $a$, $F(a;x)$ has two extrema with respect to
$x$, at
\begin{equation}
\label{eqnxpm}
x_\pm(a) \equiv a \pm \sqrt{a^2+1},
\end{equation}
The global minimum of $F(a;x)$ is located at $x_-(a)$, and the global
maximum is located at $x_+(a)$.  Note that $x_-(a)\<<0$, $x_+(a)\<>0$, and the two
extremum positions are related by $x_-(a)x_+(a)\<=-1$.  The extremal values of
$F$ are simply
\begin{equation}
\label{eqnfpm}
F[a;x_\pm(a)]= \frac{1}{2} x_\pm(a).
\end{equation}
If $a\<>0$ then $\abs{F[a;x_-(a)]}\<<\abs{F[a;x_+(a)]}$, while for $a\<<0$ the
inequality holds in the opposite sense.


\begin{thebibliography}{82}
\expandafter\ifx\csname natexlab\endcsname\relax\def\natexlab#1{#1}\fi
\expandafter\ifx\csname bibnamefont\endcsname\relax
  \def\bibnamefont#1{#1}\fi
\expandafter\ifx\csname bibfnamefont\endcsname\relax
  \def\bibfnamefont#1{#1}\fi
\expandafter\ifx\csname citenamefont\endcsname\relax
  \def\citenamefont#1{#1}\fi
\expandafter\ifx\csname url\endcsname\relax
  \def\url#1{\texttt{#1}}\fi
\expandafter\ifx\csname urlprefix\endcsname\relax\def\urlprefix{URL }\fi
\providecommand{\bibinfo}[2]{#2}
\providecommand{\eprint}[2][]{\url{#2}}

\bibitem{gilmore1978:coherent}
\bibinfo{author}{\bibfnamefont{R.}~\bibnamefont{Gilmore}}, \bibinfo{journal}{J.
  Math. Phys.} 20 (1979) 891.

\bibitem{feng1981:ibm-phase}
\bibinfo{author}{\bibfnamefont{D.~H.} \bibnamefont{Feng}},
  \bibinfo{author}{\bibfnamefont{R.}~\bibnamefont{Gilmore}}, \bibnamefont{and}
  \bibinfo{author}{\bibfnamefont{S.~R.} \bibnamefont{Deans}},
  \bibinfo{journal}{Phys. Rev. C} 23 (1981) 1254.

\bibitem{iachello1987:ibm}
\bibinfo{author}{\bibfnamefont{F.}~\bibnamefont{Iachello}} \bibnamefont{and}
  \bibinfo{author}{\bibfnamefont{A.}~\bibnamefont{Arima}},
  \emph{\bibinfo{title}{The Interacting Boson Model}}
  (\bibinfo{publisher}{Cambridge University Press},
  \bibinfo{address}{Cambridge}, \bibinfo{year}{1987}).

\bibitem{iachello1995:vibron}
\bibinfo{author}{\bibfnamefont{F.}~\bibnamefont{Iachello}} \bibnamefont{and}
  \bibinfo{author}{\bibfnamefont{R.~D.} \bibnamefont{Levine}},
  \emph{\bibinfo{title}{Algebraic Theory of Molecules}}
  (\bibinfo{publisher}{Oxford University Press}, \bibinfo{address}{Oxford},
  \bibinfo{year}{1995}).

\bibitem{dieperink1980:ibm-classical}
\bibinfo{author}{\bibfnamefont{A.~E.~L.} \bibnamefont{Dieperink}},
  \bibinfo{author}{\bibfnamefont{O.}~\bibnamefont{Scholten}}, \bibnamefont{and}
  \bibinfo{author}{\bibfnamefont{F.}~\bibnamefont{Iachello}},
  \bibinfo{journal}{Phys. Rev. Lett.} 44 (1980) 1747.

\bibitem{arima1977:ibm2-shell}
\bibinfo{author}{\bibfnamefont{A.}~\bibnamefont{Arima}},
  \bibinfo{author}{\bibfnamefont{T.}~\bibnamefont{Otsuka}},
  \bibinfo{author}{\bibfnamefont{F.}~\bibnamefont{Iachello}}, \bibnamefont{and}
  \bibinfo{author}{\bibfnamefont{I.}~\bibnamefont{Talmi}},
  \bibinfo{journal}{Phys. Lett. B} 66 (1977) 205.

\bibitem{otsuka1978:ibm2-shell}
\bibinfo{author}{\bibfnamefont{T.}~\bibnamefont{Otsuka}},
  \bibinfo{author}{\bibfnamefont{A.}~\bibnamefont{Arima}},
  \bibinfo{author}{\bibfnamefont{F.}~\bibnamefont{Iachello}}, \bibnamefont{and}
  \bibinfo{author}{\bibfnamefont{I.}~\bibnamefont{Talmi}},
  \bibinfo{journal}{Phys. Lett. B} 76 (1978) 139.

\bibitem{otsuka1978:ibm2-shell-details}
\bibinfo{author}{\bibfnamefont{T.}~\bibnamefont{Otsuka}},
  \bibinfo{author}{\bibfnamefont{A.}~\bibnamefont{Arima}}, \bibnamefont{and}
  \bibinfo{author}{\bibfnamefont{F.}~\bibnamefont{Iachello}},
  \bibinfo{journal}{Nucl. Phys. A} 309 (1978) 1.

\bibitem{vanisacker1986:ibm2-limits}
\bibinfo{author}{\bibfnamefont{P.}~\bibnamefont{{Van Isacker}}},
  \bibinfo{author}{\bibfnamefont{K.}~\bibnamefont{Heyde}},
  \bibinfo{author}{\bibfnamefont{J.}~\bibnamefont{Jolie}}, \bibnamefont{and}
  \bibinfo{author}{\bibfnamefont{A.}~\bibnamefont{Sevrin}},
  \bibinfo{journal}{Ann. Phys. (N.Y.)} 171 (1986) 253.

\bibitem{dieperink1982:ibm2-triax}
\bibinfo{author}{\bibfnamefont{A.~E.~L.} \bibnamefont{Dieperink}}
  \bibnamefont{and} \bibinfo{author}{\bibfnamefont{R.}~\bibnamefont{Bijker}},
  \bibinfo{journal}{Phys. Lett. B} 116 (1982) 77.

\bibitem{caprio2004:ibmpn-icnpls04}
\bibinfo{author}{\bibfnamefont{M.~A.} \bibnamefont{Caprio}}, in
  Ref.~\cite{proc-icnpls04}, p. \bibinfo{pages}{215}.

\bibitem{caprio2004:ibmpn}
\bibinfo{author}{\bibfnamefont{M.~A.} \bibnamefont{Caprio}} \bibnamefont{and}
  \bibinfo{author}{\bibfnamefont{F.}~\bibnamefont{Iachello}},
  \bibinfo{journal}{Phys. Rev. Lett.} 93 (2004) 242502.

\bibitem{arias2004:ibmpn-icnpls04}
\bibinfo{author}{\bibfnamefont{J.~M.} \bibnamefont{Arias}},
  \bibinfo{author}{\bibfnamefont{J.}~\bibnamefont{Dukelsky}}, \bibnamefont{and}
  \bibinfo{author}{\bibfnamefont{J.~E.} \bibnamefont{Garc\'ia-Ramos}}, in
  Ref.~\cite{proc-icnpls04}, p. \bibinfo{pages}{127}.

\bibitem{arias2004:ibmpn}
\bibinfo{author}{\bibfnamefont{J.~M.} \bibnamefont{Arias}},
  \bibinfo{author}{\bibfnamefont{J.}~\bibnamefont{Dukelsky}}, \bibnamefont{and}
  \bibinfo{author}{\bibfnamefont{J.~E.} \bibnamefont{Garc\'ia-Ramos}},
  \bibinfo{journal}{Phys. Rev. Lett.} 93 (2004) 212501.

\bibitem{deshalit1963:shell}
\bibinfo{author}{\bibfnamefont{A.}~\bibnamefont{{de-Shalit}}} \bibnamefont{and}
  \bibinfo{author}{\bibfnamefont{I.}~\bibnamefont{Talmi}},
  \emph{\bibinfo{title}{Nuclear Shell Theory}}, No.~\bibinfo{number}{14} in
  \emph{\bibinfo{series}{Pure and Applied Physics}}
  (\bibinfo{publisher}{Academic}, \bibinfo{address}{New York},
  \bibinfo{year}{1963}).

\bibitem{dieperink1983:ibm2-su3star-operators}
\bibinfo{author}{\bibfnamefont{A.~E.~L.} \bibnamefont{Dieperink}}
  \bibnamefont{and} \bibinfo{author}{\bibfnamefont{I.}~\bibnamefont{Talmi}},
  \bibinfo{journal}{Phys. Lett. B} 131 (1983) 1.

\bibitem{walet1987:ibm2-su3star-details}
\bibinfo{author}{\bibfnamefont{N.~R.} \bibnamefont{Walet}} \bibnamefont{and}
  \bibinfo{author}{\bibfnamefont{P.~J.} \bibnamefont{Brussaard}},
  \bibinfo{journal}{Nucl. Phys. A} 474 (1987) 61.

\bibitem{sevrin1987:ibm2-su3star-1}
\bibinfo{author}{\bibfnamefont{A.}~\bibnamefont{Sevrin}},
  \bibinfo{author}{\bibfnamefont{K.}~\bibnamefont{Heyde}}, \bibnamefont{and}
  \bibinfo{author}{\bibfnamefont{J.}~\bibnamefont{Jolie}},
  \bibinfo{journal}{Phys. Rev. C} 36 (1987) 2621.

\bibitem{sevrin1987:ibm2-su3star-2}
\bibinfo{author}{\bibfnamefont{A.}~\bibnamefont{Sevrin}},
  \bibinfo{author}{\bibfnamefont{K.}~\bibnamefont{Heyde}}, \bibnamefont{and}
  \bibinfo{author}{\bibfnamefont{J.}~\bibnamefont{Jolie}},
  \bibinfo{journal}{Phys. Rev. C} 36 (1987) 2631.

\bibitem{leviatan1990:ibm2-modes}
\bibinfo{author}{\bibfnamefont{A.}~\bibnamefont{Leviatan}} \bibnamefont{and}
  \bibinfo{author}{\bibfnamefont{M.~W.} \bibnamefont{Kirson}},
  \bibinfo{journal}{Ann. Phys. (N.Y.)} 201 (1990) 13.

\bibitem{ginocchio1992:ibm2-shapes}
\bibinfo{author}{\bibfnamefont{J.~N.} \bibnamefont{Ginocchio}}
  \bibnamefont{and} \bibinfo{author}{\bibfnamefont{A.}~\bibnamefont{Leviatan}},
  \bibinfo{journal}{Ann. Phys. (N.Y.)} 216 (1992) 152.

\bibitem{caprio2005:ibmpn-nmp04}
\bibinfo{author}{\bibfnamefont{M.~A.} \bibnamefont{Caprio}}, in
  \emph{\bibinfo{booktitle}{Nuclei and Mesoscopic Physics}}, edited by
  \bibinfo{editor}{\bibfnamefont{V.}~\bibnamefont{Zelevinsky}}
  \bibnamefont{\emph{et~al.}} (\bibinfo{publisher}{AIP},
  \bibinfo{address}{Melville, NY}, \bibinfo{year}{in press}).

\bibitem{gilmore1975:multilevel-coherent}
\bibinfo{author}{\bibfnamefont{R.}~\bibnamefont{Gilmore}},
  \bibinfo{author}{\bibfnamefont{C.~M.} \bibnamefont{Bowden}},
  \bibnamefont{and} \bibinfo{author}{\bibfnamefont{L.~M.}
  \bibnamefont{Narducci}}, \bibinfo{journal}{Phys. Rev. A} 12 (1975) 1019.

\bibitem{zhang1990:coherent}
\bibinfo{author}{\bibfnamefont{W.}~\bibnamefont{Zhang}},
  \bibinfo{author}{\bibfnamefont{D.~H.} \bibnamefont{Feng}}, \bibnamefont{and}
  \bibinfo{author}{\bibfnamefont{R.}~\bibnamefont{Gilmore}},
  \bibinfo{journal}{Rev. Mod. Phys.} 62 (1990) 867.

\bibitem{ginocchio1980:ibm-classical}
\bibinfo{author}{\bibfnamefont{J.~N.} \bibnamefont{Ginocchio}}
  \bibnamefont{and} \bibinfo{author}{\bibfnamefont{M.~W.}
  \bibnamefont{Kirson}}, \bibinfo{journal}{Phys. Rev. Lett.} 44 (1980) 1744.

\bibitem{ginocchio1980:ibm-coherent-bohr}
\bibinfo{author}{\bibfnamefont{J.~N.} \bibnamefont{Ginocchio}}
  \bibnamefont{and} \bibinfo{author}{\bibfnamefont{M.~W.}
  \bibnamefont{Kirson}}, \bibinfo{journal}{Nucl. Phys. A} 350 (1980) 31.

\bibitem{dieperink1980:ibm-phase}
\bibinfo{author}{\bibfnamefont{A.~E.~L.} \bibnamefont{Dieperink}}
  \bibnamefont{and} \bibinfo{author}{\bibfnamefont{O.}~\bibnamefont{Scholten}},
  \bibinfo{journal}{Nucl. Phys. A} 346 (1980) 125.

\bibitem{vanisacker1981:ibm-triax}
\bibinfo{author}{\bibfnamefont{P.}~\bibnamefont{{Van Isacker}}}
  \bibnamefont{and}
  \bibinfo{author}{\bibfnamefont{{\identity{Jin-Quan}}.}~\bibnamefont{Chen}},
  \bibinfo{journal}{Phys. Rev. C} 24 (1981) 684.

\bibitem{leviatan1987:ibm-intrinsic}
\bibinfo{author}{\bibfnamefont{A.}~\bibnamefont{Leviatan}},
  \bibinfo{journal}{Ann. Phys. (N.Y.)} 179 (1987) 201.

\bibitem{lopezmoreno1996:ibm-catastrophe}
\bibinfo{author}{\bibfnamefont{E.}~\bibnamefont{L\'opez-Moreno}}
  \bibnamefont{and}
  \bibinfo{author}{\bibfnamefont{O.}~\bibnamefont{Casta{\~n}os}},
  \bibinfo{journal}{Phys. Rev. C} 54 (1996) 2374.

\bibitem{ehrenfest1933:phase-trans}
\bibinfo{author}{\bibfnamefont{P.}~\bibnamefont{Ehrenfest}},
  \bibinfo{journal}{Proc. Amsterdam. Acad.} 36 (1933) 153.

\bibitem{dieperink1984:ibm2}
\bibinfo{author}{\bibfnamefont{A.~E.~L.} \bibnamefont{Dieperink}},
  \bibinfo{journal}{Nucl. Phys. A} 421 (1984) 189c.

\bibitem{bijker1985:ibm2-coherent}
\bibinfo{author}{\bibfnamefont{R.}~\bibnamefont{Bijker}},
  \bibinfo{journal}{Phys. Rev. C} 32 (1985) 1442.

\bibitem{vanegmond1985:ibm2-generator-coordinate}
\bibinfo{author}{\bibfnamefont{A.}~\bibnamefont{{Van Egmond}}}
  \bibnamefont{and} \bibinfo{author}{\bibfnamefont{K.}~\bibnamefont{Allaart}},
  \bibinfo{journal}{Phys. Lett. B} 164 (1985) 1.

\bibitem{bohr1998:v2}
\bibinfo{author}{\bibfnamefont{A.}~\bibnamefont{Bohr}} \bibnamefont{and}
  \bibinfo{author}{\bibfnamefont{B.~R.} \bibnamefont{Mottelson}},
  \emph{\bibinfo{title}{Nuclear Deformations}}, Vol.~\bibinfo{volume}{2} of
  \emph{\bibinfo{series}{Nuclear Structure}} (\bibinfo{publisher}{World
  Scientific}, \bibinfo{address}{Singapore}, \bibinfo{year}{1998}).

\bibitem{rose1957:am}
\bibinfo{author}{\bibfnamefont{M.~E.} \bibnamefont{Rose}},
  \emph{\bibinfo{title}{Elementary Theory of Angular Momentum}}
  (\bibinfo{publisher}{Wiley}, \bibinfo{address}{New York},
  \bibinfo{year}{1957}).

\bibitem{balantekin1983-ibm2-coherent}
\bibinfo{author}{\bibfnamefont{A.~B.} \bibnamefont{Balanktekin}},
  \bibinfo{author}{\bibfnamefont{B.~R.} \bibnamefont{Barrett}},
  \bibnamefont{and} \bibinfo{author}{\bibfnamefont{S.}~\bibnamefont{Levit}},
  \bibinfo{journal}{Phys. Lett. B} 129 (1983) 153.

\bibitem{lipas1990:ibm2-fspin}
\bibinfo{author}{\bibfnamefont{P.~O.} \bibnamefont{Lipas}},
  \bibinfo{author}{\bibfnamefont{P.}~\bibnamefont{von Brentano}},
  \bibnamefont{and} \bibinfo{author}{\bibfnamefont{A.}~\bibnamefont{Gelberg}},
  \bibinfo{journal}{Rep. Prog. Phys.} 53 (1990) 1355.

\bibitem{landau1980:statistical1}
\bibinfo{author}{\bibfnamefont{L.~D.} \bibnamefont{Landau}} \bibnamefont{and}
  \bibinfo{author}{\bibfnamefont{E.~M.} \bibnamefont{Lifschitz}},
  \emph{\bibinfo{title}{Statistical Physics, \textup{Part 1}}},
  Vol.~\bibinfo{volume}{5} of \emph{\bibinfo{series}{Course of Theoretical
  Physics}} (\bibinfo{publisher}{Butterworth Heinemann},
  \bibinfo{address}{Oxford}, \bibinfo{year}{1980}), \bibinfo{note}{translated
  by J. B. Sykes and M. J. Kearsley}.

\bibitem{jolie2001:ibm-o6-phase}
\bibinfo{author}{\bibfnamefont{J.}~\bibnamefont{Jolie}},
  \bibinfo{author}{\bibfnamefont{R.~F.} \bibnamefont{Casten}},
  \bibinfo{author}{\bibfnamefont{P.}~\bibnamefont{von Brentano}},
  \bibnamefont{and} \bibinfo{author}{\bibfnamefont{V.}~\bibnamefont{Werner}},
  \bibinfo{journal}{Phys. Rev. Lett.} 87 (2001) 162501.

\bibitem{gilmore1981:catastrophe}
\bibinfo{author}{\bibfnamefont{R.}~\bibnamefont{Gilmore}},
  \emph{\bibinfo{title}{Catastrophe Theory for Scientists and Engineers}}
  (\bibinfo{publisher}{Wiley}, \bibinfo{address}{New York},
  \bibinfo{year}{1981}).

\bibitem{talmi1993:shell-ibm}
\bibinfo{author}{\bibfnamefont{I.}~\bibnamefont{Talmi}},
  \emph{\bibinfo{title}{Simple Models of Complex Nuclei: The Shell Model and
  Interacting Boson Model}}, Vol.~\bibinfo{volume}{7} of
  \emph{\bibinfo{series}{Contemporary Concepts in Physics}}
  (\bibinfo{publisher}{Harwood Academic Publishers}, \bibinfo{address}{Chur,
  Switzerland}, \bibinfo{year}{1993}).

\bibitem{novoselsky1986:ibm2-o6-xe-ba-pt}
\bibinfo{author}{\bibfnamefont{A.}~\bibnamefont{Novoselsky}} \bibnamefont{and}
  \bibinfo{author}{\bibfnamefont{I.}~\bibnamefont{Talmi}},
  \bibinfo{journal}{Phys. Lett. B} 172 (1986) 139.

\bibitem{dobaczewski1988:hf-pn-quadrupole}
\bibinfo{author}{\bibfnamefont{J.}~\bibnamefont{Dobaczewski}},
  \bibinfo{author}{\bibfnamefont{W.}~\bibnamefont{Nazarewicz}},
  \bibinfo{author}{\bibfnamefont{J.}~\bibnamefont{Skalski}}, \bibnamefont{and}
  \bibinfo{author}{\bibfnamefont{T.}~\bibnamefont{Werner}},
  \bibinfo{journal}{Phys. Rev. Lett.} 60 (1988) 2254.

\bibitem{scholten1982:ibm2-like-qq}
\bibinfo{author}{\bibfnamefont{O.}~\bibnamefont{Scholten}},
  \bibinfo{journal}{Phys. Lett. B} 119 (1982) 5.

\bibitem{scholten1983:ibm2-microscopic-majorana}
\bibinfo{author}{\bibfnamefont{O.}~\bibnamefont{Scholten}},
  \bibinfo{journal}{Phys. Rev. C} 28 (1983) 1783.

\bibitem{vanegmond1984:ibm2-microscopic}
\bibinfo{author}{\bibfnamefont{A.}~\bibnamefont{{Van Egmond}}}
  \bibnamefont{and} \bibinfo{author}{\bibfnamefont{K.}~\bibnamefont{Allaart}},
  \bibinfo{journal}{Nucl. Phys. A} 425 (1984) 275.

\bibitem{scholten1985:ibm2-mss}
\bibinfo{author}{\bibfnamefont{O.}~\bibnamefont{Scholten}},
  \bibinfo{author}{\bibfnamefont{K.}~\bibnamefont{Heyde}},
  \bibinfo{author}{\bibfnamefont{P.}~\bibnamefont{{Van Isacker}}},
  \bibinfo{author}{\bibfnamefont{J.}~\bibnamefont{Jolie}},
  \bibinfo{author}{\bibfnamefont{J.}~\bibnamefont{Moreau}},
  \bibinfo{author}{\bibfnamefont{M.}~\bibnamefont{Waroquier}},
  \bibnamefont{and} \bibinfo{author}{\bibfnamefont{J.}~\bibnamefont{Sau}},
  \bibinfo{journal}{Nucl. Phys. A} 438 (1985) 41.

\bibitem{hartmann1987:ibm2-scissors-systematics}
\bibinfo{author}{\bibfnamefont{U.}~\bibnamefont{Hartmann}},
  \bibinfo{author}{\bibfnamefont{D.}~\bibnamefont{Bohle}},
  \bibinfo{author}{\bibfnamefont{T.}~\bibnamefont{Guhr}},
  \bibinfo{author}{\bibfnamefont{K.}~\bibnamefont{Hummel}},
  \bibinfo{author}{\bibfnamefont{G.}~\bibnamefont{Kilgus}},
  \bibinfo{author}{\bibfnamefont{U.}~\bibnamefont{Milkau}}, \bibnamefont{and}
  \bibinfo{author}{\bibfnamefont{A.}~\bibnamefont{Richter}},
  \bibinfo{journal}{Nucl. Phys. A} 465 (1987) 25.

\bibitem{davydov1958:arm-intro}
\bibinfo{author}{\bibfnamefont{A.~S.} \bibnamefont{Davydov}} \bibnamefont{and}
  \bibinfo{author}{\bibfnamefont{G.~F.} \bibnamefont{Filippov}},
  \bibinfo{journal}{Nucl. Phys.} 8 (1958) 237.

\bibitem{heyde1984:ibm-cubic-triax}
\bibinfo{author}{\bibfnamefont{K.}~\bibnamefont{Heyde}},
  \bibinfo{author}{\bibfnamefont{P.}~\bibnamefont{{Van Isacker}}},
  \bibinfo{author}{\bibfnamefont{M.}~\bibnamefont{Waroquier}},
  \bibnamefont{and} \bibinfo{author}{\bibfnamefont{J.}~\bibnamefont{Moreau}},
  \bibinfo{journal}{Phys. Rev. C} 29 (1984) 1420.

\bibitem{casten1985:ibm-o6-triax}
\bibinfo{author}{\bibfnamefont{R.~F.} \bibnamefont{Casten}},
  \bibinfo{author}{\bibfnamefont{P.}~\bibnamefont{von Brentano}},
  \bibinfo{author}{\bibfnamefont{K.}~\bibnamefont{Heyde}},
  \bibinfo{author}{\bibfnamefont{P.}~\bibnamefont{{Van Isacker}}},
  \bibnamefont{and} \bibinfo{author}{\bibfnamefont{J.}~\bibnamefont{Jolie}},
  \bibinfo{journal}{Nucl. Phys. A} 439 (1985) 289.

\bibitem{smirnov2000:ibm-triax}
\bibinfo{author}{\bibfnamefont{Y.~F.} \bibnamefont{Smirnov}},
  \bibinfo{author}{\bibfnamefont{N.~A.} \bibnamefont{Smirnova}},
  \bibnamefont{and} \bibinfo{author}{\bibfnamefont{P.}~\bibnamefont{{Van
  Isacker}}}, \bibinfo{journal}{Phys. Rev. C} 61 (2000) 041302(R).

\bibitem{warner1983:cqf}
\bibinfo{author}{\bibfnamefont{D.~D.} \bibnamefont{Warner}} \bibnamefont{and}
  \bibinfo{author}{\bibfnamefont{R.~F.} \bibnamefont{Casten}},
  \bibinfo{journal}{Phys. Rev. C} 28 (1983) 1798.

\bibitem{otsuka1985:npbos-manual}
\bibinfo{author}{\bibfnamefont{T.}~\bibnamefont{Otsuka}} \bibnamefont{and}
  \bibinfo{author}{\bibfnamefont{N.}~\bibnamefont{Yoshida}},
  \emph{\bibinfo{title}{User's manual of the program {NPBOS}}},
  \bibinfo{name}{JAERI-M 85-094} (\bibinfo{year}{1985}).

\bibitem{elliott1958:su3-shell-part1}
\bibinfo{author}{\bibfnamefont{J.~P.} \bibnamefont{Elliott}},
  \bibinfo{journal}{Proc. R. Soc. London A} 245 (1958) 128.

\bibitem{otsuka1992:ibm2-mss-t-symmetry}
\bibinfo{author}{\bibfnamefont{T.}~\bibnamefont{Otsuka}},
  \bibinfo{journal}{Hyperfine Interact.} 75 (1992) 23.

\bibitem{bes1959:gamma}
\bibinfo{author}{\bibfnamefont{D.~R.} \bibnamefont{B\`es}},
  \bibinfo{journal}{Nucl. Phys.} 10 (1959) 373.

\bibitem{eisenberg1987:v1}
\bibinfo{author}{\bibfnamefont{J.~M.} \bibnamefont{Eisenberg}}
  \bibnamefont{and} \bibinfo{author}{\bibfnamefont{W.}~\bibnamefont{Greiner}},
  \emph{\bibinfo{title}{Nuclear Models: Collective and Single-Particle
  Phenomena}}, \bibinfo{edition}{3rd} ed., Vol.~\bibinfo{volume}{1} of
  \emph{\bibinfo{series}{Nuclear Theory}} (\bibinfo{publisher}{North-Holland},
  \bibinfo{address}{Amsterdam}, \bibinfo{year}{1987}).

\bibitem{leschber:su3-rotor}
\bibinfo{author}{\bibfnamefont{Y.}~\bibnamefont{Leschber}} \bibnamefont{and}
  \bibinfo{author}{\bibfnamefont{J.~P.} \bibnamefont{Draayer}},
  \bibinfo{journal}{Phys. Lett. B} 190 (1987) 1.

\bibitem{harter1985:ibm2-projection}
\bibinfo{author}{\bibfnamefont{H.}~\bibnamefont{Harter}},
  \bibinfo{author}{\bibfnamefont{A.}~\bibnamefont{Gelberg}}, \bibnamefont{and}
  \bibinfo{author}{\bibfnamefont{P.}~\bibnamefont{von Brentano}},
  \bibinfo{journal}{Phys. Lett. B} 157 (1985) 1.

\bibitem{garciaramos1998:ibm-two-phonon}
\bibinfo{author}{\bibfnamefont{J.~E.} \bibnamefont{Garc{\'i}a-Ramos}},
  \bibinfo{author}{\bibfnamefont{C.~E.} \bibnamefont{Alonso}},
  \bibinfo{author}{\bibfnamefont{J.~M.} \bibnamefont{Arias}},
  \bibinfo{author}{\bibfnamefont{P.}~\bibnamefont{{Van Isacker}}},
  \bibnamefont{and} \bibinfo{author}{\bibfnamefont{A.}~\bibnamefont{Vitturi}},
  \bibinfo{journal}{Nucl. Phys. A} 637 (1998) 529.

\bibitem{iachello2003:y5}
\bibinfo{author}{\bibfnamefont{F.}~\bibnamefont{Iachello}},
  \bibinfo{journal}{Phys. Rev. Lett.} 91 (2003) 132502.

\bibitem{iachello2000:e5}
\bibinfo{author}{\bibfnamefont{F.}~\bibnamefont{Iachello}},
  \bibinfo{journal}{Phys. Rev. Lett.} 85 (2000) 3580.

\bibitem{casten2000:134ba-e5}
\bibinfo{author}{\bibfnamefont{R.~F.} \bibnamefont{Casten}} \bibnamefont{and}
  \bibinfo{author}{\bibfnamefont{N.~V.} \bibnamefont{Zamfir}},
  \bibinfo{journal}{Phys. Rev. Lett.} 85 (2000) 3584.

\bibitem{vanisacker1988:ibm2-m1-rare-earth}
\bibinfo{author}{\bibfnamefont{P.}~\bibnamefont{{Van Isacker}}},
  \bibinfo{author}{\bibfnamefont{P.~O.} \bibnamefont{Lipas}},
  \bibinfo{author}{\bibfnamefont{K.}~\bibnamefont{Helim{\"a}ki}},
  \bibinfo{author}{\bibfnamefont{I.}~\bibnamefont{Koivistoinen}},
  \bibnamefont{and} \bibinfo{author}{\bibfnamefont{D.~D.}
  \bibnamefont{Warner}}, \bibinfo{journal}{Nucl. Phys. A} 476 (1988) 301.

\bibitem{kim1996:ibm2-pd}
\bibinfo{author}{\bibnamefont{\identity{Ka-Hae} Kim}},
  \bibinfo{author}{\bibfnamefont{A.}~\bibnamefont{Gelberg}},
  \bibinfo{author}{\bibfnamefont{T.}~\bibnamefont{Mizusaki}},
  \bibinfo{author}{\bibfnamefont{T.}~\bibnamefont{Otsuka}}, \bibnamefont{and}
  \bibinfo{author}{\bibfnamefont{P.}~\bibnamefont{{von Brentano}}},
  \bibinfo{journal}{Nucl. Phys. A} 604 (1996) 163.

\bibitem{heyde1983:ibm-g-boson-104ru}
\bibinfo{author}{\bibfnamefont{K.}~\bibnamefont{Heyde}},
  \bibinfo{author}{\bibfnamefont{P.}~\bibnamefont{{Van Isacker}}},
  \bibinfo{author}{\bibfnamefont{M.}~\bibnamefont{Waroquier}},
  \bibinfo{author}{\bibfnamefont{G.}~\bibnamefont{Wenes}},
  \bibinfo{author}{\bibfnamefont{Y.}~\bibnamefont{Gigase}}, \bibnamefont{and}
  \bibinfo{author}{\bibfnamefont{J.}~\bibnamefont{Stachel}},
  \bibinfo{journal}{Nucl. Phys. A} 398 (1983) 235.

\bibitem{vanisacker1980:ibm2-ru-pd}
\bibinfo{author}{\bibfnamefont{P.}~\bibnamefont{{Van Isacker}}}
  \bibnamefont{and} \bibinfo{author}{\bibfnamefont{G.}~\bibnamefont{Puddu}},
  \bibinfo{journal}{Nucl. Phys. A} 348 (1980) 125.

\bibitem{duarte1998:ibm2-ru-triax}
\bibinfo{author}{\bibfnamefont{J.~L.~M.} \bibnamefont{Duarte}},
  \bibinfo{author}{\bibfnamefont{T.}~\bibnamefont{Borello-Lewin}},
  \bibinfo{author}{\bibfnamefont{G.}~\bibnamefont{Maino}}, \bibnamefont{and}
  \bibinfo{author}{\bibfnamefont{L.}~\bibnamefont{Zuffi}},
  \bibinfo{journal}{Phys. Rev. C} 57 (1998) 1539.

\bibitem{warner1997:skin-soft-scissors}
\bibinfo{author}{\bibfnamefont{D.~D.} \bibnamefont{Warner}} \bibnamefont{and}
  \bibinfo{author}{\bibfnamefont{P.}~\bibnamefont{{Van Isacker}}},
  \bibinfo{journal}{Phys. Lett. B} 395 (1997) 145.

\bibitem{leviatan1988:vibron-intrinsic}
\bibinfo{author}{\bibfnamefont{A.}~\bibnamefont{Leviatan}} \bibnamefont{and}
  \bibinfo{author}{\bibfnamefont{M.~W.} \bibnamefont{Kirson}},
  \bibinfo{journal}{Ann. Phys. (N.Y.)} 188 (1988) 142.

\bibitem{perez-bernal:INPREP}
\bibinfo{author}{\bibfnamefont{F.}~\bibnamefont{P\'erez-Bernal}}
  \bibnamefont{\emph{et~al.}} (\bibinfo{year}{in preparation}).

\bibitem{iachello1999:polymer-algebraic}
\bibinfo{author}{\bibfnamefont{F.}~\bibnamefont{Iachello}} \bibnamefont{and}
  \bibinfo{author}{\bibfnamefont{P.}~\bibnamefont{Truini}},
  \bibinfo{journal}{Ann. Phys. (N.Y.)} 276 (1999) 120.

\bibitem{marago2000:bec-scissors}
\bibinfo{author}{\bibfnamefont{O.~M.} \bibnamefont{Marag\`o}},
  \bibinfo{author}{\bibfnamefont{S.~A.} \bibnamefont{Hopkins}},
  \bibinfo{author}{\bibfnamefont{J.}~\bibnamefont{Arlt}},
  \bibinfo{author}{\bibfnamefont{E.}~\bibnamefont{Hodby}},
  \bibinfo{author}{\bibfnamefont{G.}~\bibnamefont{Hechenblaikner}},
  \bibnamefont{and} \bibinfo{author}{\bibfnamefont{C.~J.} \bibnamefont{Foot}},
  \bibinfo{journal}{Phys. Rev. Lett.} 84 (2000) 2056.

\bibitem{bohr1982:ibm-coherent-excited-168er}
\bibinfo{author}{\bibfnamefont{A.}~\bibnamefont{Bohr}} \bibnamefont{and}
  \bibinfo{author}{\bibfnamefont{B.~R.} \bibnamefont{Mottelson}},
  \bibinfo{journal}{Physica Scripta} 25 (1982) 28.

\bibitem{bijker1982:ibm-coherent-excited}
\bibinfo{author}{\bibfnamefont{R.}~\bibnamefont{Bijker}} \bibnamefont{and}
  \bibinfo{author}{\bibfnamefont{A.~E.~L.} \bibnamefont{Dieperink}},
  \bibinfo{journal}{Phys. Rev. C} 26 (1982) 2688.

\bibitem{leviatan1985:ibm-o5-coherent}
\bibinfo{author}{\bibfnamefont{A.}~\bibnamefont{Leviatan}},
  \bibinfo{journal}{Z. Phys. A} 321 (1985) 467.

\bibitem{leviatan1990:ibm-ibm2-coherent-excited}
\bibinfo{author}{\bibfnamefont{A.}~\bibnamefont{Leviatan}},
  \bibinfo{journal}{Prog. Part. Nucl. Phys.} 24 (1990) 85.

\bibitem{caprio2005:coherent}
\bibinfo{author}{\bibfnamefont{M.~A.} \bibnamefont{Caprio}},
  \bibinfo{journal}{J. Phys. A} \bibinfo{note}{(in press)}.

\bibitem{graham1994:concrete}
\bibinfo{author}{\bibfnamefont{R.~L.} \bibnamefont{Graham}},
  \bibinfo{author}{\bibfnamefont{D.~E.} \bibnamefont{Knuth}}, \bibnamefont{and}
  \bibinfo{author}{\bibfnamefont{O.}~\bibnamefont{Patashnik}},
  \emph{\bibinfo{title}{Concrete Mathematics: A Foundation for Computer
  Science}}, \bibinfo{edition}{2nd} ed. (\bibinfo{publisher}{Addison-Wesley},
  \bibinfo{address}{Reading, MA}, \bibinfo{year}{1994}).

\bibitem{proc-icnpls04}
\bibinfo{editor}{\bibfnamefont{R.}~\bibnamefont{Bijker}},
  \bibinfo{editor}{\bibfnamefont{R.~F.} \bibnamefont{Casten}},
  \bibnamefont{and} \bibinfo{editor}{\bibfnamefont{A.}~\bibnamefont{Frank}},
  eds., \emph{\bibinfo{title}{Nuclear Physics, Large and Small: International
  Conference on Microscopic Studies of Collective Phenomena}},
  \bibinfo{series}{AIP Conf. Proc.} No. \bibinfo{number}{726}
  (\bibinfo{publisher}{AIP}, \bibinfo{address}{Melville, NY},
  \bibinfo{year}{2004}).

\end{thebibliography}
\providecommand{\ELSEVIER}{}
\ELSEVIER\newcommand{\identity}[1]{{#1}}

\end{document}